\documentclass[twocolumn]{aastex631} 

\usepackage{float} 
\usepackage{amsmath}
\usepackage{epsf}
\usepackage{color}
\usepackage{graphicx}
\usepackage{color}
\usepackage{enumitem}
\usepackage{mathtools}
\usepackage[mathcal]{eucal}
 \let\mathscr\relax
\usepackage[scr]{rsfso}
\usepackage{ulem} 
\usepackage{cancel}
\usepackage{textcase}

\usepackage{tikz}
\usetikzlibrary{arrows,calc,positioning}

\tikzstyle{intt}=[draw,text centered,minimum size=6em,text width=3.25cm,text height=0.34cm]
\tikzstyle{intl}=[draw,text centered,minimum size=2em,text width=2.75cm,text height=0.34cm]
\tikzstyle{int}=[draw,minimum size=2.5em,text centered,text width=3.5cm]
\tikzstyle{intg}=[draw,minimum size=3em,text centered,text width=3.5cm]
\tikzstyle{sum}=[draw,shape=circle,inner sep=2pt,text centered,node distance=3.5cm]
\tikzstyle{summ}=[drawshape=circle,inner sep=4pt,text centered,node distance=3.cm]

\shorttitle{Observed Bar Fraction at $z \sim$ 0.5--4 with {\textit{JWST}} CEERS Data}
\shortauthors{Guo, Jogee, Wise et al.}

\def\gtsim{\lower.5ex\hbox{$\buildrel > \over\sim$}}
\def\gtrsim{\mathrel{\hbox{\rlap{\hbox{\lower4pt\hbox{$\sim$}}}\hbox{$>$}}}}
\def\lesssim{\mathrel{\hbox{\rlap{\hbox{\lower4pt\hbox{$\sim$}}}\hbox{$<$}}}}
\def\ltsim{\lower.5ex\hbox{$\buildrel < \over\sim$}}
\def\simgt{{\raise-.5ex\hbox{$\buildrel>\over\sim$}}\ } 
\def\simlt{{\raise-.5ex\hbox{$\buildrel<\over\sim$}}\ }

\def\farcs{\hbox{$.\!\!^{\prime\prime}$}} 
 
\def\deg{{$^\circ$}}
\def\sun{\mbox{$_\odot$}}

\def\e#1{\underline{#1}}  

\newcommand{\degree}{\hbox{$^\circ$}}

\usepackage{graphicx}	
\usepackage{amsmath}	
\usepackage{amssymb}	

\begin{document}
\title{The Abundance and Properties of Barred Galaxies out to $z \sim$ 4 Using \textit{JWST} CEERS Data}


\author[0000-0002-4162-6523]{Yuchen Guo}
\affiliation{Department of Astronomy, The University of Texas at Austin, Austin, TX, USA}

\author[0000-0002-1590-0568]{Shardha Jogee}
\affiliation{Department of Astronomy, The University of Texas at Austin, Austin, TX, USA}

\author{Eden Wise}
\affiliation{Department of Astronomy, The University of Texas at Austin, Austin, TX, USA}

\author{Keith Pritchett Jr.}
\affiliation{Department of Astronomy, Siena College, Loudonville, NY, USA}

\author[0000-0001-8688-2443]{Elizabeth J.\ McGrath}
\affiliation{Department of Physics and Astronomy, Colby College, Waterville, ME 04901, USA}

\author[0000-0001-8519-1130]{Steven L. Finkelstein}
\affiliation{Department of Astronomy, The University of Texas at Austin, Austin, TX, USA}

\author[0000-0001-9298-3523]{Kartheik G. Iyer}
\altaffiliation{Hubble Fellow}
\affiliation{Columbia Astrophysics Laboratory, Columbia University, 550 West 120th Street, New York, NY 10027, USA}
\affiliation{Center for Computational Astrophysics, Flatiron Institute, New York, NY 10010, USA}


\author[0000-0002-7959-8783]{Pablo Arrabal Haro}
\affiliation{NSF's National Optical-Infrared Astronomy Research Laboratory, 950 N. Cherry Ave., Tucson, AZ 85719, USA}

\author[0000-0002-9921-9218]{Micaela B. Bagley}
\affiliation{Department of Astronomy, The University of Texas at Austin, Austin, TX, USA}

\author[0000-0001-5414-5131]{Mark Dickinson}
\affiliation{NSF's National Optical-Infrared Astronomy Research Laboratory, 950 N. Cherry Ave., Tucson, AZ 85719, USA}

\author[0000-0001-9187-3605]{Jeyhan S. Kartaltepe}
\affiliation{Laboratory for Multiwavelength Astrophysics, School of Physics and Astronomy, Rochester Institute of Technology, 84 Lomb Memorial Drive, Rochester, NY 14623, USA}

\author[0000-0002-6610-2048]{Anton M. Koekemoer}
\affiliation{Space Telescope Science Institute, 3700 San Martin Dr., Baltimore, MD 21218, USA}

\author[0000-0001-7503-8482]{Casey Papovich}
\affiliation{Department of Physics and Astronomy, Texas A\&M University, College Station, TX, 77843-4242 USA}
\affiliation{George P.\ and Cynthia Woods Mitchell Institute for Fundamental Physics and Astronomy, Texas A\&M University, College Station, TX, 77843-4242 USA}

\author[0000-0003-3382-5941]{Nor Pirzkal}
\affiliation{ESA/AURA Space Telescope Science Institute}

\author[0000-0003-3466-035X]{{L. Y. Aaron} {Yung}}
\affiliation{Space Telescope Science Institute, 3700 San Martin Dr., Baltimore, MD 21218, USA}


\author[0000-0001-8534-7502]{Bren E. Backhaus}
\affil{Department of Physics and Astronomy, University of Kansas, Lawrence, KS 66045, USA}

\author[0000-0002-5564-9873]{Eric F.\ Bell}
\affiliation{Department of Astronomy, University of Michigan, 1085 S.\ University Ave., Ann Arbor MI 48109, USA}

\author[0000-0003-0883-2226]{Rachana Bhatawdekar}
\affil{European Space Agency (ESA), European Space Astronomy Centre (ESAC), Camino Bajo del Castillo s/n, 28692 Villanueva de la Ca\~{n}ada, Madrid, Spain}

\author[0000-0001-8551-071X]{Yingjie Cheng}
\affiliation{University of Massachusetts Amherst, 710 North Pleasant Street, Amherst, MA 01003-9305, USA}

\author[0000-0001-6820-0015]{Luca Costantin}
\affiliation{Centro de Astrobiolog\'ia (CSIC-INTA), Ctra de Ajalvir km 4, Torrej\'on de Ardoz, 28850, Madrid, Spain}

\author[0000-0002-6219-5558]{Alexander de la Vega}
\affiliation{Department of Physics and Astronomy, University of California, 900 University Ave, Riverside, CA 92521, USA}

\author[0000-0002-7831-8751]{Mauro Giavalisco}
\affiliation{University of Massachusetts Amherst, 710 North Pleasant Street, Amherst, MA 01003-9305, USA}

\author[0000-0001-6145-5090]{Nimish P. Hathi}
\affiliation{Space Telescope Science Institute, 3700 San Martin Dr., Baltimore, MD 21218, USA}

\author[0000-0002-4884-6756]{Benne W. Holwerda}
\affil{Physics \& Astronomy Department, University of Louisville, 40292 KY, Louisville, USA}

\author[0000-0002-8816-5146]{Peter Kurczynski}
\affil{Astrophysics Science Division Code 660, NASA Goddard Space Flight Center, Greenbelt MD 207781, USA}

\author[0000-0003-1581-7825]{Ray A. Lucas}
\affiliation{Space Telescope Science Institute, 3700 San Martin Dr., Baltimore, MD 21218, USA}

\author[0000-0001-5846-4404]{Bahram Mobasher}
\affiliation{Department of Physics and Astronomy, University of 
  California, 900 University Ave, Riverside, CA 92521, USA}

\author[0000-0003-4528-5639]{Pablo G. P\'erez-Gonz\'alez}
\affiliation{Centro de Astrobiolog\'{\i}a (CAB), CSIC-INTA, Ctra. de Ajalvir km 4, Torrej\'on de Ardoz, E-28850, Madrid, Spain}

\author[0000-0001-9879-7780]{Fabio Pacucci}
\affiliation{Center for Astrophysics $\vert$ Harvard \& Smithsonian, Cambridge, MA 02138, USA}
\affiliation{Black Hole Initiative, Harvard University, Cambridge, MA 02138, USA}


\begin{abstract}

We present {the first estimate} of the observed fraction and properties of bars out to $z \sim 4$ using $\textit{JWST}$ CEERS NIRCam images. We analyze 1770 galaxies with  
$M_\star > 10^{10} M_\odot$ at $0.5 \leq z \leq 4$ and identify barred galaxies \textcolor{black}{from 839 moderately inclined disk galaxies} via ellipse fits and visual classification of both F200W and F444W images. Our results apply mainly to bars with projected semi-major axis $a_{\rm bar}$ $> 1.5 $ kpc ($\sim$ 2 $\times$ PSF in F200W images) that can be robustly traced by ellipse fits.
For such bars, the {observed} bar fraction at $z\sim$ 2--4 is low ($\lesssim 10\%$), and they appear to be emerging at least as early as $z\sim 4$. 
\textcolor{black}{Our observed bar fraction is consistent with the bar fraction predicted by TNG50 simulations for large bars with $a_{\rm bar}$ $> 1.5$ kpc} \textcolor{black}{at $z \sim$ 0.5--4} and with the bar fraction from Auriga simulations \textcolor{black}{out to $z \sim$ 1.5}.
However, TNG50 simulations predict a large population of \textcolor{black}{smaller} bars that our data cannot robustly detect. If \textcolor{black}{such bars} exist, the true bar fraction at $z \sim$ 2--4 may be significantly higher than our results. At $z \ge 1.5$, many barred galaxies show nearby neighbors, suggesting bars may be tidally triggered. {From $z \sim$ 4--0.5,  the observed bar fraction, average projected bar length, and projected bar strength rise.} Our results highlight the early emergence and evolution of barred galaxies and the rising importance of bar-driven secular evolution from $z \sim$~4 to today.

\end{abstract}




\section{Introduction}\label{Sec: Introduction}

 Stellar bars play a crucial role in the secular evolution of disk galaxies. Through torques and shocks, bars can drive gas inflows to the central region, triggering star formation and building bulges and pseudo-bulges (e.g., \citealt{Athanassoula2003, Kormendy&Kennicutt2004, Jogee-Scoville-Kenney2005, Sellwood2016}). Many observational studies present evidence for larger molecular gas concentrations (e.g., \citealt{Sakamoto-etal-1999, Jogee-Scoville-Kenney2005}) or enhanced star formation rates (SFRs) in the circumnuclear region of barred galaxies (e.g., \citealt{Ho-etal-1997, Coelho-Gadotti-2011, Zhou-etal-2015, Lin-Li-Du2020}). Stellar bars can also redistribute mass and angular momentum by interacting with gas, stars, and the dark matter halo (e.g. \citealt{Athanassoula2002,  Saha-Naab2013, Sellwood2016, Collier-Shlosman-Heller2018, Collier-etal-2019}). The connection between bar structures and the presence of active galactic nuclei (AGN) remains unclear, as both theoretical studies (e.g., \citealt{Hopkins-etal-2010, Smethurst-etal-2019, Kataria-etal-2024}) and observational studies present mixed results (e.g., \citealt{Knapen-etal-1995, Buta-Combes1996, Cheung-etal-2015, Galloway-etal-2015, Goulding-etal-2017, Garland-etal-2023}). 
Theoretical and observational studies show that large-scale bars do not directly fuel the AGN as bar-driven gas inflows tend to stall near the { inner Lindblad resonances and do not reach the central regions to fuel the AGN } 
(e.g., \citealt{Buta-Combes1996}).  In order to drive the gas further inward, other mechanisms are needed, such as nuclear bars, dynamical friction on clumps, nuclear spirals, and feedback (e.g., \citealt{Shlosman-etal-1989}; 
see review by \citealt{Jogee2006} and references therein; \citealt{Hopkins-etal-2010}).

\par
To understand the role of bars in galaxy evolution, it is crucial to investigate the emergence and prevalence of bars in disk galaxies at different cosmic times.
Many studies have investigated barred galaxies at  $z \sim$ 0, and the majority of bar studies in the rest-frame near-infrared (NIR) report that $\sim$ 65\% of massive disk galaxies host a 
bar (e.g., \citealt{Eskridge-etal-2000, Marinova-Jogee2007,Menendez-Delmestre-etal-2007,Sheth-etal-2008, Erwin2018}). 

\par
At higher redshifts, the majority of {Hubble Space Telescope ({\textit{HST}})} studies have explored bars in the rest-frame optical light out to $z\sim$~1 (e.g., \citealt{Abraham-etal-1999, Elmegreen-Elmegreen-Hirst2004, Jogee-etal-2004, Sheth-etal-2008, Cameron-etal-2010, Melvin-etal-2014}), and results on how the observed bar fraction (defined as the fraction of disk galaxies that are barred) \textcolor{black}{varies} from little variation to a strong decline. The variations on the observed bar fraction out to  $z \sim$ 1 depend on the stellar mass range selected (\citealt{Cameron-etal-2010, Erwin2018}). The study by \cite{Simmons-etal-2014} represents a first attempt to push the explorations of bars in the rest-frame optical out to $z \sim$~2 and finds the observed bar fraction decreases to $\sim$ 5--10\% at $z\sim$ 2. However, the study faced difficulties identifying bars at  $z>$ 1.5 due to the limited depth and spatial resolution of {\textit{HST} Wide Field Camera 3 (WFC3)} F160W images. Furthermore, \cite{Erwin2018} stresses that the decreasing bar sizes at higher redshifts can help explain some of the decreasing trends of the observed bar fraction.

\par
The sensitive, high-resolution NIRCam images from the {\textit{James Webb Space Telescope}} ({\textit{JWST}}; \citealt{Gardner-etal-2023}) have spawned unprecedented progress in the exploration of bars at  $z \gtrsim$~1.
In our earlier work \cite{Guo-etal-2023}, we presented the first evidence of \textcolor{black}{well-developed} bars above $z\sim$ 2 and showed six examples of robustly identified bars at  $z>1$ in the rest-frame NIR light from F444W images, including two bars with the highest spectroscopic redshifts known at the time ($z\sim$~2.136  and 2.312). 
The six example barred galaxies shown in \cite{Guo-etal-2023} have a range of  SFRs, and most of them have possible nearby companions, raising the tantalizing possibility of tidally triggered bars at early epochs. Using $\textit{JWST}$ NIRCam imaging, \cite{Costantin-etal-2023} 
report a barred spiral galaxy at a photometric redshift $z_{phot}$ $\sim$ 3 when the Universe was only two billion years old. \cite{Huang-etal-2023} report a discovery of a barred dusty star-forming spiral galaxy with a spectroscopic redshift $z=2.467$, confirmed both by NIRCam imaging and ALMA detection. \textcolor{black}{\cite{McKinney-etal-2024} study the morphology of heavily dust-attenuated sub-millimeter galaxies in using \textit{JWST} NIRCam imaging from COSMOS-web (\citealt{Casey-etal-2023}). They identify possible bar candidates out to $z \sim 3.4$.} Recent studies also report ``bar-like'' structures in lensed galaxies: \cite{Smail-etal-2024} report an ultramassive lensed galaxy at $z_{spec}  = 4.26$ that has a ``bar-like'' structure in the central regions of the galaxy, and \cite{Amvrosiadis-etal-2024} report detection of the onset of bar formation in a gas-rich lensed galaxy at $z_{phot} \sim 3.8$.

\par
While the above individual cases of barred galaxies at early epochs are exciting, it is now time to systematically quantify the abundance of barred galaxies by measuring the observed bar fraction at $z>1$. The recent paper by \cite{Le-Conte-etal-2024} (hereafter LC24) explores the rest-frame NIR bar fraction at $z \sim$ 1--3 using {\textit{JWST}} NIRCam F444W images. \textcolor{black}{We note that two recently submitted studies, G\'eron et al. (in prep.) and \cite{Huertas-Company-etal-2025}, also present the observed bar fraction out to $z \sim 4$ or $z \sim 5$ using $\textit{JWST}$ data. Our results are consistent with their observed bar fraction. In this paper, we focus on comparing our results with LC24. }

\par
Complementing LC24, in this work, we attempt to study the abundance and properties of bars at $z \sim$ 0.5--4. Specifically, LC24 mainly used four {\textit{JWST}} NIRCam pointings from the Cosmic Evolution Early Release Science Survey (CEERS; \citealt{Finkelstein-etal-2022}) and visually inspected F444W images and radial profiles of ellipticity and position angle to identify barred galaxies at $1 \le z \le3$.  Our study explores the bars across a wider redshift range of $z\sim$ 0.5--4 and uses all 10 CEERS NIRCam pointings to derive a large sample of 1770 galaxies with stellar mass $M_{\star}>10^{10}M_{\odot}$. We identify barred galaxies by applying two methods -- visual classification and {quantitative criteria applied to ellipse fits} -- to both F200W and F444W images. This allows us to leverage the sharper spatial resolution in F200W images to identify smaller bars and the longer rest-frame wavelength traced by F444W images to unveil obscured dusty bars and other stellar features.

\par
Alongside the observational breakthroughs, there have been major advances on the theoretical front. New high-resolution cosmological simulations (e.g., \citealt*{Kraljic-Bournaud-Martig2012}; \citealt{Scannapieco-Athanassoula2012, Bonoli-etal-2016, Spinoso-etal-2017, Algorry-etal-2017, Fragkoudi-etal-2021, Rosas-Guevara-etal-2020, Rosas-Guevara-etal-2022, Bi-Shlosman-Romano-Diaz2022, Fragkoudi-etal-2025}) are now probing the growth of bars and their impact on galaxy evolution out to at least $z \sim $~4. {The predictions on the bar fraction are widely different: some simulations find that bars only appear after $z \sim 1$ (e.g., \citealt{Kraljic-Bournaud-Martig2012, Algorry-etal-2017}), while others find that bars already exist at $z \gtrsim 3$ (e.g., \citealt{Rosas-Guevara-etal-2022, Bi-Shlosman-Romano-Diaz2022, Fragkoudi-etal-2025}).} The question of whether bars can grow in early disks that might be dynamically hot, and the impact of the dark matter fraction, feedback model, and tidal interactions on bar formation are topics of active discussion (e.g., \citealt{Ghosh-etal-2023};  \citealt{Rosas-Guevara-etal-2022, Bi-Shlosman-Romano-Diaz2022}; \citealt{Bland-Hawthorn-etal-2023};\citealt{Ceverino-etal-2023, Bland-Hawthorn-etal-2024}; Dekel, A., private communication).

\par
In this paper, we push the studies of barred galaxies for the first time out to $z\sim$ 4 when the Universe was \textcolor{black}{$\sim$ 11\%} of its present age. We provide a comprehensive estimate of the observed bar fraction and the observed properties of stellar bars at $z\sim$ 0.5--4.0, using \textit{JWST} NIRCam F200W and F444W images from the CEERS survey. {Our results only apply to bars with projected semi-major axis $a_{\rm bar}$ $> 1.5$ kpc ($\sim$ 2 $\times$ the full width at half-maximum (FWHM) of PSF in F200W images) that can be traced by ellipse fits of our data.} The observational data and sample selection are outlined in \S~\ref{Sec: observation and data}. In \S~\ref{method} we describe our multi-pronged methodology to identify disk galaxies and bars using ellipse fits, and visual classification. 
In \S~\ref{sec:results}, we present our results on the observed bar fraction using two different bar identification methods (ellipse fits and visual classification) applied to both F200W and F444W images. In section \S~\ref{sec:bar-measurement}, we outline the properties (e.g., strength, sizes) of the observed bars. {In \S~\ref{sec:vs-theory}, we compare the observed bar fraction to results from the TNG50 hydrodynamical simulations \textcolor{black}{and the Auriga simulations} to see how theoretical predictions compare to empirical results.}
In \S~\ref{sec:towards-intrinsic}, we discuss how the observed bar fraction provides a lower limit to the intrinsic bar fraction due to systematic effects causing us to miss a subset of bars.  
In \S~\ref{discussion}, we discuss the implications of our results for the early emergence of barred galaxies, theoretical models of disk galaxies, and bar-driven secular evolution of galaxies. While this paper focuses mainly on {the observed bar fraction and properties}, future work will explore the relationship between bars and galaxy properties (e.g., SF, bulges, AGN, and presence of companions) using a control sample of unbarred galaxies.

\par
In this paper, we assume the latest {\textit Planck} flat $\Lambda$CDM cosmology with H$_{0}=$67.36, $\Omega_m=$0.3153, and $\Omega_{\Lambda}=$0.6847 \citep{Planck-etal-20}.  All magnitudes are in the absolute bolometric system \citep[AB; ][]{Oke-Gunn1983}.

\section{Observational data and sample selection} \label{Sec: observation and data}

\subsection{CEERS NIRCam Imaging}\label{Sec: CEERS Imaging}

\par
 CEERS (\citealt{Finkelstein-etal-2022}) is an early-release science program that surveys 50\% of the Extended Groth Strip (EGS) previously covered by the Cosmic Assembly Near-infrared Deep Extragalactic Legacy Survey (CANDELS;  \citealt{Grogin-etal-2011, Koekemoer-etal-2011}). 
 We focus on the 10 NIRCam pointings covering roughly {88} square arcminutes {(corresponding to a comoving volume of $\sim$ 960000 Mpc$^3$ at $z \sim$ 0.5--4)} of the area of the CANDELS EGS in seven bands: F115W, F150W, F200W, F277W, F356W, F410M, and F444W. The total exposure time for pixels observed in all three dithers was typically 2835 s per band, corresponding to a 5$\sigma$ depth ranging from 28.8 to 29.7 \citep{Bagley-etal-2023}. 

\par
 CEERS latest public data release 0.5 (including NIRCam pointings 1, 2, 3, 6) and data release 0.6 (including NIRCam pointings 4, 5, 7, 8, 9, 10) are used in this study. The reduction of the NIRCam images is performed using version 1.8.5 of the \textit{JWST} Calibration Pipeline with some custom modifications. For details on the reduction
steps, see \citep{Bagley-etal-2023}. The final mosaics for each pointing in all bands have pixel scales of 0\farcs03/pixel. The publicly released mosaics are available at \url{ceers.github.io/dr06.html} and on MAST via \dataset[10.17909/z7p0-8481]{\doi{10.17909/z7p0-8481}}. {We used an internal photometry catalog in this study, and the catalog} was produced by using \textsc{Source Extractor} \citep[][]{Bertin-Arnouts96} v2.25.0 in two image mode, with an inverse-variance weighted combination of the PSF-matched F277W and F356W images as the detection image, and photometry measured on all seven bands. 

\subsection{Sample Selection}\label{sec:sample selection}
\par
We start with a sample of galaxies with stellar mass $M_{\star} > 10^{10} M_{\odot}$ at 0.5~$\le z \le $~4.
{Sources are identified in \textcolor{black}{an internal photometry catalog with photometric redshifts provided by the CEERS team}. The same catalog is used and discussed in \cite{Finkelstein-etal-2024-internal-catalog}}. 
The authors of the catalog estimate photometric redshifts were estimated with \texttt{eazy} (\citealt{Brammer-et-al-2008}), following the methodology in \cite{Finkelstein-et-al-2023}, and including the additional templates from \cite{Larson-et-al-2023}.

\par
We use the stellar mass measurements provided by the CEERS team (Iyer et al. in prep.). We briefly summarize below the approach used by the authors: the authors of the catalog estimate stellar masses are estimated for all objects in the catalog using the \textsc{Dense Basis} \footnote{\href{https://dense-basis.readthedocs.io/}{https://dense-basis.readthedocs.io/}} \citep{Iyer-etal-2019}. 
The code performs a fully Bayesian inference of the {star formation history (SFH)}, dust attenuation, and chemical enrichment for each galaxy using a fully non-parametric Gaussian process-based description for the SFH described in \cite{Iyer-etal-2019}. The model uses a Chabrier IMF \citep{Chabrier-2003}, Calzetti dust attenuation law \citep{Calzetti-etal-2001}, and Madau IGM absorption \citep{Madau-1995}. Setting explicit priors in SFH space using non-parametric SFHs has been shown to be robust against outshining due to younger stellar populations that otherwise bias estimates of masses and star formation rates (e.g., \citealt{Iyer-Gawiser-2017,Leja-etal-2019,Lower-etal-2020}). 

\par
The deep CEERS NIRCam imaging has 5$\sigma$ depth ranging from 28.8 to 29.7 AB magnitude across all bands and pointings (\citealt{Bagley-etal-2023}).
Using the approach described in \cite{Pozzetti-etal-2010}, {we estimate that $10^{10}$ $M_{\odot}$ is well above the 90\% stellar mass complete limit at $0.5\le z \le 4$}. Hence, our sample of  $M_{\star} > 10^{10} M_{\odot}$ galaxies is mass-complete. 

\par
The stellar mass and redshifts selection results in a sample of 1770 massive galaxies
with $M_{\star} > 10^{10} M_{\odot}$ at  redshifts  $z\sim$~0.5--4, with 328 (0.5~$\le z < $~1.0), 448 (1.0~$\le z < $~1.5), 395 (1.5~$\le z < $~2.0), 393 (2.0~$\le z < $~3.0), and  206  (3.0~$\le z \le $~4.0) galaxies in each redshift bin. 

\par
{We supplement
photometric redshifts with available published spectroscopic
redshifts in EGS (N. Hathi 2022, private communication) and spectroscopic redshifts derived from \textit{JWST} NIRSpec observations (\citealt{Arrabal-Haro-etal-2023}; Arrabal Haro et al. 2024, in prep.). We find 
$\sim$~55\% of sources have an available spectroscopic redshift. }
If a source has more than one spectroscopic redshift measurement, we choose the
one with the highest quality. Using the sources with known spectroscopic redshifts, we estimate the normalized median absolute deviation ($\sigma_{\rm NMAD}$\footnote{The normalized median absolute deviation assesses the quality of the photometric redshifts, with $\sigma_{\rm NMAD}$= $1.48\times \rm{median}(|\Delta z- \rm{median}(\Delta z)|/(1+z_{spec}))$ and $\Delta z  = z_{\rm{phot}} - z_{\rm{spec}}$ (\citealt{Brammer-et-al-2008}).}) of photometric redshifts to be $\sim$~0.04. 
We note that sources with spectroscopic redshifts are mainly $z \lesssim 3$ sources and are, on average, $\sim$ 1 magnitude brighter (in the F356W band) compared to the rest of the sources that only have photometric redshifts.

\section{Methodology}\label{method}

\begin{figure}
    \begin{tikzpicture}[
      >=latex',
      auto
    ]
      \node [intg] (kp)  {All Galaxies};
      \node [int]  (ki1) [node distance=1.25cm,below =of kp] {Identifying Disk Galaxies via Visual Classification};
      \node [intg] (ki3) [node distance=5cm,below of=kp] {Remove Inclined Disk Galaxies (i $>$ 65$^\circ$)}; 
      \node [intg] (ki4) [node distance=2cm and -1.5cm,below left=of ki3] {Identifying Barred Galaxies using Ellipse Fits};
      \node [intg] (ki5) [node distance=2cm and -1.5cm,below right=of ki3] {Identifying Barred Galaxies via Visual Classification};
      \draw[->] (kp) -- ($(kp.south)+(0,-0.75)$) -| (ki1) node[above,pos=0.25] {} ;
      \draw[->] (ki1) -- ($(ki1.south)+(0,-0.75)$) -| (ki3) node[above,pos=0.25] {} ;
      \draw[->] (ki3) -- ($(ki3.south)+(0,-0.75)$) -| (ki4) node[above,pos=1.4] {} ;
      \draw[->] (ki3) -- ($(ki3.south)+(0,-0.75)$) -| (ki5) node[above,pos=0.4] {} ;
    \end{tikzpicture}
    \caption{Methodology for identifying barred disk galaxies. See \S~\ref{method} for details.}
    \label{fig: flowchart-strategy}
  \end{figure}
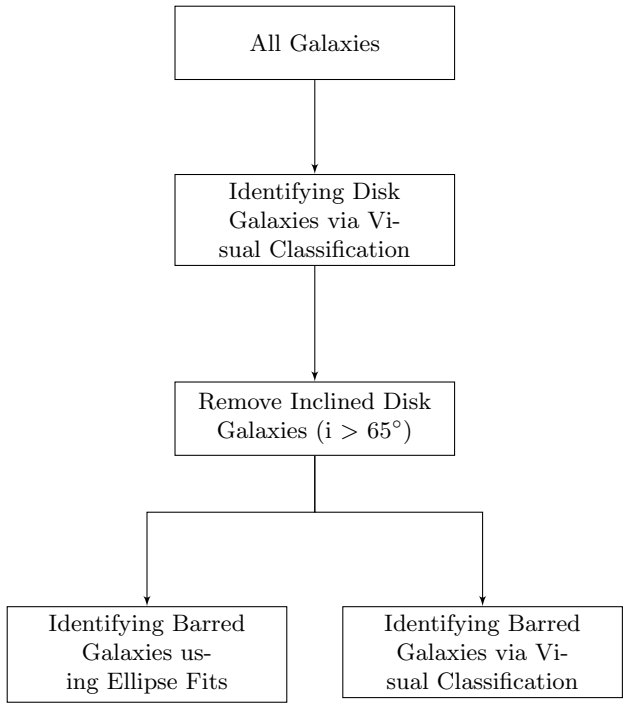
\par
In this work, we refer to the bar fraction directly measured from observations as the ``observed bar fraction" to distinguish it from the true or ``intrinsic bar fraction" discussed in \S~\ref{sec:towards-intrinsic} of this paper. The intrinsic and observed bar fractions may differ due to many reasons, including systematic effects that cause a subset of bars to be missed in the observations. The observed bar fraction in this paper is analogous to the ``bar fraction" reported in previous observational studies.
\par
The observed bar fraction $f_{\rm bar}$ is defined as the fraction of disk galaxies that are observed to have stellar bars:
\begin{equation}
\centering
    f_{\rm bar} = \frac{N_{\rm bar}}{N_{\rm disk}}
    \label{eq:fbar}
\end{equation} where $N_{disk}$ is the sample size of the moderately inclined disk galaxies and $N_{bar}$ is the number of barred galaxies identified {in} the sample of moderately inclined disk galaxies. 

Disk galaxies are considered to be systems with an extended outer disk. In the local $z \sim 0 $ universe, disk galaxies can be further classified into Hubble types, including S0, Sa, Sb, Sc, Sd, and Sm/ Irr. At higher redshifts, Hubble types are not well-defined, and therefore we do not further classify disk galaxies into  Hubble types. 
We discuss the identification of disk galaxies in \S~\ref{sec:sel-disk} and the identification of bars in \S~\ref{sec:Method-bar-efit}.

\subsection{Sample of Moderately Inclined Disk Galaxies}\label{sec:sel-disk}
\par 
We first select disks of all inclinations from visual classification in \S~\ref{sec:visual classification} and then apply the axis ratios obtained in \S~\ref{sec:structual parameters} to select the sample of moderately inclined ($i \le 65^\circ$) disk galaxies. 

\subsubsection{Identification of Disk Galaxies via Visual Classification}\label{sec:visual classification}
\par
The primary goal of visual classification is to identify disk galaxies (including disks of all inclinations). 

\par
Three authors of this study (Y.G., E.W., K.P.) classified all the 1770 galaxies. Before visual classifications, we trained our classifiers with detailed materials, including the latest results from recent \textit{JWST} studies (e.g., \citealt{Ferreira-etal-2022a, Jacobs-etal-2022, Kartaltepe-etal-2023}).
Each classifier was asked to visually inspect
postage stamps of seven 4\farcs 0 $\times$ 4\farcs 0 cutouts in NIRCam F115W, F150W,  F200W, F277W, F356W, F410M and F444W images, for each of our 1770 sample galaxies at  0.5~$\le z \le $~4. {At 0.5~$\le z \le $~4, postage stamps show the rest-frame optical and NIR morphology for each source, allowing the classifiers to utilize as much information as possible.}
The classifiers are asked to select one main morphology class for each galaxy from the following options:

\begin{itemize}

    \item \textbf{Disk Galaxies}: This category contains all galaxies that have an extended outer disk component. The outer disk is identified as an outer component with a slowly declining surface brightness profile. It surrounds the central region of the galaxy, which can often host bulges with a more steeply declining surface brightness profile.
    Galaxies in this category can contain other components (e.g., a spheroid component, such as a classical bulge) or be weakly interacting (disks are slightly distorted but are still identifiable and well-resolved). Inclined disk galaxies are included in this category. 
    \vspace{1mm}

    \item \textbf{Pure Spheroidal Galaxies}: The category contains galaxies that are only spheroids without an outer disk. An example at $z \sim 0$ would be a classical elliptical galaxy.
    \vspace{1mm}
    
    \item \textbf{Strongly Asymmetric/Distorted Galaxies}: The category contains galaxies that are dominated by strongly distorted or asymmetric features. These are typically indicative of strongly interacting galaxies at these epochs.
    \vspace{1mm}
    
    \item \textbf{Unclassifiable Sources}: This category includes unresolved sources, sources located in regions with image defects in F200W or F444W images, over-deblended sources, or artifacts.
    
\end{itemize}

\par
A galaxy is considered to be a visually classified disk galaxy {if at least two out of three} classifiers classify it in the ``Disk Galaxies'' category. The number of disks selected from visual classifications are shown in Figure \ref{fig:disk fraction} (dark blue bars), 
corresponding to a disk fraction {of $\sim$ 74\%, 75\%, 70\%, 66\%, and 63\% at $z\sim$ 0.5--1.0, 1.0--1.5, 1.5--2.0, 2.0--3.0 and 3.0--4.0, respectively.}
 
These values are reasonably consistent with the fraction of disk galaxies reported in previous studies {(e.g., \citealt{Tamburri-etal-2014, Kartaltepe-etal-2023, Ferreira-etal-2022b, Lee-etal-2023, Huertas-Company-etal-2024})},
taking { into account the different stellar mass ranges used and the variations in criteria for classifying a disk galaxy in the aforementioned studies}.

\subsubsection{Characterizing Disk Galaxies via S\'{e}rsic Fits} \label{sec:structual parameters}

\par
We use the structural measurements measured using {\sc galfit} \citep{Peng-et-al-2010} from McGrath et al. (in prep.). Here, we briefly summarize the approach used by the authors: {\sc galfit}  v3.0.5 was run individually on each of the six broadband bands using the background-subtracted version of the CEERS imaging mosaics.
Empirical PSFs, generated by stacking stars across all CEERS fields \citep{Finkelstein-etal-2022}, were used as input to {\sc galfit}. Corresponding thumbnails of the ERR array (see \citet{Bagley-etal-2023} for details) are used as input noise maps (i.e., ``sigma images'') for each source in {\sc galfit}. 
Single component S\'{e}rsic fitting was performed for all selected sources with stellar mass $\gtrsim$ 10$^7 M_{\odot}$. During fitting, the background value was fixed at zero, and constraints were applied.

\par
\textcolor{black}{We use the axis ratio measurements ($b/a$) from \textcolor{black}{single component S\'{e}rsic fits ({McGrath et al. (in prep.)})} to estimate the inclination ($i$) of each galaxy, using the conversion $cos^2 i = (b/a)^2$.  We assume the intrinsic {axis ratio of a disk galaxy to be 1}. We define a galaxy to have a moderate inclination ($i \le 65^{\circ}$) in our sample if the disk galaxy has an axis ratio $>$ 0.42 in at least one of the F277W, F356W, and F444W fits. Previous bar studies commonly adopt $i > 65^{\circ}$ (e.g., \citealt{Diaz-Garcia-etal-2016, Erwin2018} or $i > 60^{\circ}$ (e.g., \citealt{Jogee-etal-2004, Sheth-etal-2008, Guo-etal-2023}) as the inclination cut.}

\par
\begin{figure}
    \includegraphics[width=0.47\textwidth]{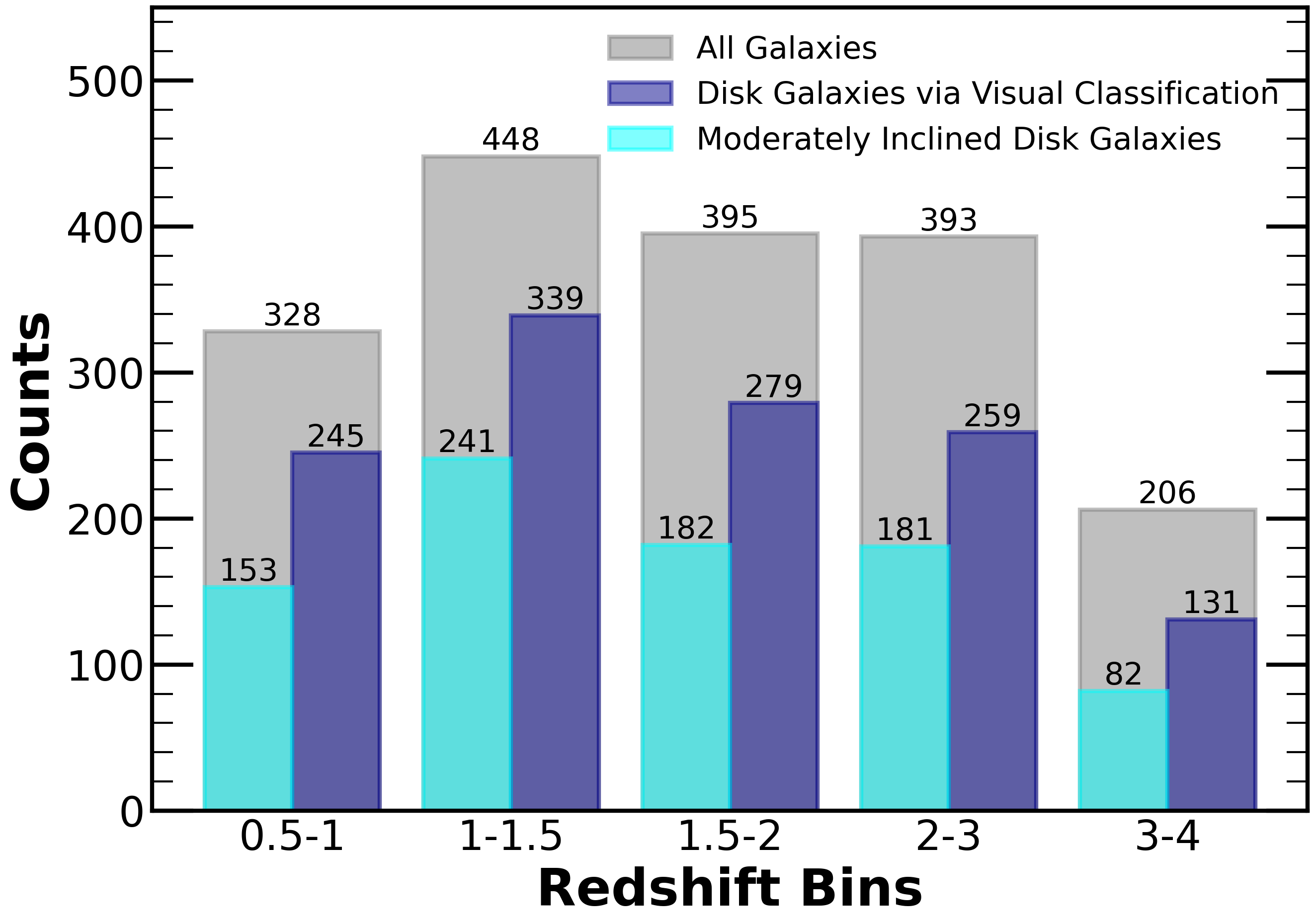}

    \caption{Number of disk galaxies and moderately inclined disk galaxies in each redshift bin. This figure illustrates the classification of disk galaxies following the methodology in \S~\ref{sec:sel-disk} and shows the number of galaxies during different stages of the classification process. We start with all galaxies (\textcolor{black}{gray}), visually identify disk galaxies (dark blue), and sub-select moderately inclined disk galaxies with $i \le 65{\degree}$ (cyan). In each redshift bin, about one-third of disk galaxies are removed 
    because of high inclination ($i >65{\degree}$). }

\label{fig:disk fraction}
\end{figure}

Using the \textcolor{black}{fits}, we exclude disk galaxies with inclination $i >$ 65$^{\circ}$ from the visually classified disk galaxies sample. This results in the exclusion of approximately one-third of disk galaxies in each redshift bin. 
The resulting number of moderately inclined disk galaxies in each redshift bin is shown in Figure \ref{fig:disk fraction} (cyan bars). We obtain a sample of 839 moderately inclined disk galaxies.

\par

\textcolor{black}{In this paper, we identify disks via visual classification (\S\ref{sec:visual classification}). We perform an additional test on our methodology by inspecting the distribution of S\'{e}rsic indices from McGrath et al. (in prep.) for our sample of disk galaxies. We find the distribution is similar to that in previous studies }(e.g., \citealt{Kartaltepe-etal-2023}) with the median S\'{e}rsic index at {n $\sim$ 1.6$^{+0.8}_{-0.6}$, with the range defined by the 25th and 75th percentiles.}

\par

\subsection{Bar Identification}\label{sec:Method-bar-efit}
\par
Stellar bars can be identified via different methods such as visual classification (e.g., \citealt{RC3, Masters-etal-2010,Simmons-etal-2014}), Fourier decomposition (e.g., \citealt{Athanassoula-etal-1990, Rosas-Guevara-etal-2020,Rosas-Guevara-etal-2022}), 2-D structural decomposition of the galaxy surface brightness distribution (e.g., \citealt{Laurikainen-Salo-Buta2005, Weinzirl-etal-2009, Gadotti-etal-2015}) or ellipse fits of galaxy images (e.g., \citealt{Jedrzejewski1987, Wozniak-etal-1995, Jogee-Kenney-Smith1999,Jogee-etal-2002a, Jogee-etal-2004, Knapen-Shlosman-Peletier2000,Elmegreen-Elmegreen-Hirst2004,Erwin2005,Marinova-Jogee2007,Menendez-Delmestre-etal-2007,Sheth-etal-2008, Guo-etal-2023}). {In this paper, we adopt two methods, ellipse fits and visual classification, and we discuss the details for each method in \S~\ref{sec:bar-efit-criteria} and \S~\ref{sec:bar-vis-criteria} respectively.}

\subsubsection{Leveraging Both F200W and F444W Images} \label{sec: levarage-200and444}
\par
For this work, we use both F200W and F444W images as the sharper spatial resolution in F200W images allows us to identify smaller bars while the longer rest-frame wavelength traced by F444W images \textcolor{black}{can unveil dust-obscured stellar features.}

\par
\par
At $z \sim$ 0.5 to 4, the rest-frame wavelengths traced by F444W images range from $\sim$ 0.9 $\rm \mu m$ to $\sim$ 3 $\rm \mu m$ (see Table \ref{tab:rest-frame-and-psf}). The rest-frame optical-red light and NIR light are effective tracers of stellar mass and are less affected by dust and star formation, {so they can better trace stellar structures than rest-frame UV or optical-blue light (e.g., \citealt{Frogel-Quillen-Pogge1996, Meidt-etal-2014, Suess-etal-2022, Ren-etal-2024})}. Figure 1 in \cite{Guo-etal-2023} illustrates the effect of changing the rest-frame wavelength, in which the bar structure appears more evident in its F444W image compared to shorter wavelength bands.
Furthermore, previous studies at  $z \sim$ 0 have shown that the NIR bar fraction is higher than the optical bar fraction (e.g., \citealt{Marinova-Jogee2007, Menendez-Delmestre-etal-2007}).

\begin{deluxetable}{ccccc}
\tablewidth{800pt}
\tablenum{1}
\tablecaption{Rest-frame wavelengths and physical sizes of PSF in F200W and F444W images. 
\label{tab:barprop}}

\decimalcolnumbers
\tablehead{
\colhead{ Redshift} & \multicolumn{2}{c}{F200W}  &
\multicolumn{2}{c}{F444W} \\
\cline{1-3} \cline{4-5} 
\colhead{ $z$} & \colhead{ $\lambda_{\rm rest}$ (\AA)} & \colhead{PSF (kpc) } &
\colhead{$\lambda_{\rm rest}$ (\AA)} &  \colhead{PSF (kpc)  }}

\decimalcolnumbers
\startdata
0.75 & 11428.57 &0.61 & 25371.43& 1.23\\
1.25 &  8888.89 & 0.69&  19733.33 & 1.4\\
1.75 & 7272.73  & 0.70& 16145.45 &1.41\\
2.5 &  5714.29  & 0.67& 12685.71&1.35 \\
3.5 &  4444.44  & 0.60& 9866.67& 1.22\\
\enddata

\tablecomments{Columns are (1) The median redshift in each redshift bin; (2) The rest-frame wavelength of the F200W image 
It ranges from the rest-frame NIR at $z \sim$ 0.75 to the rest-frame optical-blue at $z \sim$ 3.5; (3) The physical size corresponding to the PSF FWHM ($\sim$ 0\farcs 08) of the F200W image; 
(4) The rest-frame wavelength of the F444W image, corresponding to the rest-frame NIR at $z\sim$ 0.75--3.5; (5) The physical size corresponding to the PSF FWHM ($\sim$ 0\farcs {163}) of the F444W image. It ranges from 1.2 to 1.4 kpc, {showing the challenge of detecting
{$\lesssim 2.4$ kpc} bars in the F444W images.} \textcolor{black}{In each band, the physical size of the PSF FWHM shows small variances across $z \sim 0.5-4$, mitigating the resolution effects caused by angular size -- redshift relation.}}
    \label{tab:rest-frame-and-psf}
\vspace{-1cm}
\end{deluxetable}


\par
However, the FWHM of the PSF in F444W images is 0\farcs163, providing a spatial resolution of $\sim$ 1.2--1.4 kpc across $z \sim$ 0.5 to 4 (see Table \ref{tab:rest-frame-and-psf}), which implies that short stellar bars will likely be missed in our classifications as {roughly two resolution elements ($\gtrsim$ 2.4 kpc)} are generally required along the semi-major length of the bar for a robust \textcolor{black}{bar identification (e.g., \citealt{Erwin2018, Liang-etal-2023})}.  
Disks are found to have smaller sizes at higher redshifts (e.g., \citealt{Van-ver-Wel-etal-2014, Suess-etal-2022}), and if bars exist in such disks,  they are expected to have shorter lengths. Consequently, a fraction of bars may be missed in the F444W images, and we will underestimate the observed bar fraction. Therefore, as a complementary approach, we choose to estimate the bar fraction in the F200W images, in which the PSF FWHM is only 0\farcs08 (corresponding to $\sim$ 0.6--0.7 kpc across $z \sim$ 0.5--4, see Table \ref{tab:rest-frame-and-psf}). {As ellipse fits can robustly detect bars with projected semi-major axis $a_{\rm bar}\gtrsim$ 2 $\times$ PSF FWHM, with F200W images, we expect to robustly detect $a_{\rm bar}>$ 1.5 kpc.} We note that across the redshift range $z \sim$ 4 to 0.5,  the rest-frame wavelength in F200W images changes from $\gtrsim$ 4000\AA~to NIR, thereby tracing different stellar populations. We do not attempt to go beyond $z\sim 4$ because at these higher redshifts,  our \textit{JWST} NIRCam F200W images would trace the rest-frame UV light, which is not an effective tracer of the stellar potential and stellar bars, and primarily traces hot young massive stars.

\par

\subsubsection{Bar Identification via Ellipse {Fits}}\label{sec:bar-efit-criteria}

\par
The detailed methodology of identifying barred galaxies via ellipse fits is described in
\cite{Jogee-etal-2004} and \cite{Marinova-Jogee2007}. This method has already been proven effective for bars at higher redshifts in our pilot study \cite{Guo-etal-2023}. In this work, we followed the methodology based on \cite{Guo-etal-2023}. Here we briefly summarize the main steps: 
\par
\begin{figure*}
    \centering
    \includegraphics[width=0.8\textwidth]{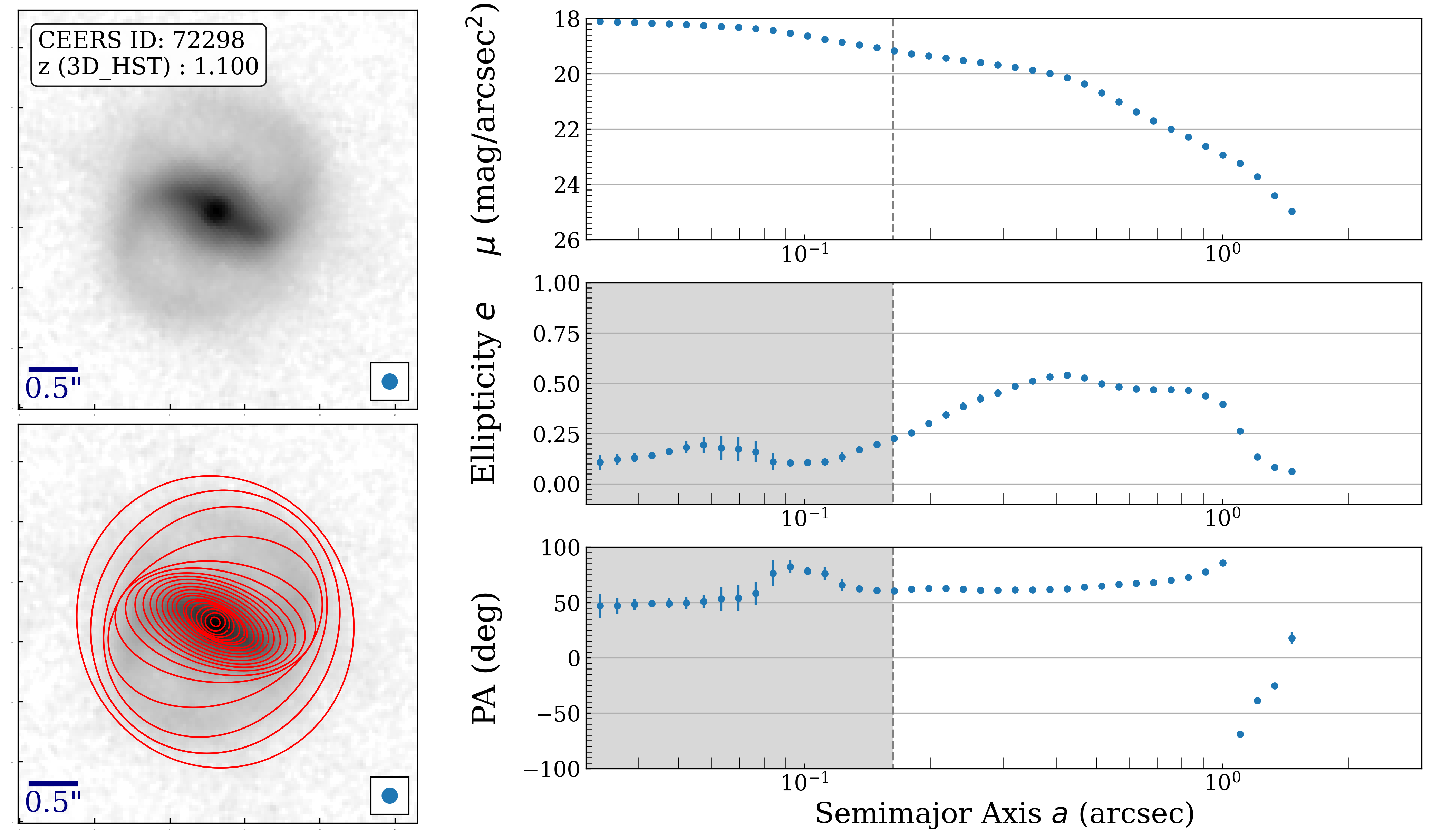}
    \caption{Ellipse fits to the {\textit{JWST}} NIRCam F444W image of CEERS 72298 at $z \sim 1.1$. 
The left panel for each galaxy shows 
the F444W image alone (top) and then with the ellipse fits superposed
(bottom). 
The blue circle at the bottom right of each image represents
the PSF FWHM (0\farcs163 corresponding to $\sim$1.4 kpc  at
$z\sim$~1.1), and the horizontal bar shows a 0\farcs5 scale for reference. 
The right panel for each galaxy shows
the radial profiles of surface brightness ($\mu$), ellipticity ($e$),
and position angle (PA)  versus semi-major axis $a$ derived from
ellipse fits. 
PA goes from 0 to -90 clockwise (from North to West) and goes from 0
to 90 counter-clockwise (from North to East). The vertical dashed line represents the F444W image PSF FWHM. {Fits with semi-major axis smaller than PSF FWHM (shaded region) are not considered.} The figure illustrates the bar signature described in \S~\ref{sec:bar-efit-criteria}: (i) along the bar in the bar-dominated region, at {$a \sim 0\farcs15$ to $0\farcs4$}, $e$ smoothly increases while the PA stays fairly constant; (ii) from {$a \sim 0\farcs5$ to $1\farcs0$}, as we transition from the bar end to a region of the disk dominated by the spiral arms, the $e$ stays fairly high while the PA twists; (iii) at $a > 1\farcs0$, further out in the outer disk where the spiral arms are less prominent, the $e$ decreases and PA changes.}

    \label{fig:efit}
    \vspace{6mm}
\end{figure*}

 We fitted ellipses to the F200W image and F444W image of each of the 839 moderately inclined disk galaxies. The cutout images being fitted are 6\farcs0 $\times$ 6\farcs0.
For specific cases, when necessary, nearby sources were masked. Then we ran ``isophote.Ellipse.fit\_image'' \textsc{Photutils} from Python's astropy package \citep{photutils} twice, following the two steps outlined in \S~4.1 of \cite{Guo-etal-2023}. From the successful ellipse fits, we generated radial profiles of
surface brightness (SB), ellipticity ($e$), and position angle (PA)
plotted versus the ellipse semi-major axis $a$ (see Figure \ref{fig:efit} as an example).
  Not all sources had successful {fits} with the initial guess of $e$, PA, and $a$. 
  To fit as many sources as possible, we use a grid of initial guesses, which encompasses a range of values for $e$, PA, and $a$, each divided into 20 steps.
  However, even after exploring all combinations of our initial guesses, a fraction of galaxies did not yield successful ellipse fits. Overall, for the sample of moderately inclined ($i \leq 65^\circ$) disk galaxies at $0.5 \leq z \leq 4.0$, $\sim$ 92\% (\textcolor{black}{770} out of 839 sources) have successful ellipse fits in their F444W images, and $\sim$ \textcolor{black}{77}\% (\textcolor{black}{642} out of 839 sources) in their F200W images (see Table \ref{tab:Ndisk} for details). Ellipse fits tend to fail in galaxies that have a low average signal-to-noise ratio (SNR), clumpy structures inside the disks, or peculiar features in the outer region of the disk. We note that at $z \sim$ 3--4, only $\sim 60\%$ of moderately inclined disk galaxies have successful ellipse fits in their F200W images. This is expected because, at these redshifts, F200W images are more sensitive to dust and star formation and more likely to show clumpy structures ({see \S~\ref{sec:bar-efitobs} and \S~\ref{sec:towards-intrinsic} for a more detailed discussion.)}

\begin{deluxetable}{cccc}
\tablewidth{1\textwidth}
\tablenum{2}
\tablecaption{Number of moderately inclined disk galaxies based on ellipse fits and visual classification.
\label{tab:Ndisk}}

\decimalcolnumbers
\tablehead{
\colhead{ Redshift Bin} &\colhead{$N_{disk}$}  &\colhead{$N_{disk}$} &\colhead{$N_{disk}$}\\\colhead{ } &\colhead{[efit, F200W]}  &\colhead{[efit, F444W]} &\colhead{[vis, all bands]}}

\decimalcolnumbers
\startdata
0.5-1.0 & \textcolor{black}{126} & \textcolor{black}{145} & 153\\
1.0-1.5 &  \textcolor{black}{207} & 223 & 241\\
1.5-2.0 &  \textcolor{black}{139} & 162 & 182\\
2.0-3.0 &  \textcolor{black}{121}  & 167 & 181 \\
3.0-4.0 & 49 & 73 & 82\\
\enddata

\tablecomments{Columns are (1) Redshift bin; (2) Number of moderately inclined disks that have successful ellipse fits in F200W images; (3) Number of moderately inclined disks that have successful ellipse fits in F444W images; (4) Number of moderately inclined disks that are visually classified as disks.}
    \label{tab:Ndisk}

\vspace{-1cm}
\end{deluxetable}

\par
With the ellipse fits, we consider a galaxy to be a barred galaxy 
only if it satisfies the two criteria below. Note that regions where the semi-major axis is less than the PSF are not considered because, inside this unresolved region, the quality of ellipse fits is poor. {As ellipse fits can robustly detect bars with projected semi-major axis $a_{\rm bar}\gtrsim$ 2 $\times$ PSF FWHM, with F200W images, we expect to robustly detect $a_{\rm bar}>$ 1.5 kpc.} 
We apply two quantitative criteria to the profiles of $e$ and PA:

\begin{enumerate}

\item 
In the bar-dominated region,  we require the ellipticity $e$ to rise smoothly to a maximum value $e_{\rm
{bar}} \ge$~0.25, while the PA stays fairly constant along the bar, with some small variation $\Delta \theta_{\rm 1} $ allowed. At a redshift of $z \sim 0$, a value of 20$^{\circ}$ is used for $\Delta \theta_{\rm 1} $, but for the high-redshift $z \sim $ 1--4 regime in this study, we allow values as large as 30$^{\circ}$ due to the lower spatial resolution and complicating effects of dust and recent SF. Most clear-cut bar cases in our study still meet the 20$^{\circ}$ criterion, but some borderline cases align with a slightly larger $\Delta \theta_{\rm 1} $. We define $a_{\rm \epsilon}$ as the length of the semi-major axis corresponding to $e_{\rm {bar}}$.

\item In the region dominated by the outer disk, we require the ellipticity
to drop by at least 0.1  from the bar's maximum ellipticity $e_{\rm
{bar}} $ and
the PA to change by some variation $\Delta \theta_{\rm 2} $  ($\gtrsim$ 10$^{\circ}$)
~from the associated bar PA. 
\end{enumerate}

\begin{figure*}
    \includegraphics[width=1\textwidth]{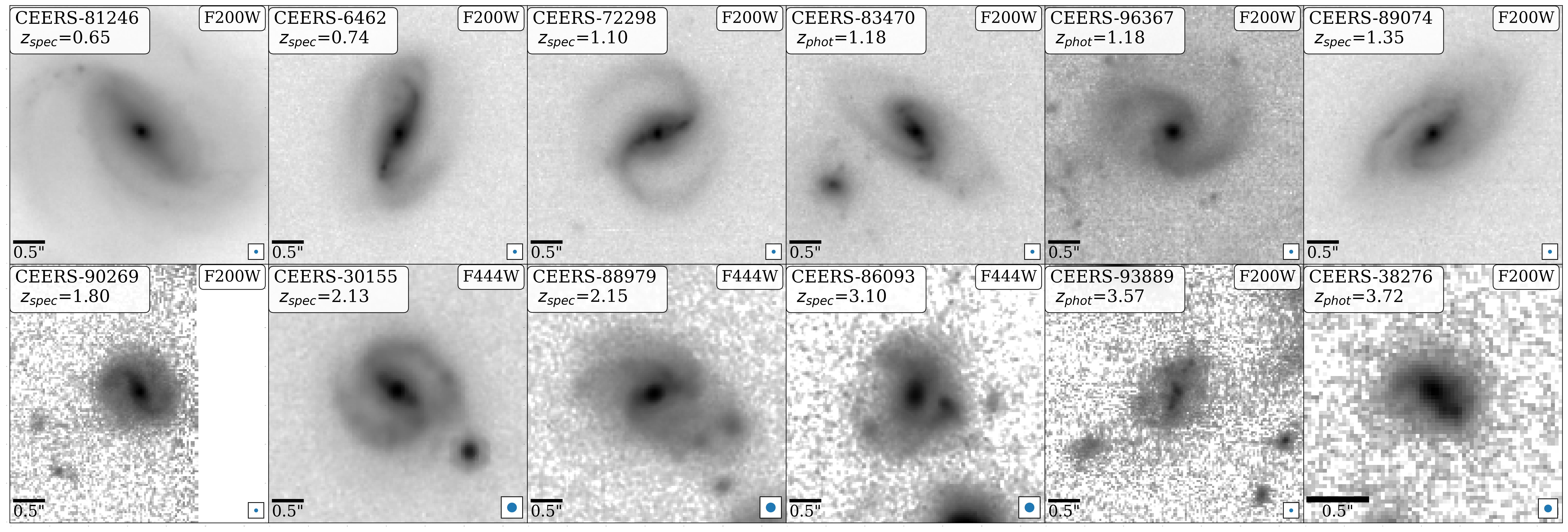}
    \caption{{Montage of {\textit{JWST}} \textcolor{black}{images} showing the morphology of 
    the selected example barred galaxies (identified via ellipse fits or/and visual classification). 
    The labels show the CEERS ID and photometric redshift ($z_{phot}$) or spectroscopic redshift ($z_{spec}$) of each galaxy. The circle at the lower right corner represents the PSF FWHM, and the horizontal bar shows a 0\farcs5 scale for reference. 
     CEERS-38276 is  1\farcs0 $\times$ 1\farcs0 in size, and the other images are 4\farcs0 $\times$ 4\farcs0 in size. CEERS-38276, with a photometric redshift at $z \sim 3.7$, is {one of the highest-redshift barred galaxy candidates} known to date, corresponding to epochs when the Universe was \textcolor{black}{$\sim$ 12\%} of its present age. We also note that disks and bars appear smaller at higher redshifts.}}
    
    \label{fig:Example-barred-galaxies}
    \vspace{6mm}
\end{figure*}

Similar criteria are also used in other studies using ellipse fitting (e.g., \citealt{Jogee-etal-2004, Marinova-Jogee2007, Olguin-Iglesias-Kotilainen-Chavushyan2020}). 
However, for this study on high-redshift bars, it is crucial to also verify which features in the images are leading to the bar signatures (e.g., a rise in $e$, a flat PA, a drop in $e$) by carefully inspecting the fitted ellipses overlaid on the image (e.g., see Figure \ref{fig:efit}). This inspection allows us to weed out cases whereby artifacts, such as spiral arms, clumps, or inclined rings, might mimic some of the bar signatures. Examples of robustly identified bars are shown in Figure \ref{fig:Example-barred-galaxies}. 

\subsubsection{Bar Identification via Visual Classification}\label{sec:bar-vis-criteria}
We employ the visual classification of bars as a method that is complementary to the identification of bars via ellipse fits. In rest-frame optical and NIR images, a stellar bar visually appears as an elongated bright feature that extends from the central region of the galaxy into the outer disk. While visual classification may be less objective and quantitative, it leverages the power of the eye to capture a wide range of details in the image that can impact bar and disk identification, such as asymmetric features, clumps, inclined rings, and companions.

 \par We asked the classifiers to visually identify barred galaxies in F200W and F444W images separately. 
 When classifying a galaxy as unbarred or barred, we assign a confidence score ranging from 0 to 3 to characterize the robustness and degree of confidence of the classification. The scores are defined as 3 (clearly barred),  2 (likely barred),  1 (likely unbarred), and 0 (clearly unbarred):
 \begin{itemize}
     \item A score of 3 (clearly barred) is assigned if an image shows a clear bar residing in a low-to-moderately inclined disk galaxy.
     \item A score of 2 (likely barred) rather than a score of 3 (clearly barred) is assigned if an image shows a likely bar, but complicating factors prevent a foolproof identification. Examples of complicating factors include the bar feature being small, the presence of nearby clumps, or the disk being close to the inclination limit.
     \item A score of 1 (likely unbarred) rather than a score of 2 (likely barred) or a score of 0 (clearly unbarred) is assigned if nearly all visual indicators suggest there is no bar, but there are some mildly elongated central structures that we cannot completely rule out as potentially being weak bars.
     \item A score of 0 (clearly unbarred) is assigned if an image does not show any evidence of a bar.

 \end{itemize}

We combined the visual classification results from all three classifiers and took the average score. In this work, we report a source as a barred galaxy robustly identified via visual classification if the average score 
{$\ge 1.33$, corresponding to two out of the three classifiers deciding the source has a score of 2 or higher.}

\section{Results}\label{sec:results}

\subsection{Observed Bar Fraction}\label{sec:Observed Bar Fraction}


\subsubsection{Observed Bar Fraction Derived from Ellipse Fits}\label{sec:bar-efitobs}

\par
In this section, we report three observed bar fractions ($f_{\rm bar}$) based on ellipse fits: $f_{\rm bar}$ based on F200W images, $f_{\rm bar}$ based on F444W images, and the overall $f_{\rm bar}$ based on bars identified in either band.

Each of the $f_{\rm bar}$ is derived following $f_{\rm bar}$ = $N_{\rm bar}$ /$N_{\rm disk}$ (Equation \ref{eq:fbar}; see \S~\ref{method}), while $N_{\rm bar}$ and $N_{\rm disk}$ are different for each measurement. For the bar fraction based on the F200W or F444W images, $N_{\rm bar}$ is based on bars identified via fits in that band, and $N_{\rm disk}$ is the number of moderately inclined disk galaxies that have successful fits in that band, respectively. The overall bar fraction is based on bars identified via fits of either band and $N_{\rm disk}$ is the number of disk galaxies that have successful fits in either band. 
\par
As stated earlier in \S~\ref{sec: levarage-200and444}, the sharper F200W images (Table \ref{tab:rest-frame-and-psf}) are effective at tracing small bars, while the longer-wavelength F444W images may more effectively trace bars that are obscured by dust or impacted by recent star formation. Therefore, the overall $f_{\rm bar}$ is the best measure because it accounts for the advantages in the individual measurements and captures most bars. 

\par
One limitation of using ellipse fits to derive the observed bar fraction is that not all disk galaxies had successful fits.  Columns 2 and 3 in Table \ref{tab:Ndisk} present the $N_{\rm disk}$ that have successful fits in each redshift bin for each band, and column 4 shows the $N_{\rm disk}$ identified via visual classification. Fortunately, $\sim$ 92\% (\textcolor{black}{770}) of the 839 visually classified moderately inclined disk galaxies have successful ellipse fits in their F444W images, and $\sim$ \textcolor{black}{77}\% (\textcolor{black}{642}) of them have successful ellipse fits in their F200W images. However, we note that at $z \sim$ 2--4, only $\sim$ 60--67\% of the moderately inclined disk galaxies have successful ellipse fits in their F200W images. The ellipse fits tend to fail in galaxies that have a low average signal-to-noise ratio (SNR), clumpy structures inside the disks, or peculiar features in the outer region of the disk. 
At $z > 2$, the shorter-wavelength F200W images are more sensitive to dust and star formation and, hence, are more likely to show clumpy, discontinuous light distribution in star-forming galaxies. 
Therefore, even if a galaxy can be visually classified as a disk galaxy, it may not be successfully ellipse-fitted, 
especially in the F200W images. The inability to fit all galaxies has been reported in some previous studies using the ellipse fitting technique (e.g., \citealt{Jogee-etal-2004, Marinova-Jogee2007, Le-Conte-etal-2024}). {Barred galaxies identified via visual classification can be missed in the selection of bars identified via ellipse fits due to the failure of ellipse fitting.}

\par

\begin{figure}[ht!]
    \includegraphics[width=0.47\textwidth]{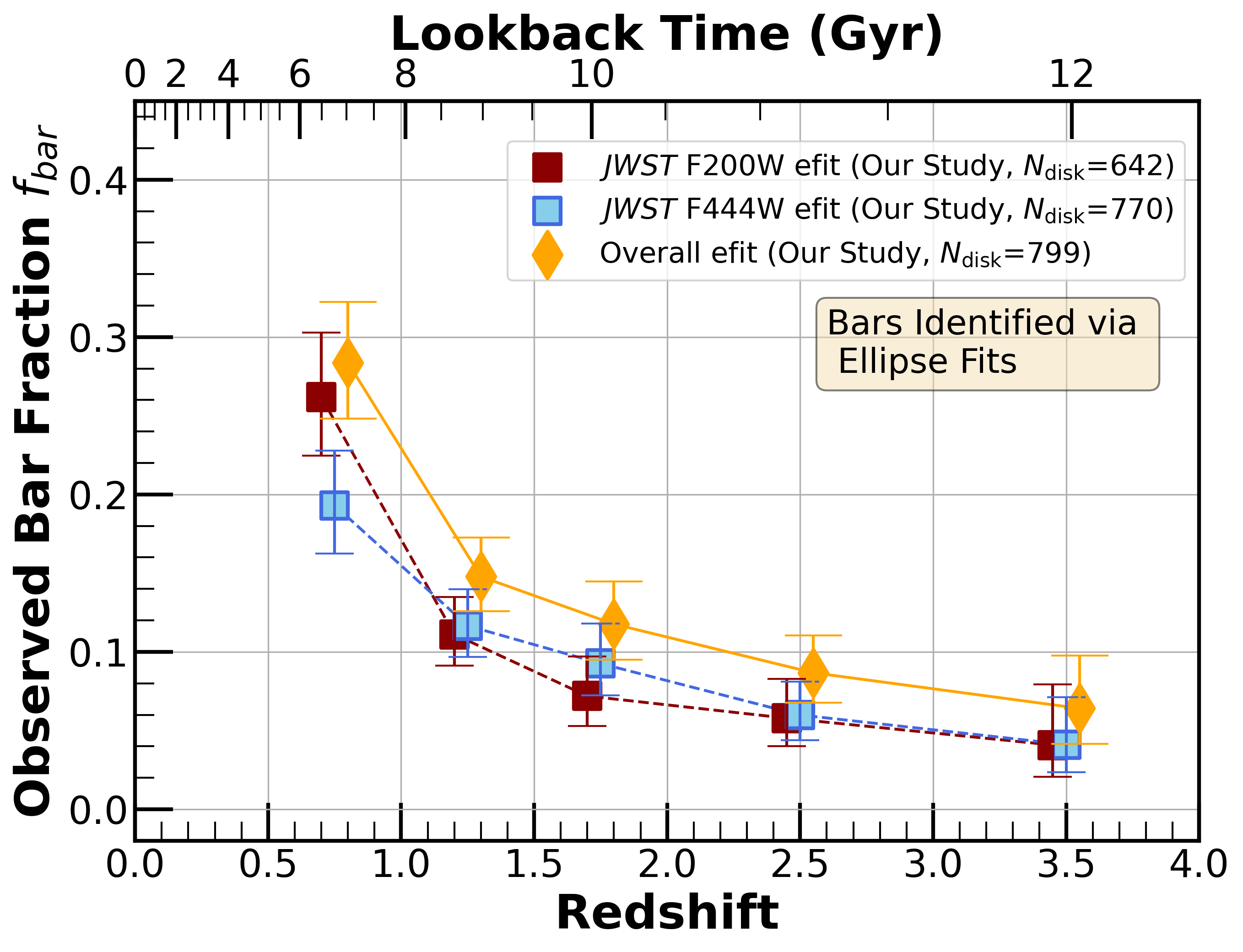}
    \caption{Observed bar fraction derived from ellipse fits. We present three measures of the bar fraction based on identifying bars via ellipse fits in the F444W band (blue squares), F200W band (red squares),  or in either band (orange diamonds).  The error bars on each data point are the propagated binomial errors. The orange diamonds represent the overall bar fraction. All three observed bar fractions $f_{\rm bar}$ show declining trends. 
    \textcolor{black}{The red squares and orange diamonds are slightly shifted from the center of each redshift bin to prevent overlap.} 
 Table \ref{tab:fbar} shows values corresponding to each point. See \S~\ref{sec:bar-efitobs} for details.}
\label{fig:observed-fbar-efit}
\end{figure}

Figure \ref{fig:observed-fbar-efit} presents the evolution of $f_{\rm bar}$ as a function of redshift in 5 redshift bins: 0.5--1.0, 1.0--1.5, 1.5--2.0, 2.0--3.0, and 3.0--4.0. Each data point is plotted at the midpoint of each redshift bin. Table \ref{tab:fbar} and Figure \ref{fig:observed-fbar-efit}  show the three measurements of $f_{\rm bar}$ using ellipse fits: $f_{\rm bar}$ in F200W images (dark red squares), $f_{\rm bar}$ in F444W images (blue squares), and the overall $f_{\rm bar}$ based on bars in either band (orange diamonds). The overall $f_{\rm bar}$ is the best measure because it captures most bars, primarily reflecting the $f_{\rm bar}$ of bars with projected semi-major axis $a_{\rm bar}$ $> 1.5 $ kpc ($\sim$ 2 $\times$ PSF in F200W images) that can be robustly traced by our ellipse fits. \textcolor{black}{The error bars represent the statistical uncertainty in the $f_{\rm bar}$, are the 68\% binomial proportion confidence interval estimated with Wilson score interval.} They do not reflect systematic effects, which might lead the intrinsic bar fraction to differ from the observed bar fraction (see \S~\ref{sec:towards-intrinsic}).

\par 
{All three measures of the observed bar fraction from ellipse fits show declining trends from  $z \sim$ 0.5--1 (corresponding to look-back times of $\sim$ 5.2--7.9 Gyr) to $z \sim$ 3--4 (corresponding to look-back times of $\sim$ 10.5--11.6 Gyr). {The $f_{\rm bar}$ in F444W images decreases from \textcolor{black}{$19.3^{+3.5}_{-3.1}$\%} at $z \sim$ 0.75 to \textcolor{black}{$4.1^{+3.0}_{-1.8}$\% } at $z \sim$ 3.5. The $f_{\rm bar}$ in F200W images decreases from \textcolor{black}{$26.2^{+4.1}_{-3.7}$\%} at $z \sim$ 0.75 to \textcolor{black}{$4.1^{+3.9}_{-2.0}$\%}  at $z \sim$ 3.5. The overall $f_{\rm bar}$ decrease from \textcolor{black}{$28.4^{+3.8}_{-3.6}$\%} at $z \sim$ 0.75 to \textcolor{black}{$6.4^{+3.4}_{-2.3}$\%} at $z \sim$ 3.5.} }
\par
The observed bar fraction derived using ellipse fits $f_{\rm bar}$ in the F200W band is $\sim7$\% higher at $z\sim$ 0.75 compared to that in the F444W band. At $z\sim$ 0.5--1, both F200W and F444W images trace rest-frame NIR light. We find that, at $z\sim$ 0.5--1, 11 robustly identified barred galaxies with bar length $<$ 2.4 kpc (roughly corresponding to $2 \times$ PSF FWHM in F444W images) are only identified in F200W images. Figure \ref{fig:abar-only} illustrates the comparison of the distribution of projected bar lengths (see \S~\ref{sec:bar-measurement} for details). Hence, the difference we observed is {mostly the} effect of higher spatial resolution in F200W images (PSF FWHM $\sim$ 0\farcs08) that allows identification for smaller bars.  
\par

\begin{deluxetable*}{ccccccc}
\tablewidth{1\textwidth}
\tablenum{3}
\tablecaption{Observed bar fractions in this study.  
}

\decimalcolnumbers

\tablehead{
\colhead{ Redshift Bin} & \multicolumn{2}{c}{F200W}  &
\multicolumn{2}{c}{F444W}&
\multicolumn{2}{c}{F200W+F444W} \\
\cline{2-3} \cline{4-5} \cline{6-7} 
 & \colhead{$f_{\rm bar}$ (\%)} & \colhead{$f_{\rm bar}$ (\%) } &
\colhead{$f_{\rm bar}$ (\%)} &  \colhead{$f_{\rm bar}$ (\%)} & \colhead{$f_{\rm bar}$ (\%)} &  \colhead{$f_{\rm bar}$ (\%) }\\
 & \colhead{(efit)} & \colhead{(visual)} &
\colhead{(efit)} &  \colhead{(visual)} & \colhead{(efit)} &  \colhead{(visual) }}

\decimalcolnumbers

\textcolor{black}{
\startdata
0.5--1.0 & $26.2^{+4.1}_{-3.7}$ (33/126) & $26.1^{+3.7}_{-3.4}$ (40/153)  & $19.3^{+3.5}_{-3.1}$ (28/145) & $19.6^{+3.4}_{-3.0}$ (29/153)  & $28.4^{+3.8}_{-3.6}$ (42/148)& $28.1^{+3.8}_{-3.5}$ (42/153)\\
1.0--1.5 & $11.1^{+2.4}_{-2.0}$ (23/207) & $13.3^{+2.3}_{-2.0}$ (32/241) & $11.7^{+2.3}_{-2.0}$ (26/223) & $11.6^{+2.2}_{-1.9}$ (28/241) & $14.8^{+2.5}_{-2.2}$ (34/230) & $15.8^{+2.5}_{-2.2}$ (39/241)\\
1.5--2.0 & $7.2^{+2.5}_{-1.9}$ (10/139) &  $6.0^{+2.0}_{-1.5}$ (11/182) & $9.3^{+2.5}_{-2.0}$ (15/162) & $4.4^{+1.8}_{-1.3}$ (8/182)  & $11.8^{+2.7}_{-2.3}$ (20/170) & $7.1^{+2.2}_{-1.7}$ (13/182)\\
2.0--3.0 & $5.8^{+2.5}_{-1.8}$ (7/121) & $4.4^{+1.8}_{-1.3}$ (8/181) & $6.0^{+2.1}_{-1.6}$ (10/167)& $6.1^{+2.0}_{-1.5}$ (11/181)  &  $8.7^{+2.4}_{-1.9}$ (15/173) & $8.3^{+2.3}_{-1.8}$ (15/181)\\
3.0--4.0  &  $4.1^{+3.9}_{-2.0}$ (2/49) & $2.4^{+2.4}_{-1.2}$ (2/82) &  $4.1^{+3.0}_{-1.8}$ (3/73) & $0.0^{+1.2}_{-0.0}$ (0/82) & $6.4^{+3.4}_{-2.3}$ (5/78) & $2.4^{+2.4}_{-1.2}$ (2/82)\\
\enddata
}

\tablecomments{ Columns are (1) Redshift bins; (2) Observed bar fraction in F200W band derived from ellipse fits; (3) Observed bar fraction in F200W band derived from visual classifications; (4) and (5) are the same as columns (2) and (3) in F444W band; (6) and (7) are the same as columns (2) and (3) of the overall bar fraction based on bars identified in either the F200W or F444W images. For columns (2)-(7),  lists the bar fraction (\%) and its error (\%), followed by the ratio ($N_{\rm bar}$/ $N_{\rm disk}$) (see Equation \ref{eq:fbar} in \S~\ref{method}). We note that the bar fraction derived via ellipse fits and via visual classification are mostly consistent, even if they are derived from slightly different samples.}

    \label{tab:fbar}
\vspace{-0.3cm}
\end{deluxetable*}

\subsubsection{Observed Bar Fraction Derived from Visual Classification}\label{sec:bar-visobs}
\par
In this section, we report three observed bar fractions ($f_{\rm bar}$) based on visual classification: $f_{\rm bar}$ based on F200W images, $f_{\rm bar}$ based on F444W images, and the overall $f_{\rm bar}$ based on bars identified in either band. The observed bar fraction based on the visual identification of barred galaxies is calculated following Equation \ref{eq:fbar} in \S~\ref{method}. For the three measurements, $N_{\rm disk}$ is the number of moderately inclined disk galaxies (column 4 in Table \ref{tab:Ndisk}), and $N_{\rm bar}$ is the number of barred galaxies visually identified in F200W images, in F444W images, and in either band respectively. Galaxies are considered to be visually identified as barred if they have an average score of $\ge$ 1.33, as described in \S~\ref{sec:visual classification}.

\par

\begin{figure}[hbt!]
    \includegraphics[width=0.47\textwidth]{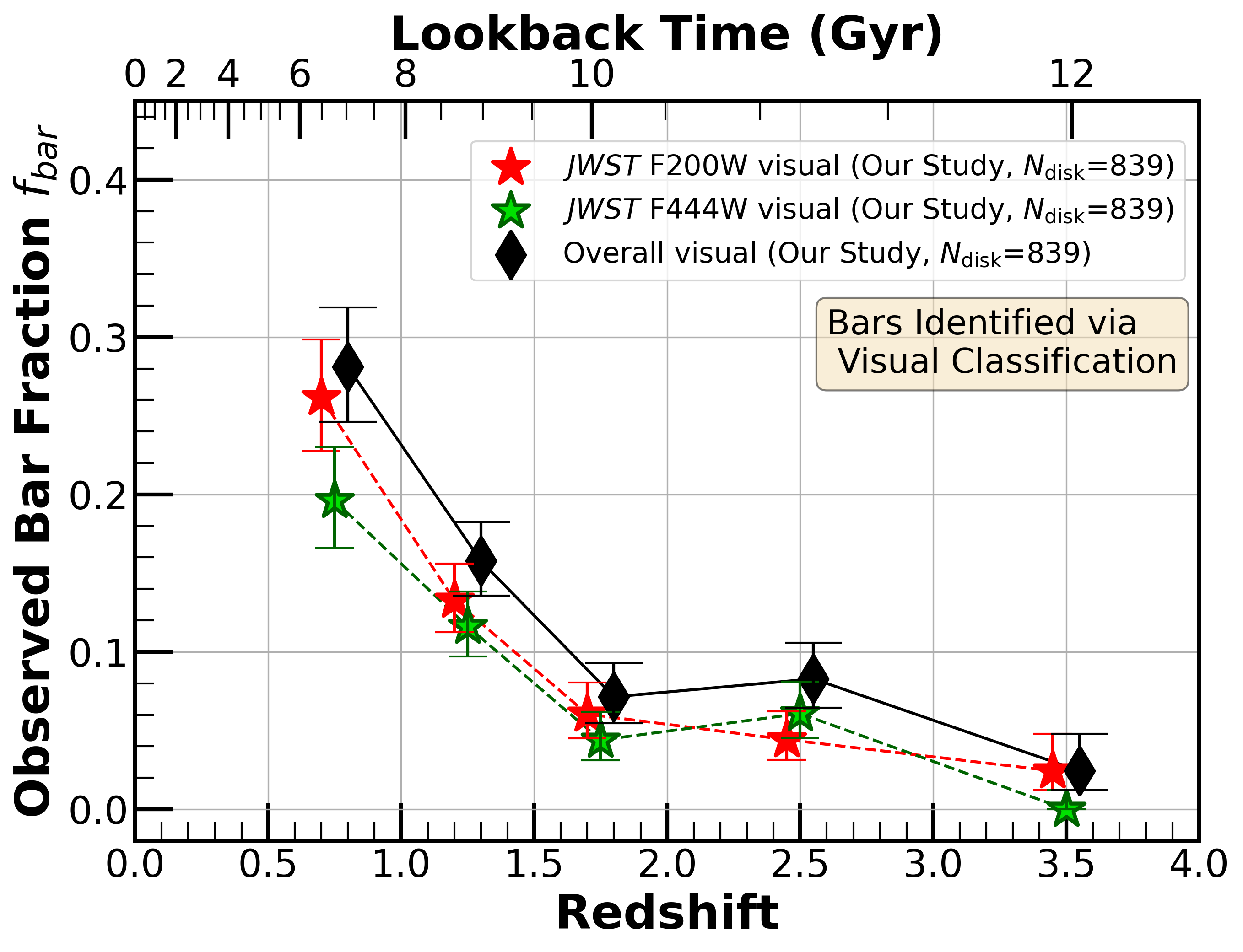}
     \caption{Observed bar fraction derived from visual classifications. We present three measures of the bar fraction based on identifying bars via \textcolor{black}{visual classification} in the F444W band 
 (green stars)  F200W band (red stars),  or in either band (black diamonds). The error bars on each data point are the propagated binomial errors. The black diamonds represent the overall bar fraction. All three observed bar fractions $f_{\rm bar-vis}$ show declining trends. \textcolor{black}{The red stars and black diamonds are slightly shifted from the center of each redshift bin to prevent overlap.} Table \ref{tab:fbar} shows values corresponding to each point. See \S~\ref{sec:bar-visobs} for details.}
\label{fig:vis-bar-frac}
\end{figure}

We present three measures of the bar fraction from visual classification in Table \ref{tab:fbar} and Figure \ref{fig:vis-bar-frac}: $f_{bar}$ in F200W images (red stars), $f_{bar}$ in F444W images (green stars), and overall $f_{bar}$ based on bars identified in either band (black diamonds). All three measures of the observed bar fraction based on visual classification show a declining trend from $z \sim$ 0.5--1 to $z \sim$ 3--4. {The observed bar fraction based on F444W images decreases from \textcolor{black}{$19.6^{+3.4}_{-3.0}$}\% at $z \sim$ 0.75 to \textcolor{black}{$\sim 0^{+1.2}_{0.0}$}\% at $z \sim$ 3.5. The observed bar fraction based on F200W images decreases from \textcolor{black}{$26.1^{+3.7}_{-3.4}$}\% at $z \sim$ 0.75 to \textcolor{black}{$2.4^{+2.4}_{-1.2}$} at $z \sim$ 3.5. The overall bar fraction based on bars identified in either image is slightly higher than individual measures, decreasing from \textcolor{black}{$28.1^{+3.8}_{-3.5}$}\% at $z \sim$ 0.75 to \textcolor{black}{$2.4^{+2.4}_{-1.2}$} at $z \sim$ 3.5.}%

\subsubsection{\textit{HST} and \textit{JWST} Results on Observed Bar Fraction}\label{sec:HST+JWST}

\begin{figure*}[ht!]

    \centering
    \includegraphics[width=0.8\textwidth]{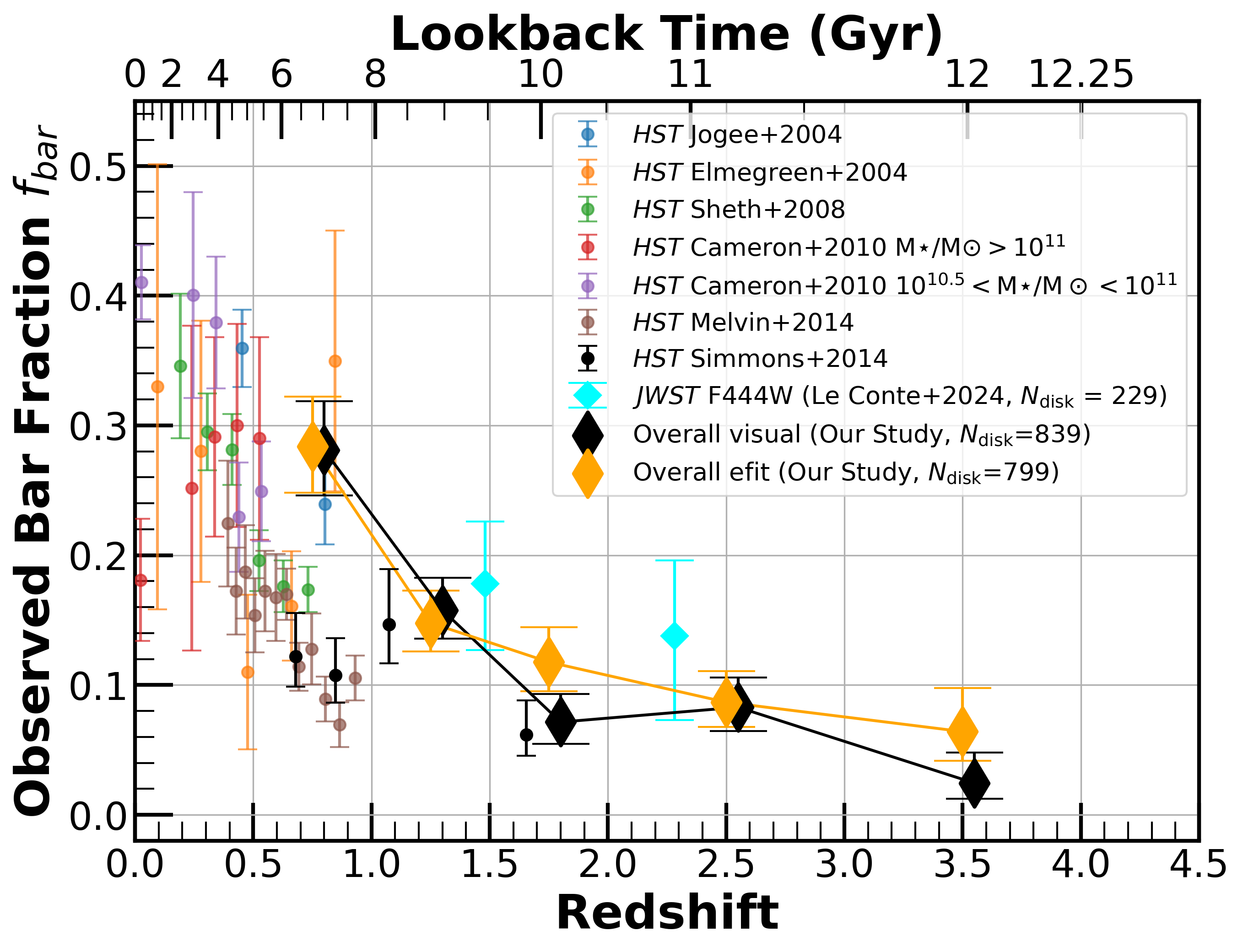}
    \caption{
    Observed bar fraction measured across nearly two decades (2004 to 2024): \textcolor{black}{$\textit{HST}$ results mainly go out to $z \sim 1.2$ \textcolor{black}(\citealt{Jogee-etal-2004, Elmegreen-Elmegreen-Hirst2004, Sheth-etal-2008, Cameron-etal-2010, Melvin-etal-2014, Simmons-etal-2014})}  
    and are mainly based on rest-frame optical images. \textcolor{black}{The \textit{HST}-based results plotted on this figure are taken from Figure 9 in \cite{Melvin-etal-2014}}. At $z > 1.5$, we \textcolor{black}{present} two $\textit{JWST}$ studies: LC24 going out to $z \sim 3$  based on visual {classification using $\textit{JWST}$ F444W images} and our studies going out to $z \sim 4$ based on visual classification and ellipse fits of both F200W and F444W $\textit{JWST}$ images. The error bars in LC24 results are taken from the original study, taking into account {both statistical errors and systematic errors}.  
    The $\textit{JWST}$ results demonstrate the existence of well-developed barred galaxies at $z \sim$ 1.5–4.0 in the young Universe. 
    We caution that all plotted points represent the observed bar fraction, which is a lower limit to the true intrinsic bar fraction. See \S~\ref{sec:HST+JWST} for details. \textcolor{black}{The black diamonds are slightly shifted from the center of each redshift bin to prevent overlap. Same shift is applied to Figures \ref{fig:TNG_Vs_observed_Fbar}, \ref{fig: Appen-D}, \ref{fig: Appen-B}, and \ref{fig: Appen-A}.}
\label{fig:fbar-all}}
\end{figure*}
\par
We compile the observed bar fractions reported in some of the previous $\textit{HST}$ studies and the current $\textit{JWST}$ studies, including the observed bar fractions reported in this work.

\par
Figure \ref{fig:fbar-all} summarizes two decades of bar studies from 2004 to 2024, based on $\textit{HST}$ and $\textit{JWST}$. $\textit{HST}$ results go out to $z \sim 1.2$ and are mainly based on rest-frame optical images. At $z > 1.5$, \textcolor{black}{we present} two $\textit{JWST}$ studies: {LC24 based on visual inspection of F444W galaxy images, as well as radial profiles of ellipticity and position angle, and our study going out to $z \sim 4$ based on visual classification and {quantitative criteria applied to ellipse fits} of both F200W and F444W images.} Our work complements LC24 in several ways. 

\par  Our methodology is different from the methodology used in LC24. {LC24 visually inspected F444W images and profiles of $e$ and PA to identify bars. In this work, in addition to F444W images, we also used the sharper-spatial-resolution F200W images to try to capture smaller stellar bars that may have been missed by F444W images and utilize both visual classifications and ellipse fits with quantitative criteria to identify barred galaxies independently.} 

\par 
{As shown in Figure \ref{fig:fbar-all}, the bar fraction from LC24 is slightly higher than our bar fraction by $\sim$ 4--6\%, 
but are consistent within the uncertainties. 
\textcolor{black}{It is to be noted that our error bars represent 68\% binomial proportion confidence interval that tied to the statistical uncertainties.} In contrast, the error bars on LC24 results include the systematic and statistical uncertainties. 
We explore the uncertainties in our bar identification and present the results in Appendix \ref{APPEN-bar-classification-uncertainty}. We also discuss below the potential reasons for the slightly different results of LC24 and our study. }

\par
{To explore the potential impact of different classification schemes on the slightly different results of LC24 and our study, we explore how the classification of bars in LC24 compares to our work by investigating the 21 barred galaxies identified in LC24 that are also included in our initial sample of disk galaxies. LC24 considers a galaxy as a strongly barred galaxy if at least three out of five classifiers vote the source as “barred”. Conversely, they classify a galaxy as weakly barred if two out of five votes select “barred” or at least three out of five votes select “maybe barred”. It is to be noted that the definition of strong and weak bars in LC24 is significantly different from that used in most bar studies where the properties of the bar (e.g., ellipticity, shape, length, mass, Qb) are used to define its strength, rather than the votes of classifiers. For the 8 strongly barred galaxies in LC24, we consider 7/8 or 87\% of cases as robustly identified bars and only disagree on one source at $z \sim$ 1.46.  This shows that there is significant consistency across our two studies when it comes to identifying the most prominent bars that have a large impact on galaxy evolution. Out of the 13 weakly barred candidates, in LC24, our study classifies 6 or 46\% of them as robustly barred. Thus, one reason for the difference between LC24 and our result could be due to differences in classification in the cases of ambiguous bars. }
\par
{Two other potential reasons for the slightly different results of LC24 and our study could be differences in the sample used. LC24 uses the first epoch of CEERS data (4 NIRCam pointings; \citealt{Bagley-etal-2023}) and part of the initial public observations for the Public Release Imaging for Extragalactic Research (PRIMER; PI: Dunlop, ID=1837), and uses photometric redshifts and stellar masses from the CANDELS catalogs based on \textit{HST} data. In contrast, we analyzed data from 10 pointings of CEERS and used photometric redshifts and stellar masses derived using the CEERS catalog based on \textit{JWST} data {(see \S~\ref{Sec: observation and data})}. The CEERS catalog is deeper than the CANDELS catalog and probes rest-frame NIR wavelengths for sources at $z > 1$, resulting in different samples. Additionally, our results on the bar fraction in Figure \ref{fig:fbar-all} apply to galaxies with stellar mass $> 10^{10} M_{\odot}$, while LC24’s bar fraction is based on galaxies with a wider stellar mass range ($\gtrsim 10^{9} M_{\odot}$).  These differences in samples (number of galaxies, stellar mass, etc.) could lead to differences in bar fraction between LC24 and our study.}

\par
{Previous studies have shown that the evolution of the observed bar fraction with redshift at $z<1$ depends on the stellar mass range selected (e.g., \citealt{Cameron-etal-2010, Melvin-etal-2014, Erwin2018, Erwin-2023, Mendez-Abreu-etal-2023}). {In Appendix \ref{APPEN-mass}, we explore how our bar fraction evolution over $z \sim$ 0.5--4, as shown in Figure \ref{fig:fbar-all}, depends on the stellar mass.} As shown in Figure \ref{fig: Appen-D}, we find no significant difference in the bar fraction for galaxies in two stellar mass ranges:  $M_{\star} \sim 10^{10}-10^{10.5} M_{\odot}$ and  $M_{\star} \sim 10^{10.5}-10^{11} M_{\odot}$. There are too few barred galaxies with $M_{\star} > 10^{11} M_{\odot}$ in our sample, so we do not discuss this mass range.  Since there is no strong mass dependence of the bar fraction within our selected stellar mass range, we do not separate our results in multiple stellar mass bins in the rest of the paper.}

\par
{ Using the \textit{JWST} data, we showed that barred galaxies come into existence in massive disk galaxies at least as early as $z \sim 3.5$ when the Universe was only $\sim$ 13\% of its present age. Disk galaxies at those epochs may be significantly different from present-day disks in terms of their gas fraction, turbulence, and the extent to which they are dynamically cold, and it is notable that bars can form in them. As time progresses from $z \sim 3.5$ to $z \sim 0.75$, the observed bar fraction rises as more bars form or become longer lived (see \S~\ref{discussion}), implying that bar-driven secular evolution plays a more important role at later times ($z<1$).}

\par
Figure \ref{fig:fbar-all} presents many of the $\textit{HST}$ results that go out to $z \sim 1.2$. We stress that given the different methods and different sample selections (e.g., {based on different rest-frame imaging, limiting magnitudes, and stellar mass ranges}), we can not make a point-to-point comparison of our results with the earlier studies. However, we are interested in the general trend and an overview of what we know about the observed bar fraction using the {\textit{HST}} and {\textit{JWST}} over the last two decades. The combination of past {\textit{HST}}-based studies and our {\textit{JWST}}-based study suggests a general trend where the observed bar fraction continues to rise from $z \sim 4$ to $z \sim 0$.
{
It is also interesting to note that the observed bar fraction at $z \sim 0.75$ from our {\textit{JWST}}-based studies {is} $\sim$ 28\%,  which is approximately two times higher than the bar fraction reported by most \textit{HST} studies.  We suggest that this difference arises because the higher spatial resolution and longer rest-frame NIR wavelengths of \textit{JWST} NIRCam images can unveil a population of bars that were previously missed in \textit{HST} studies. We test this 
 hypothesis in Appendix \ref{APPEN-a-vs-HST-tests} where we show that the \textit{JWST} bar fraction is lowered to values close to the \textit{HST} bar fraction when we remove the short and weak bars that are more robustly captured by \textit{JWST} data than \textit{HST} imaging.}

\subsection{Distribution of Bar Length and Bar Strength}\label{sec:bar-measurement} 

\subsubsection{Distribution of Projected Bar Length}
\begin{figure}[h!]

    \centering
    \includegraphics[width=0.47\textwidth]{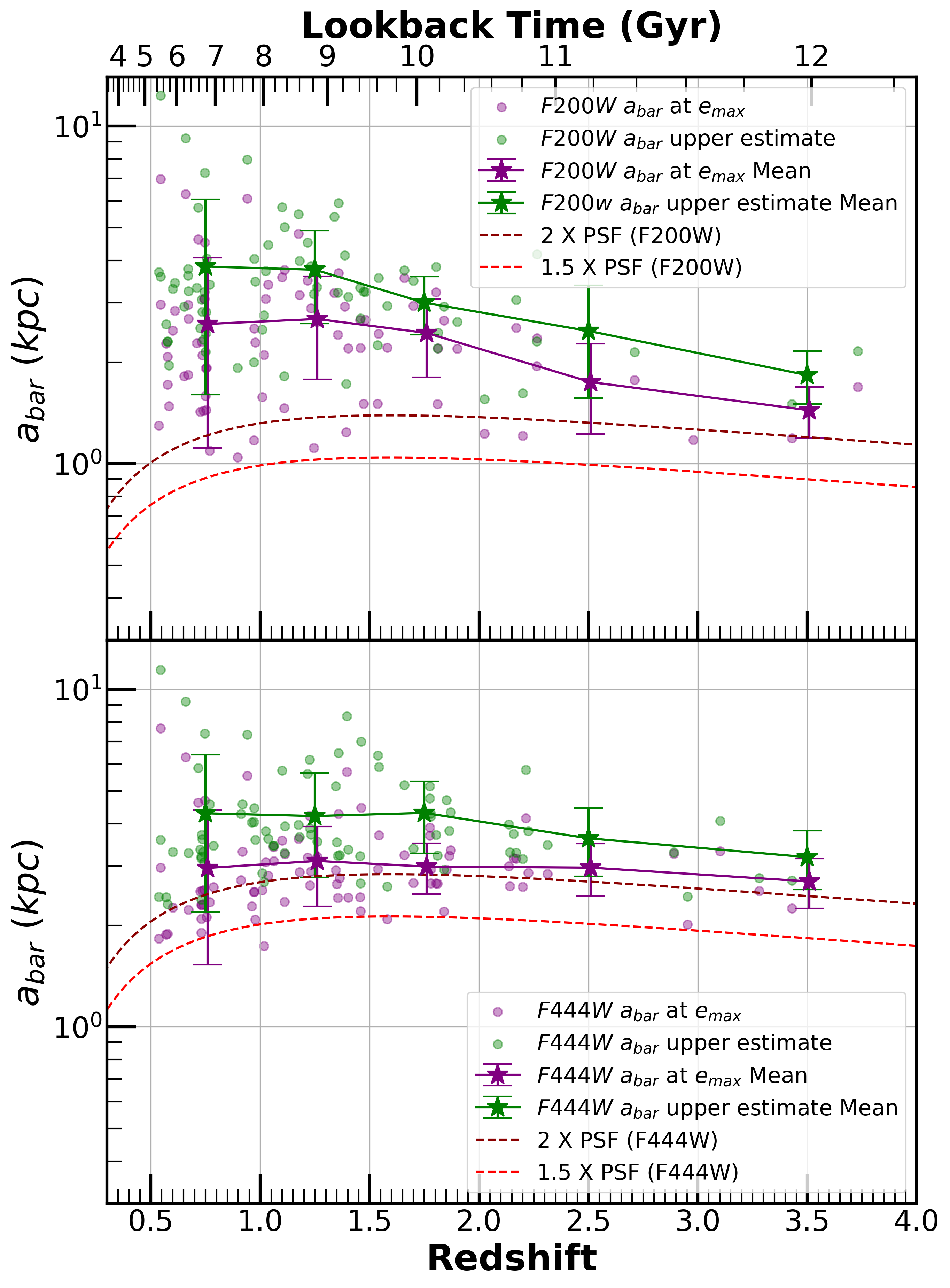}

    \caption{ Evolution of the projected bar lengths in F200W images (upper panel) and in F444W images (lower panel). In each panel, the green (purple) line represents the mean of the upper estimates $a_{\rm upper}$ (lower estimates $a_{\epsilon}$) of the projected bar length in each redshift bin, measured in each image. Individual values are plotted in the background using faint dots. The dark-red (red) dashed line shows the size of  2 (1.5) $\times$ PSF FWHM. Most of the sources have bar lengths greater than 2 $\times$ PSF FWHM. The error bar represents the standard deviation (1$\sigma$) in the given redshift bin. The lower panel is the same as the upper panel, except it is for results measured in F444W images. We observe a slight evolution of bar lengths measured in F200W images. See \S~\ref{sec:bar-measurement} for details.}
\label{fig:abar-vs-redshift}
\end{figure}

Different methods of defining bar length have been used in previous stellar bar studies, using observed or simulated images of bars. 
Across the different methods, a $\sim$ 15\% to 35\% difference in bar length is 
common (e.g., \citealt{Athanassoula-Misiriotis2002}). In this study, we only present the bar lengths and strengths of the bars selected via ellipse fits {in F200W and/or F444W images}. To address the uncertainty in bar length measurements, we present two bar length measurements representing the lower and upper limits of the measured bar length, following the methodology outlined in \cite{Erwin2005}.
\par
Following the strategy used in our pilot paper, we first adopt the widely used definition that the bar length is the length of the semi-major axis (sma)  $a_{\rm \epsilon}$  where  ellipticity 
smoothly rises to a maximum value along the bar  (e.g., \citealt{Athanassoula-Misiriotis2002, Jogee-etal-2004, Marinova-Jogee2007, Menendez-Delmestre-etal-2007,Guo-etal-2023}). This quantitative method can be unambiguously applied to many galaxies and be reproducible. The presence of spiral arms or rings can stretch the ellipses, either producing a plateau in ellipticity where the maximum is hard to define or leading the maximum ellipticity to fall outside of the bar (see Figure 2 and Figure 3 in \cite{Guo-etal-2023} for examples of each case). In order to take into account this effect, we define $a_{\rm \epsilon}$ as the location where the ellipticity \textit{first} rises to a maximum along the bar. Therefore, given the nature of this definition, we note that this method can underestimate the bar length because the maximum ellipticity we select is always close to the end of the bar but still inside the bar. Also, some studies (e.g., \citealt{Athanassoula-Misiriotis2002, Martinez-Valpuesta-Shlosman-Heller2006}) find that this definition can underestimate the true bar length. We, therefore, take $a_{\rm \epsilon}$ as the lower limit of the bar length measurement.

\par
For the upper limit of the bar length measurement, we adopt the method used in \cite{Erwin2005}, where the projected bar length $a_{\rm upper}$ is defined as the minimum of two 
measures: the first minimum in ellipticity outside the bar’s peak ellipticity, or the point at which the position angles of the fitted ellipses differ by $\ge$ 10$^\circ$ from the bar’s position angle. This method combines two common bar length definitions that use a drop of ellipticity and the change of PA to define the bar length. While $a_{\rm \epsilon}$ is more sensitive to the transition from the bulge-dominated region to the bar-dominated region, the other is more sensitive to the transition from the end of the bar to the outer disk. As shown in \cite{Erwin2005}, this definition can sometimes overestimate the bar length, especially when there are no spiral arms present. However, this definition well served the purpose of giving an upper limit for measuring the bar length in our study.
\par
We stress that bar lengths measured using different definitions in this study are the \textit{projected bar lengths}. The exploration of deprojected bar lengths is beyond the scope of the current study. 
The deprojection process requires 
{improved knowledge (e.g., on the thickness of the galactic disk and the thickness of the bar)} and tests in this new redshift regime and will be the topic of future work.

\begin{figure}
    \centering
    \includegraphics[width=0.47\textwidth]{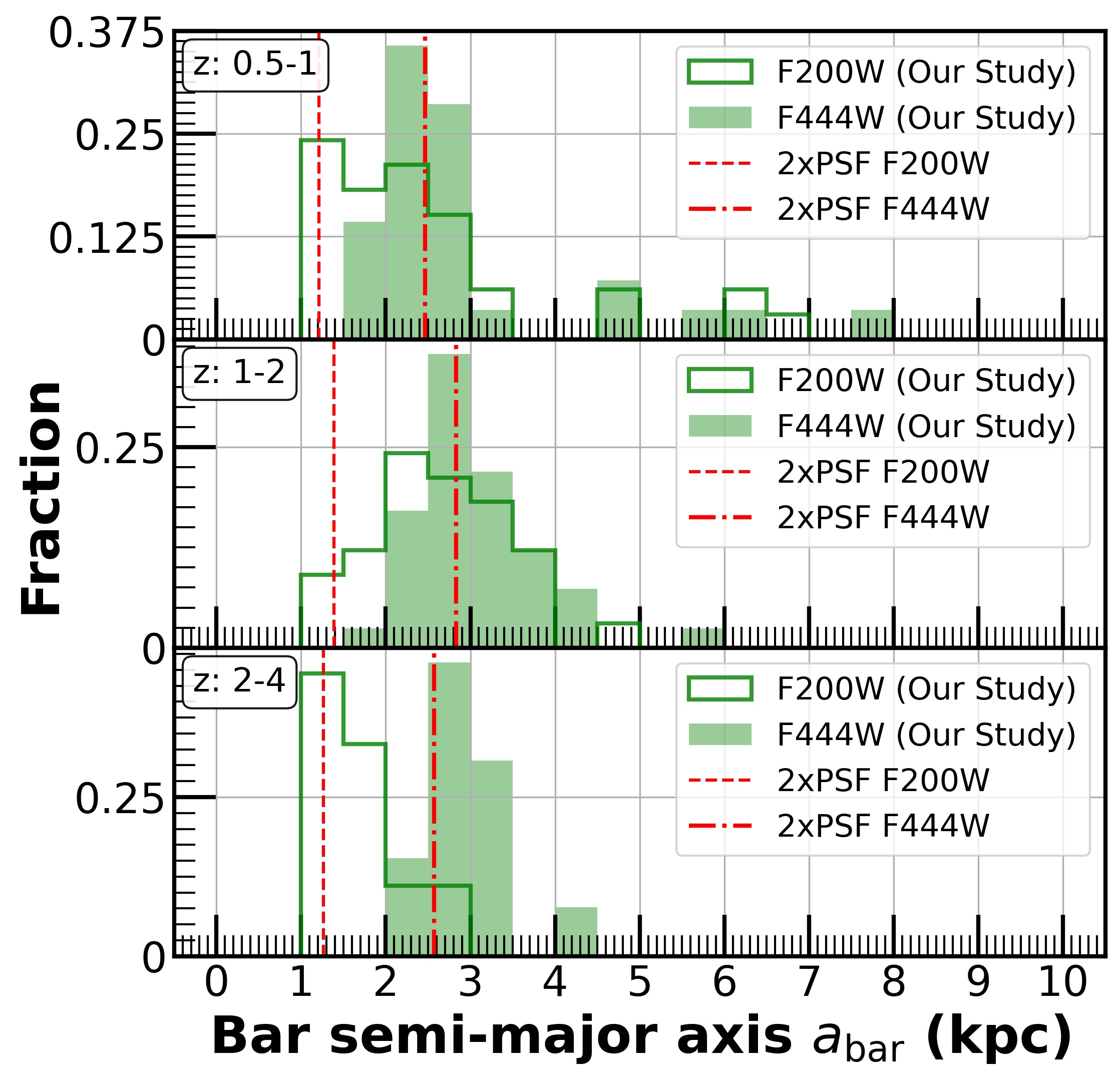}
    \caption{
    For bars identified and characterized via ellipse fits, we show the distribution of projected bar semi-major axis $a_{bar}$ for bars detected in F200W images (unfilled green histograms) and in F444W images (filled green histograms) in three redshift bins. 
    The vertical lines to the LHS of the plot represent the PSF, and bars with size 2 times the PSF can be reliably identified via ellipse fits. 
    Note that the F200W images detect a {significant} fraction of small bars with $ a_{\rm bar} \sim$ 1--2 kpc, which is undetected in F444W images and that this fraction rises in the higher redshift bin ($z \sim$ 2--4). See \S~\ref{sec:bar-measurement} for details.
    }
    \label{fig:abar-only}
\end{figure}

Both the upper estimates and the lower estimates of bar lengths are present in Figure \ref{fig:abar-vs-redshift}. 
{On average, the upper estimates can be $\sim$ 0.4 to 1.3 kpc longer than the lower estimates. We observe a slight evolution of bar lengths measured in F200W images. The average projected bar length rises from  $\sim$ 1.7--2.5 kpc at $z \sim$ 2.5 to {$\sim$ 2.6--3.8 kpc at $z \sim$ 0.75}. In the highest redshift bin, most bars are likely smaller and undetected, and only a few galaxies are identified in the highest redshift bin. Therefore, we do not discuss the trends from $z \sim$ 2.5 to $z \sim$ 3.5.} On the other hand, we do not observe an obvious evolution of bar lengths in  F444W images, which may be due to the fact that our detection of bars is limited by the larger PSF.

\par

In Figure \ref{fig:abar-only}, we compare the distribution of lower bar length estimates $a_{\epsilon}$ in F200W images (open histograms) and in F444W images (filled histograms). At $z \sim$ 2--4, the distribution of $a_{\epsilon}$ in F200W images is much smaller than that in F444W images. This suggests that at $z \sim$ 2--4, F200W images are likely to pick up smaller bars that can hardly be identified in F444W images, while F444W images can find bars that may be affected by dust and star formation and missed in F200W images.

\subsubsection{Distribution of Projected Bar Strength}

\begin{figure}[hbt!]

    \centering
    \includegraphics[width=0.47\textwidth]{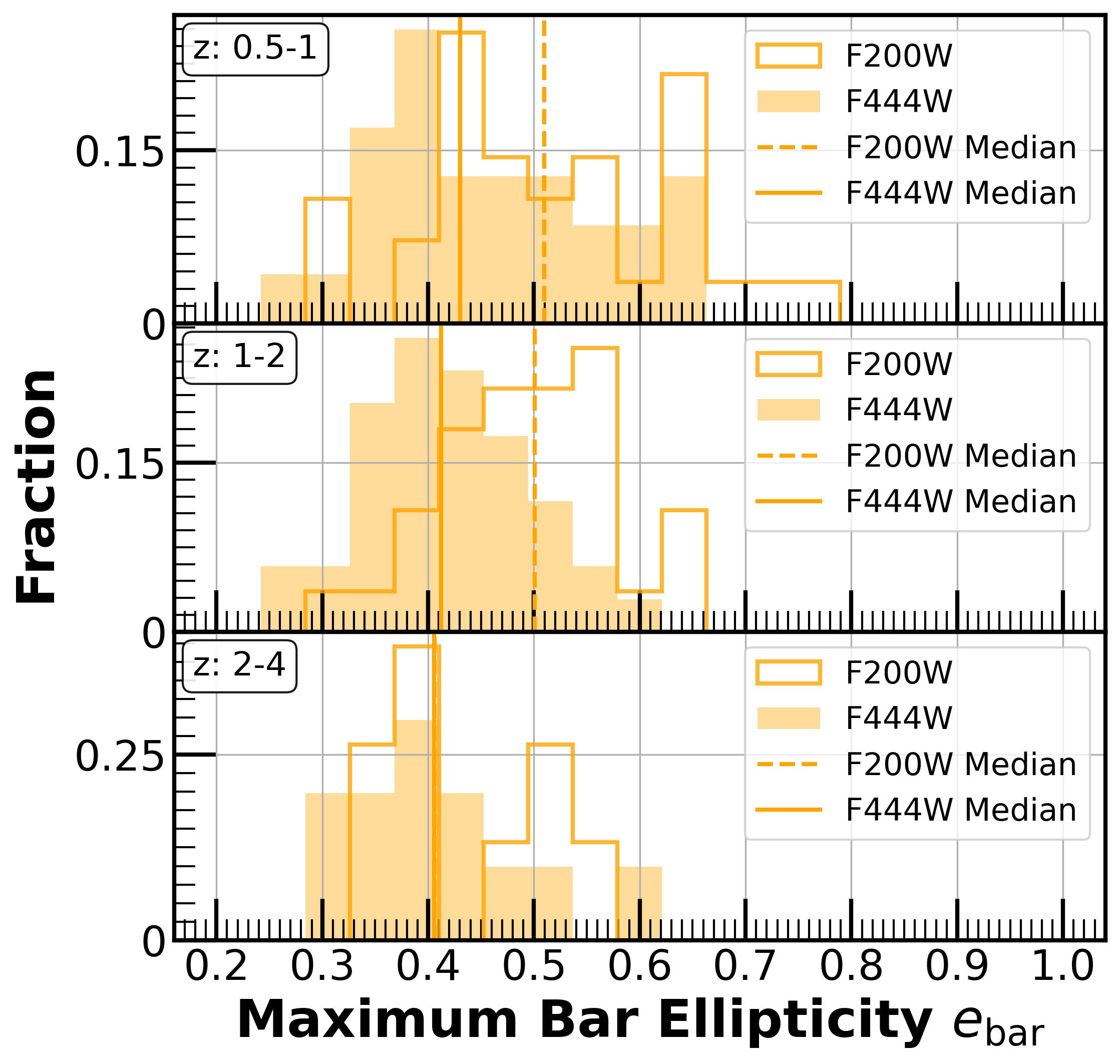}
    \caption{ 
    For bars identified and characterized via ellipse fits, we show the distribution of projected bar strength as characterized by the maximum projected ellipticity of the bar ($e_{\rm bar}$). Values measured for bars detected in F200W images (unfilled orange histograms) and F444W images (filled orange histograms) are shown for three redshift bins. The two vertical lines represent the median values of $e_{\rm bar}$ in each band.  The tail of strong bars with $e_{\rm bar}$ $\sim$ 0.6 to 0.8 becomes increasingly prominent in the two lower redshift bins ($z\sim$ 1--2 and $z\sim$ 0.5--1). See \S~\ref{sec:bar-measurement} for details.    }
\label{fig:ebar_distribution}
\end{figure}
We consider the maximum projected ellipticity of the bar $e_{\rm bar}$ ($e_{\rm bar}$ = $1 - b/a$)
as one measure of bar strength as it correlates  with empirical measures of bar
strength, such as the gravitational bar torque \citep{Laurikainen-Salo-Rautianinen2002} and
with theoretical measures of bar strength based on the relative
amplitude of the bi-symmetric (m=2) Fourier component of the mass
density distribution \citep{Shen-Sellwood2004}. 
In the radial profile of projected ellipticity from the ellipse fits
(see Figure~\ref{fig:efit}), the ellipticity
rises smoothly to a maximum value  in the bar-dominated 
region and we take this maximum value as  $e_{\rm bar}$. 
\par
Figure \ref{fig:ebar_distribution} shows the distribution of $e_{\rm bar}$ measured in F200W images and in F444W images. The projected $e_{\rm bar}$ in our sample ranges from 0.25 to 0.8. Note that for the histogram of bars, we use three redshift bins: 0.5--1, 1--2, and 2--4. In each redshift bin, the distribution of $e_{\rm bar}$ is skewed towards higher values in the F200W images. This can be partially explained by the fact that light from the bar component is less spread in the F200W image due to its sharper resolution. Therefore, for the intrinsically shorter and thinner bars (e.g., $\sim$ 2.4 kpc, corresponding to 2 $\times$ the PSF in F444W images), they will appear rounder in F444W images compared to F200W images. Interestingly, we notice that the tail of strong bars with high $e_{\rm bar}$ ~0.6 to 0.8 becomes more prominent at later times. 

\subsection{Comparison to Barred Galaxies in Simulations }\label{sec:vs-theory}
\subsubsection{Comparison to Barred Galaxies in TNG50 Simulations}
\vspace{-2mm}
In this work, we compare our results to the bars identified in the {Illustris TNG50 cosmological simulations (\citealt{Marinacci-etal-2018,  Naiman-etal-2018, Nelson-etal-2018,  Springel-etal-2018, Pillepich-etal-2018, Pillepich-etal-2019, Nelson-etal-2019, Nelson-etal-2019-Release})}, which utilize a combination of Adaptive Mesh Refinement (AMR) and Smoothed Particle Hydrodynamics (SPH) techniques to model the dark matter and gas content. \textcolor{black}{TNG50 simulations {encompass a volume of $51.7^3$ (Mpc/h)$^3$} with a high resolution of $\sim$ 0.29--0.58 kpc over time, and incorporate stellar and black hole feedback. }

\begin{figure}[t!]

    \centering
    \includegraphics[width=0.47\textwidth]{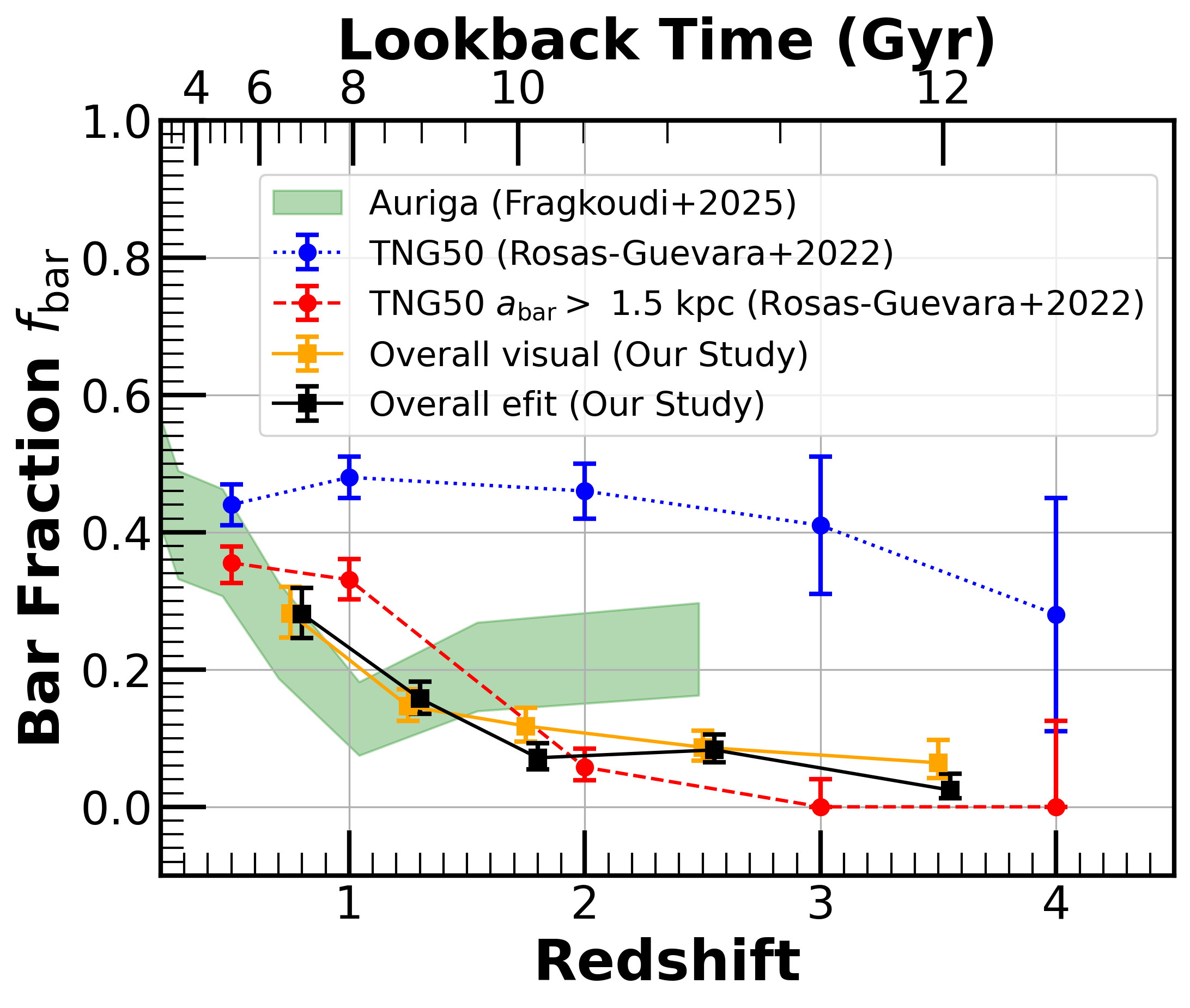}
    \caption{Our observed bar fractions based on bars detected in either the \textit{JWST} F200W or F444W image from ellipse fits (black) and visual classification (orange) are compared to bar fractions measured in TNG50 simulations (\citealt{Rosas-Guevara-etal-2022}) \textcolor{black}{and Auriga simulations (\citealt{Fragkoudi-etal-2025})}. Two measures of the TNG50 bar fraction are shown: dark blue circles (counting all bars) and red circles (only counting bars with semi-major axis $a_{\rm bar} > 1.5$ kpc).
    At $z \sim$ 2--4, the total TNG50 bar fraction is significantly larger than our observed bar fraction because TNG50 simulations show a large population of small bars with $a_{\rm bar}$ $< 1.5$ kpc that our data cannot robustly detect. 
    \textcolor{black}{Results from Auriga (\citealt{Fragkoudi-etal-2025}) are shown as the green-shaded region.  Unlike the TNG50 simulations, the Auriga simulations predict a decline in the bar fraction from $z \sim$ 0 to $z \sim$ 2.5. } \textcolor{black}{Specifically, their bar fraction decreases from $z \sim$ 0--1 and remains at approximately 20\% out to $z \sim$ 2.5}. 
    Our observed bar fraction is consistent with the bar fraction predicted by TNG50 simulations for large bars with $a_{\rm bar} >$ 1.5 kpc \textcolor{black}{at $z\sim$ 0.5--4} and with the Auriga simulations \textcolor{black}{out to $z \sim 1.5$}.}

\label{fig:TNG_Vs_observed_Fbar}
\end{figure}
\par

\par

In Figure \ref{fig:TNG_Vs_observed_Fbar}, we compare our results to the TNG50 {bar fraction} results from \cite{Rosas-Guevara-etal-2022}, who studied the evolution of bars in simulated galaxies with stellar mass  $\ge 10^{10}$ $M_\odot$. {Two measurements of the TNG50 bar fraction are shown: the total bar fraction counting all bars and the bar fraction counting only bars with deprojected semi-major axis $a_{\rm bar} > 1.5$ kpc}. 

\par
At redshifts of $z \sim$ 0.5, 1, 2, 3, and 4, \cite{Rosas-Guevara-etal-2022} identified 141, 125, 48, 10 and 2 barred galaxies and obtained a total bar fraction of  $44\pm3\%$, $48\pm3\%$, $46\pm4\%$, $41\pm10\%$, and $28\pm17\%$.
Due to the \textcolor{black}{limited} volume of the simulations, only a few disk galaxies, as well as barred galaxies, were found in the high-redshift  $z\sim$ 3--4 bin (\citealt{Rosas-Guevara-etal-2022}). 

\par
\begin{figure}[t!]

    \centering
    \includegraphics[width=0.47\textwidth]{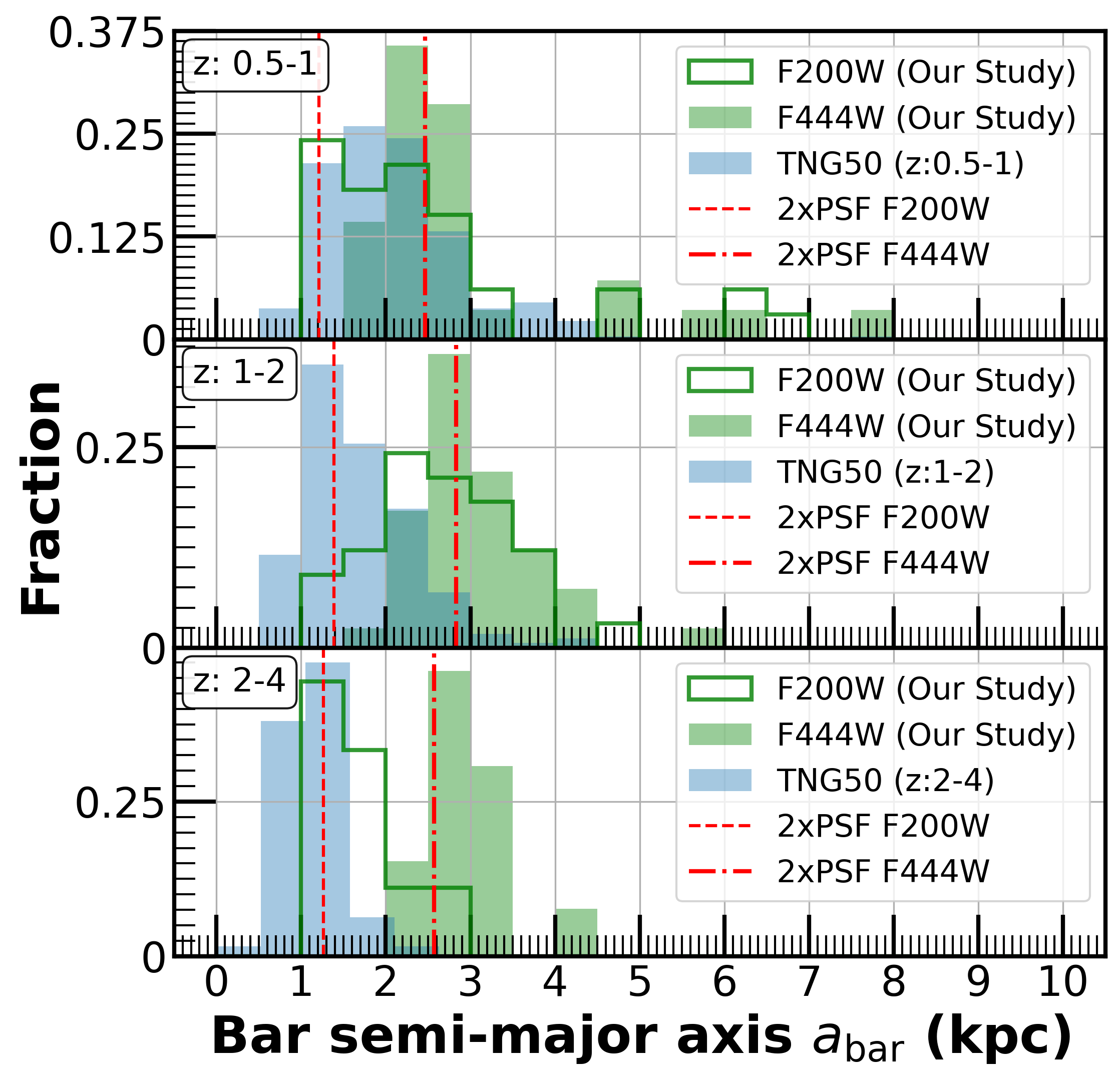}
    \caption{ Distribution of deprojected bar semi-major axis $a_{\rm bar}$  measured in TNG50 simulations (\citealt{Rosas-Guevara-etal-2022}; shaded blue histograms) compared to the distribution of projected lengths for bars detected via ellipse fits in F200W images (unfilled green histograms) and in F444W images (filled green histograms) in our study. {Measurements of $a_{\epsilon}$ are used as $a_{\rm bar}$ in this figure for our study.} For the TNG50 results, the plots at $z \sim$~0.5--1, $z \sim$~1--2, and $z \sim$~2--4,  combine the results from TNG50 snapshots at $z=$0.5 and $z=$1, $z=$1 and $z=$2, and $z=$2, $z=$3, and $z=$4, respectively. Bars with a size 2.0 times the PSF can be robustly identified via ellipse fits, and the vertical red dashed lines represent this limit.  When comparing TNG50 results to our observational study, it should be noted that bar sizes in TNG50 simulations are deprojected, while those in our study are projected values.  At $z \sim$ 2--4, compared to our results, TNG50 simulations show a much larger population of small bars with $a_{\rm bar}$ $< 1.5$ kpc and a lack of larger bars with $a_{\rm bar} >  2$ kpc. This explains the TNG bar fractions in Fig \ref{fig:TNG_Vs_observed_Fbar}. }
\label{fig:abar-TNG}
\end{figure}

{Figure \ref{fig:abar-TNG} shows the distribution of bar lengths in the simulations overlaid on the bar lengths of our bars detected in F200W and F444W images in our study.} In the TGN50 simulations, the median bar size decreases from 2.1 kpc at $z\sim~0.5$ to 1.1 kpc at $z\sim$ 2--4. The {bottom panel} of Figure \ref{fig:abar-TNG} shows that at $z\sim$ {2}--4, TNG50 simulations show a large population of small bars with $a_{\rm bar} < 1.5$ kpc that our data cannot robustly detect. This explains why the total TNG50 bar fraction is significantly larger than our observed bar fraction in Figure \ref{fig:TNG_Vs_observed_Fbar}. {When we only include \textcolor{black}{TNG50 results for larger bars} with $a_{\rm bar} > 1.5$ kpc, the TNG50 bar fraction becomes more similar to our observed bar fraction in Figure \ref{fig:TNG_Vs_observed_Fbar}.}
\par
{The bottom panel of Figure \ref{fig:abar-TNG} also shows that at $z\sim$ 2--4, TNG50 simulations not only produce a large population of small bars (with $a_{\rm bar} < 1.5$ kpc), but they also fail to produce the larger bars with $a_{\rm bar} > 2$ kpc, which are {observed in our study (see Figure \ref{fig:TNG_Vs_observed_Fbar})}. 
Besides our study, large bars are also detected at $z >2$ in other studies, including \cite{Huang-etal-2023}, \cite{Guo-etal-2023}, \cite{Costantin-etal-2023}, and \cite{Le-Conte-etal-2024}. For example, \cite{Huang-etal-2023} reported a bar with a bar length of $\sim$ 7.5 kpc at $z \sim 2.46$ and \cite{Costantin-etal-2023} reported  a bar at $z_{phot} \sim 3$ with a bar length of 3.3 kpc.}

\par
\begin{figure}
    \centering
    \includegraphics[width=0.45\textwidth]{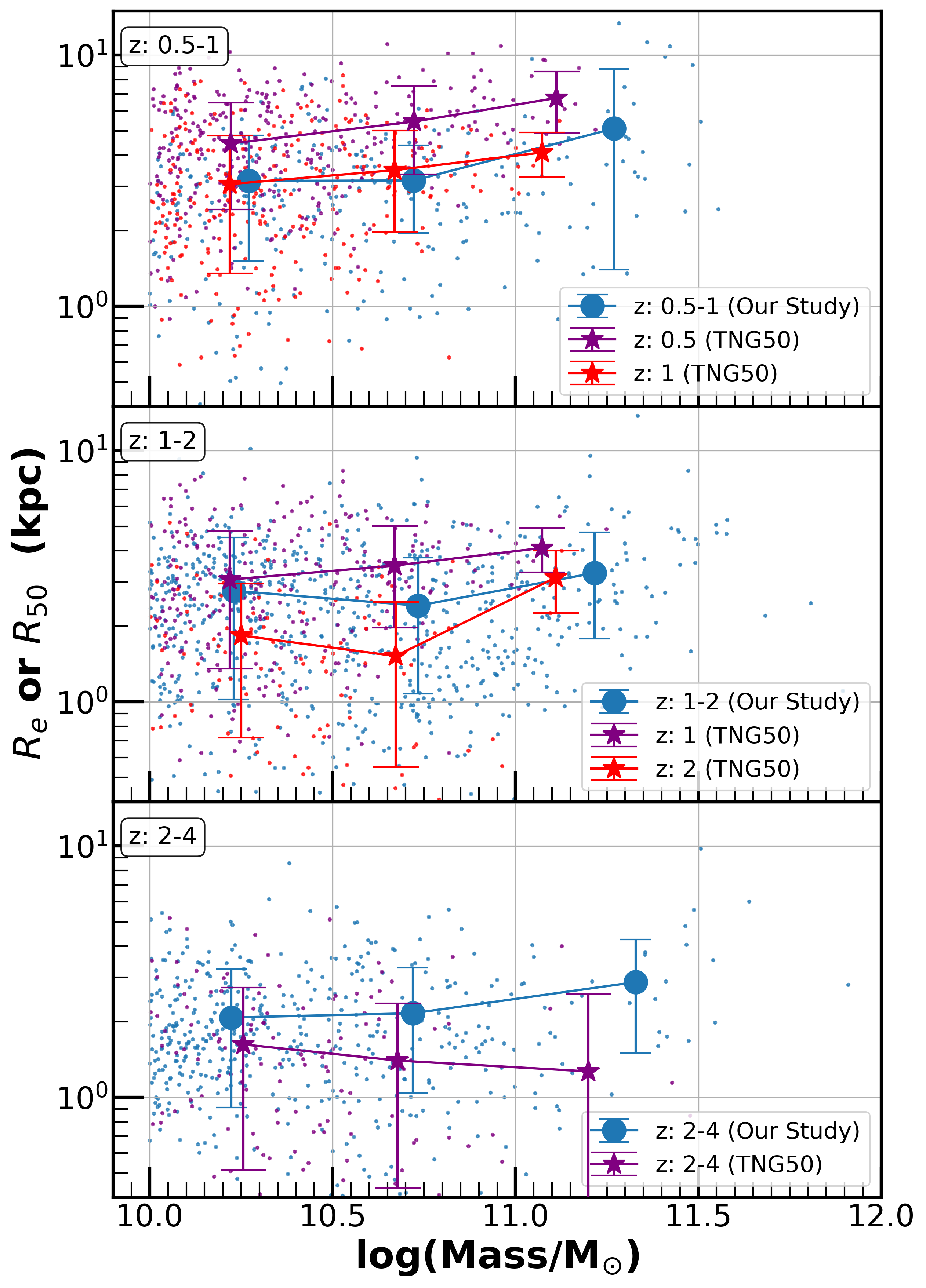}

\caption{{The {galaxy} size is plotted versus stellar mass for the moderately inclined disk galaxies in our data (blue dots) and disk galaxies in TNG50 simulations (\citealt{Rosas-Guevara-etal-2022}) (purple and red squares) in three redshift bins ($z \sim$ 0.5--1 in top panel, $z \sim$ 1--2 in middle panel,  and $z \sim$ 2--4 in bottom panel). For the TNG50 simulations, the galaxy size plotted is $R_{50}$, defined as the radius within which 50 percent of the total stellar mass is contained (\citealt{Rosas-Guevara-etal-2022}; Y. Rosas-Guevara, private communication). For our data, the galaxy size plotted is the effective radius $R_{e}$ in the F356W images (McGrath et al., in prep.). In the top two panels at $z \sim$ 0.5--2, the average mass-size relation is qualitatively similar in the data and TNG50 simulations. However, at $z \sim$ 2--4, sizes of disk galaxies are, on average, smaller in TNG50 simulations than in the data.}}

\label{fig: Appen-C}
\end{figure}

{It is legitimate to wonder whether the large population of small bars and the apparent lack of large bars in TNG50 simulations at $z\sim$ 2--4 could potentially be due to these simulations producing smaller galactic disks than those in the data.  We explore this possibility in Figure \ref{fig: Appen-C}, which shows the {galaxy} size versus stellar mass for the {moderately inclined disk galaxies in our data and the disk galaxies in TNG50 simulations (\citealt{Rosas-Guevara-etal-2022}) in three redshift bins}.  This is not a strict apples-to-apples comparison, as slightly different definitions of sizes are used in the data and simulations. For the TNG50 simulations, the {galaxy} size shown is $R_{50}$, defined as the radius within which 50 percent of the total stellar mass is contained (\citealt{Rosas-Guevara-etal-2022}; Y. Rosas-Guevara, private communication). For our data, the {galaxy}  size shown is the effective radius $R_e$ in the F356W images (McGrath et al., in prep.), defined as the semi-major axis from within which half of the galaxy light is contained. We find that at $z \sim$ 0.5--2 (the top two panels of Figure \ref{fig: Appen-C}), the average mass-size relation is qualitatively similar in the data and TNG50 simulations. However, at $z \sim$ 2--4 (bottom panel of Figure \ref{fig: Appen-C}), disk sizes are, on average, smaller in TNG50 simulations than in the data. This bias of TNG50 simulations toward smaller disk sizes could partly explain their bias towards small bar sizes, assuming scenarios where bar and disk growth are interrelated. 
\par 
\textcolor{black}{In summary, our observed bar fraction from $z \sim$ 0.5--4 is consistent with the bar fraction predicted by TGN50 simulations for bars with semi-major axis $a_{\rm bar} > 1.5$ kpc.  However, TNG50 simulations predict a large population of small bars with $a_{\rm bar} < 1.5$ kpc that our data cannot robustly detect.  If such a population of small bars really exists, the true bar fraction may be significantly higher than our results.}

\par

\subsubsection{\textcolor{black}{Comparison to Barred Galaxies in Auriga Simulations }}
\par 
\textcolor{black}{We compare our results with the predicted bar fraction from the recent study \cite{Fragkoudi-etal-2025} that used a suite of 39 Milky-Way-mass halos from Auriga cosmological simulations. The results from \cite{Fragkoudi-etal-2025} are shown as green shaded region in Figure \ref{fig:TNG_Vs_observed_Fbar}.}

\par 
\textcolor{black}{Auriga is a suite of gravo-magnetohydrodynamical
zoom-in cosmological simulations running with the moving-mesh code \textsc{AREPO} (\citealt{Springel-2010, Pakmor-etal-2016}), including a galaxy formation physical model that couples cosmic evolution of dark matter, gas, stars, and supermassive black holes (\citealt{Grand-etal-2018, Grand-etal-2019, Grand-etal-2024}). The gas and stars have a mass resolution down to $\sim 5 \times 10^4 $ M$_\odot$, with a high spatial resolution of $\le$ 0.37 kpc. Galaxies formed in Milky-Way-mass halos are massive with $M_\star >$ 10$^{10}$ M$_\odot$ (\citealt{Grand-etal-2024}). }

 \par
\textcolor{black}{
\cite{Fragkoudi-etal-2025} investigate the bar fraction in a suite of 39 Milky-Way-mass halos from Auriga cosmological simulations. Unlike the TNG50 simulations, the Auriga simulations predict a decline in the bar fraction from $z \sim$ 0 to $z \sim$ 2.5. Specifically, their bar fraction decreases from $z \sim$ 0--1 and remains at approximately 20\% out to $z \sim$ 2.5 (Figure \ref{fig:TNG_Vs_observed_Fbar}.)  Our observed bar fraction at $z \sim$ 0.5--2.5 seems broadly consistent with the bar fraction in Auriga simulations} \textcolor{black}{out to $z \sim 1.5$} 
(Figure \ref{fig:TNG_Vs_observed_Fbar}.) }
 
 \par
\textcolor{black}{However, the comparison between the Auriga and TNG50 simulations, or between Auriga and our data, is not an apples-to-apples comparison for several reasons. Firstly, the Auriga simulations focus on a select sample of 39 Milky-Way-mass halos, while the TNG50 sample and our sample are a mass-complete sample for $M_{\star} > 10^{10} M_{\odot}$.  Another important factor is that most of the bars in Auriga simulations at $z \sim$ 2--3 are short-lived features. For example, at  $z \sim$ 2--3, the bar fraction predicted in Auriga simulations is $\sim$ 23\% (shown in Figure 4 in \cite{Fragkoudi-etal-2025}), corresponding to 9 out of  39 galaxies are barred, but only one of them is a long-lived bar that is documented in Table 1 in \cite{Fragkoudi-etal-2025}. We discuss how the lifetime of the bar can complicate our bar fraction interpretation in \S\ref{sec:5.3}. In particular, if the lifetime of the bar is shorter than $\sim$ 1.1 Gyr (time between $z \sim$ 2--3), then we are likely to miss some of those short-lived bars in our observations.}

\par
\textcolor{black}{
Another complicating factor is that the \cite{Fragkoudi-etal-2025} paper does not provide bar sizes for the short-lived bars that dominate the Auriga bar fraction at higher redshifts and, instead, only provides the sizes for the long-lived bars at the time of formation. Therefore, it is not possible for us to evaluate whether our observed bar fraction and the Auriga bar fraction are sampling similarly sized bars. It is worth noting that Auriga simulations successfully produce large bars with bar lengths as large as a few kpcs at $z \sim $ 2--3. The large bars are likely formed after significant mergers at high redshifts (\citealt{Fragkoudi-etal-2025}). The presence of large bars in Auriga simulations at $z \gtrsim 2$ is consistent with our study and other $JWST$ studies (e.g., \citealt{Huang-etal-2023, Costantin-etal-2023, Le-Conte-etal-2024}).}

\subsection{Impact of Redshift-Dependent Systematic Effects on Observed Bar Fraction}\label{sec:towards-intrinsic}

We presented the overall observed bar fraction derived from ellipse fits and from visual classification in \S~\ref{sec:HST+JWST}, and found the observed bar fraction declines from $z \sim$  0.5 to $z \sim$ 4. {If taken at face value, this would suggest that disk galaxies hosting bars become increasingly rare at early epochs (with only $\sim$ 2.4\%--6.4\% of disk galaxies being barred at $z \sim$ 3.5)}. However, we stress that the observed bar fraction {may be} a lower limit to the intrinsic bar fraction due to observation-limited redshift-dependent systematic effects (e.g., impact of PSF, cosmological SB dimming of the outer disk, impact of dust and recent SF, and high-redshift features that complicate traditional bar identification) that cause us to miss bars in our analyses.

\par
The spatial resolution, which is limited by the PSF FWHM, is one of the main factors. The limited spatial resolution will cause both visual classification and ellipse fits to miss small bars. Studies of bars at $z \sim$ 0 are less affected by the PSF and often can reach a spatial resolution of {$\sim$ {0.2--0.5 kpc}} (e.g., \citealt{Marinova-Jogee2007, Menendez-Delmestre-etal-2007, Buta-etal-2015}). \cite{Erwin2018} has carefully compared the observed bar fraction at $z \sim$ 0 from SDSS (e.g., \citealt{Skibba-etal-2012,Cheung-etal-2013}) and from \textit{Spitzer} based studies (S$_4$G) (e.g., \citealt{Buta-etal-2015, Diaz-Garcia-etal-2016}) and found that the different spatial resolution of different instruments accounts for the discrepancy in the observed bar fractions.  
In our analyses, we try to mitigate the effect of PSF by looking at the observed bar fraction in F200W images, whose PSF FWHM corresponds to $\sim$ 0.6--0.7 kpc at $z\sim$ 0.5--4. {
Ellipse fits can effectively identify bars with $a_{\rm bar}>$  1.5--2 times the PSF, and the F200W images allow us to robustly identify bars  with $a_{\rm bar}$ {$\gtrsim 1.5$ kpc.}
}
\par
If we assume that the sizes of bars at  $z \sim 0$ (e.g., \citealt{Erwin2018}) were unchanged at higher redshifts, then in our sample containing disk galaxies at $M_{\star}> 10^{10} M_{\odot}$ we would expect over 90\% of bars would have bar lengths {greater than $\sim$ 1.5 kpc} (see Figure 8 in \cite{Erwin2018}) and such bars would likely to be picked up by the NIRCam F200W images. 

\par
However, given the evolution in disk sizes (e.g., \citealt{Van-ver-Wel-etal-2014, Ward-etal-2024}),  with disks becoming smaller from $z \sim 0.5$ to $z \sim$ 3--4 by $\sim$ 40--60\%, we might reasonably expect bars to become smaller at earlier epochs. It is hard to predict the exact impact of disk evolution on bar lengths as the {ratio of bar length to disk size has a wide range of values at $z \sim 0$ ($a_{bar}/R_{25} \sim 0.1 - 0.8$ and $a_{bar}/h \sim 0.2 - 1$; see \citealt{Marinova-Jogee2007}, \citealt{Gadotti-2011} and \citealt{Erwin2018})}. %
For the sake of argument, {we use the $R_e$ measurements from McGrath et al. (in prep.) and assume $R_e$ measurements are about one-third of the $R_{\rm 25}$ measurements (e.g., \citealt{Leroy-etal-2021}) to get the distribution of $R_{\rm 25}$ for our disk galaxies at $z \sim$ 2--4. Then, we assume a median bar-to-disk length ($a_{bar}/R_{25}$) of $\sim 0.4$ to estimate the distribution of bar lengths at $z \sim$ 2--4. According to this roughly estimated distribution, $\sim$ 50\% of the bars would have $a_{\rm bar}<$ 1.5 kpc at $z \sim$ 2--4 and thus not be robustly detected by F200W images.}

\par
Another factor that can affect bar identification at high redshifts is the cosmic surface brightness (CSB) dimming of the outer disk. CSB dimming can cause the outer disk to be increasingly dimmed (factor of $\sim (1+z)^{-4}$) at higher redshifts, and if this prevents the outer disk from being clearly identifiable, the bar-disk transition will not be evident and the barred galaxy may be misidentified (e.g., \cite{Jogee-etal-2002a}). We try to restrict this effect by applying a stellar mass cut where we can focus on more massive sources, which are, in general, brighter according to the mass-luminosity relation. Nevertheless, we note that cases of barred galaxies can be missed because of the faint disks. One example is the barred galaxy at $z \sim 3$ reported in \cite{Costantin-etal-2023}. The galaxy is not classified as a barred galaxy during visual classifications (\citealt{Kartaltepe-etal-2023}) but is later identified when using quantitative methods like stacking, ellipse fitting, and multi-component image fitting.  \cite{Liang-etal-2023} explored 
a set of artificially redshifted DESI images observed in \textit{JWST} CEERS and pointed out that the impact of dimming and noise on bar identification is not significant when the disk galaxy in CEERS has $M_{\star}>10^{9.75}$$M_\odot$ at $z<3$. Thus, we might not expect surface brightness dimming of the outer disk to have a large effect on bar identification in massive galaxies at $z<3$, but this effect may get increasingly important at higher redshifts. Nonetheless, more investigation on the effect of CSB dimming is needed.  

\par
The effect of dust and star formation has been addressed to some extent by our strategy of independently deriving observed bar fractions in F200W and F444W images. Even though the rest-frame optical light can be 
blocked by dust 
the rest-frame NIR light can penetrate through the dust, revealing the stellar structure that is predominately composed of lower-mass stars (e.g., \citealt{Frogel-Quillen-Pogge1996,Meidt-etal-2014, Suess-etal-2022}). However, due to the loss of resolution in the F444W images, it is unavoidable that dust-obscured short bars ($\lesssim$ 2.4 kpc) fail to be identified in both F200W and F444W images. 
However, we note that the impact of recent SF may not be a dominant factor in missing strong bars as both observational studies and theoretical simulations show that SF does not tend to happen along strong bars (e.g., \citealt{James-etal-2009, James-Percival-2016, Donohoe-Keyes-etal-2019}).

\par 
While we have discussed many systematic effects that can cause us to miss bars and underestimate the bar fraction, it is important to note that some factors can also cause us to overestimate the bar fraction.   For example, both ellipse fitting and visual classification can lead to falsely classified bars by mistaking other features, such as clumps or merging events for bars, due to the limited resolution and wavelength range of the observations we have. To firmly confirm the existence of high-redshift bars, follow-up observations on galaxy kinematics are needed.

  \subsection{Nearby Neighbors of Barred Galaxies} \label{sec:nearby neighbors}
  \par
 Bars can form spontaneously in isolated disks or be induced by minor mergers or tidal interactions (e.g., \citealt{Hernquist-Mihos1995,Izquierdo-Villalba-etal-2022, Bi-Shlosman-Romano-Diaz2022, Ansar-etal-2023}).
 In the pilot paper \cite{Guo-etal-2023}, we noted that the majority of the six barred galaxies at $1.1 \le z \le 2.3$ shown in the paper appear to have nearby sources that could be potential companions. 
 \par
 Similarly, many barred galaxies identified in this paper also appear to have nearby neighbors (e.g., CEERS-86093 in Figure \ref{fig:Example-barred-galaxies}). For a sample of barred galaxies (robustly identified via at least one method in at least one band; see \S~\ref{sec:bar-efitobs} and \S~\ref{sec:bar-visobs}) and a control sample of moderately inclined unbarred disk galaxies, {we calculate the proportion of galaxies that show at least one nearby source within an angular separation of  2\farcs5 (corresponding to $\sim$ 20--23 kpc at $0.5 \le z \le 4$) and with $|\Delta z/(1+z)| <$~$\sigma_{\rm NMAD}(z_{phot})\sim 0.04$.}

 \par 
{At $z<1.5$, we find that the proportion of barred galaxies with nearby sources \textcolor{black}{($43.0\pm 3.1\%$)} is not statistically different from that of unbarred disk galaxies \textcolor{black}{($38.9\pm 1.8\%$)}. However, at $z \ge 1.5$, the proportion of barred galaxies with nearby sources \textcolor{black}{($55.6\pm 5.0\%$)} appears to be statistically different from that of unbarred disk galaxies \textcolor{black}{($37.1\pm 1.5\%$)}, as indicated by the chi-squared test resulting in a p-value $\sim$ \textcolor{black}{0.01}. }
 
 \par 
 These preliminary results raise the tantalizing possibility that at early epochs  ($z \ge 1.5$), bars may preferentially form in tidally interacting galaxies. This scenario would be consistent with the idea that at early epochs, it may be harder for bars to form spontaneously in isolated disk galaxies as these early disk galaxies may have properties that are not conducive to bar formation, such as being dynamically hot and hosting gas and stars with high turbulence and velocity dispersions.
However, a more detailed analysis of tidally induced features and supplement spectroscopic redshifts may be needed to conclude that bars at high redshifts tend to form via tidal interaction. 

 \par
 At present, we present these results as preliminary until we are able to conduct a more thorough analysis. This would include a more detailed comparison of the target and control samples, {multiple methods to identify potential companions,} an inspection of potential tidally induced features in the putative galaxy pairs, and the inclusion of spectroscopic redshifts for the faint neighbors to confirm that they are real companions.

\par

\section{Discussion}\label{discussion}
 In this section, we discuss the potential implications of the results presented in section \S~\ref{sec:results}.

\subsection{The Emergence of Barred Galaxies at $z\sim$ 3--4}
\par
Our observational results apply mainly to bars with projected semi-major axis $a_{\rm bar}$ $> 1.5 $ kpc ( $\gtrsim$ 2 $\times$ PSF of F200W images) that can be robustly traced by ellipse fits of our data. For such bars, the bar fraction at $z\sim$ 2--4 is low, with values  $\sim 9\%$ at $z\sim 2.5$  and $\lesssim 6\%$ at $z \sim 3.5$. However, redshift-dependent systematic effects (e.g., PSF, SB dimming, dust, SF; \S~\ref{sec:towards-intrinsic}) might cause us to miss a subset of bars. Furthermore, at $z\sim$ 2--4, compared to our results, TNG50 simulations show a significantly larger bar fraction due to a large population of small bars with $a_{\rm bar}$ $< 1.5$ kpc that our data cannot robustly trace. If such a population exists, the true bar fraction may be significantly higher than our observed bar fraction.

\par
If the true bar fraction is low at $z > 2$, it would support the idea that most of the disks at $z>2$ 
have properties that are not conducive to bar formation or survival. These properties include being dynamically hot, turbulent, and having thick disks (e.g., \citealt{Genzel-etal-2011, Pillepich-etal-2019, Wisnioski-etal-2019, Birkin-etal-2023, Hamilton-Campos-etal-2023}). Inside dynamically hot (thick) disks, bar formation can be suppressed or delayed (e.g., \citealt{Kraljic-Bournaud-Martig2012,Sheth-etal-2012,Aumer-etal-2017, Ghosh-etal-2023}), leading to a low intrinsic bar fraction at $z>2$, when the Universe was $\lesssim$ 4 billion years old.  

\par
Conversely, if the true bar fraction is high (e.g., as suggested by the TNG50 simulations; see Fig \ref{fig:TNG_Vs_observed_Fbar}), then it would suggest early disks are quite susceptible to bar formation. It is notable that some high-redshift massive dynamically cold disk galaxies have been reported (e.g., \citealt{Neeleman-etal-2023,Kohandel-etal-2024}). 

\par
The very existence of bars at these early epochs challenges and constrains many theoretical models of disk and bar evolution. The predictions on bar fraction are widely different: some simulations find that bars only appear after $z\sim$ 1 (e.g., \citealt{Kraljic-Bournaud-Martig2012,Algorry-etal-2017}), while others find that bars already exist {at $z \gtrsim 3$} (e.g., \citealt{Rosas-Guevara-etal-2022,Bi-Shlosman-Romano-Diaz2022, Fragkoudi-etal-2025}). Our observations support the models where at least some bars form earlier in the disks.

\par
\cite{Bland-Hawthorn-etal-2024} discussed that when the baryonic mass is dominant in the inner disk, turbulent gas can lead to a shorter bar formation timescale, but they also find that the presence of the high fraction of gas weakens the strength and length of the bar, or even \textcolor{black}{transforms the bar to a central bulge}. The fraction of baryonic mass can also play an important role in the early formation of bars. {\cite{Bland-Hawthorn-etal-2023} studied the effect of disk mass fraction, defined as ``the ratio of disk mass to total
galaxy mass within the radius at which the rotation curve
roughly peaks,''  on bar formation. Their findings imply that if the disk mass fraction $\gtrsim$ 0.3, barred galaxies would be common at $z \sim$ 1--3 if the host disk galaxies are not dominated by massive central bulges. \textcolor{black}{\cite{Fragkoudi-etal-2025} also pointed out the importance of baryon dominance in bar formation, with barred galaxies on average being more baryon-dominated from high redshift down to $z \sim 0$ in the Auriga simulations.}

\subsection{Formation Mode of Bars}\label{sec: Formation Mode of Bars}

\par
Simulations and theoretical models show that bars can form through spontaneous m=2 instabilities in isolated galaxies (e.g., \citealt{Efstathiou-etal-1982, Hernquist-Mihos1995, Athanassoula-etal-1990, Kormendy-Kennicutt2004, Ansar-etal-2023}) or be induced via tidal interactions (e.g.,\citealt{Izquierdo-Villalba-etal-2022, Bi-Shlosman-Romano-Diaz2022, Ansar-etal-2023}). Rotationally supported massive cold stellar disks are prone to spontaneous instability (e.g., \citealt{Efstathiou-etal-1982, Ghosh-etal-2021}). At higher redshifts, when conditions for spontaneous bar formation are unfavorable, 
tidally induced bar formation can be particularly important. Especially, several theoretical studies find many of the bars form in interacting systems at high redshifts (e.g., \citealt{Romano-Diaz-etal-2008, Peschken-etal-2019,Bi-Shlosman-Romano-Diaz2022,Izquierdo-Villalba-etal-2022}).

\par
In \S~\ref{sec:nearby neighbors}, it is especially interesting that we found that compared to a control sample of unbarred galaxies, a larger fraction of barred galaxies at $z \gtrsim 1.5$ appear to have nearby apparent neighbors or/and show signs of tidal interaction. This raises the interesting possibility that early bars may be tidally triggered,  but more detailed future work using a control sample of unbarred galaxies and spectroscopic redshifts will be needed to confirm this result.

\subsection{Evolution and Impact of Barred Galaxies Over Time}\label{sec:5.3}

\par
{The observed bar fraction (see Figure \ref{fig:fbar-all}), the bar fraction predicted by TNG50 simulations for bars with $a_{\rm bar} > 1.5$ kpc, and \textcolor{black}{the bar fraction from the Auriga simulations} (see Figure \ref{fig:TNG_Vs_observed_Fbar}) rise over time. } 

\par
The increasing bar fraction may be the result of bars needing some time to form and grow large enough to be detectable. Different theoretical models predict bars form on a range of timescales, ranging from $\sim 1$ Gyr (e.g., \citealt{Bland-Hawthorn-etal-2024}) to  $\sim$ 2--5 Gyr (e.g., \citealt{Saha-Naab2013}). Therefore, as bars start to emerge at $z \sim$ 2--4 (corresponding to a lookback time $\sim$ 10.5--12.0 Gyr), a substantial fraction of bars should have been formed by $z \sim 0.5-1$ (corresponding to a lookback time $\sim$ 5.0--8.0 Gyr), {as observed in our study.}

\par
Additionally, the timescale of bar formation depends on whether the stellar disk is dynamically cold (e.g., \citealt{Ghosh-etal-2021}), the properties (e.g., concentration, velocity dispersion) of gas in the disk (e.g., \citealt{Bland-Hawthorn-etal-2024}), and the properties (e.g., mass, size, spin) of the dark matter halo (e.g., \citealt{Athanassoula2003,Weinberg-etal-2007, Saha-Naab2013}). As time progresses, the stellar disk becomes dynamically colder, the gas fraction decreases, and the baryon fraction in the central region increases, all leading to disks being more prone to a relatively quick formation of bars. 

\par
{The lifetime of the bars can further complicate the observed evolution in the bar fraction over time. Some simulations (e.g., \citealt{Ghosh-etal-2023, Ansar-etal-2023, Fragkoudi-etal-2025}) and observations (e.g., \citealt{Gadotti-etal-2015, deLorenzo-Caceres-etal-2019}) suggest that some bars may be long-lived. However, other simulations show that under certain conditions, bars can also be destroyed or weakened by a larger central mass concentration (e.g., \citealt{Shen-Sellwood2004, Athanassoula-Lambert-Dehnen2005}), by episodic gas accretion (e.g., \citealt{Bournaud-Combes2002}), or by major mergers or minor mergers (e.g., \citealt{Ghosh-etal-2021}).  It is possible that some of the bars we observed are transient in young disks, and the transience of bars lowers the possibility of bars being observed in relatively short time intervals  (e.g., the highest redshift bin $3 \le z \le 4$ corresponds to a mere 0.6 Gyr). }

\par
{Bars drive the secular evolution of galaxies by efficiently fueling the gas to the central region, enhancing SF, and building bulges. As the bar fraction rises over time from $z\sim4$ to today, bar-driven secular evolution of galaxies becomes an increasingly important mechanism at later times ($z<1$). At an earlier time ($z>2$), when the Universe was less than 4 billion years old, alternative mechanisms, such as gas accretion \citep[e.g.,][]{Katz-etal-2003, Keres-etal-2005, Keres-etal-2012, Dekel-Birnboim2006,Faucher-Giguere-Keres2011}, galaxy mergers and tidal interactions \citep[e.g.,][]{Conselice-etal-2003,Kartaltepe-etal-2007, Jogee-etal-2009,Lotz-etal-2010}, likely played a dominant role in the growth and morphological transformation of galaxies.}

\section{Summary}

\par 
Over two decades, studies based on $\textit{HST}$ and now with $\textit{JWST}$  have pushed the explorations of barred galaxies from $z \sim 0$ to $z \sim 4$. Most $\textit{HST}$ results go out to $z \sim 1.2$ and are mainly based on rest-frame optical images. At $z > 1.5$, \textcolor{black}{two of the earliest $\textit{JWST}$ bar fraction studies }include {LC24 and our study going out to $z \sim 4$ based on visual classification and ellipse fits of both F200W and F444W images with quantitative criteria.}

\par 
In this work, we use {\textit{JWST}} CEERS data to present {the first estimate} of the observed bar fraction and bar properties out to $z\sim$ 4 when the Universe was \textcolor{black}{$\sim$ 11\%} of its present age. We analyze a sample of \textcolor{black}{1770 galaxies with $M_{\star}>10^{10}M_{\odot}$} and identify barred galaxies \textcolor{black}{from 839 moderately inclined disk galaxies} via ellipse fits and visual classification using the F200W and F444W images. The two bands are selected to leverage both the sharper spatial resolution of F200W images (PSF $\sim 0\farcs08$ corresponding to $\sim$ 0.6--0.7 kpc at $0.5 \le z \le 4$) and the longer rest-frame wavelength traced by F444W ($\lambda_{\mathrm{rest}} \gtrsim 1 \mu$m out to $z \sim 3.5$). 

Our main findings are described below:

\begin{enumerate}
    
\item  
{For bars with projected semi-major axis $a_{\rm bar}$ $> 1.5 $ kpc ($\sim$ 2 $\times$ the PSF in the F200W images), the bar fraction at $z\sim$ 2--4 is low ($<10\%$), and they appear to be emerging at least as early as $z\sim 4$ when the Universe was \textcolor{black}{$\sim$ 11\%} of its present age.} The very existence of bars at these early epochs challenges and constrains many theoretical models of disk and bar evolution. Disk galaxies at those epochs may be significantly different from present-day disks in terms of their gas fraction, turbulence, and the extent to which they are dynamically cold, and it is notable that bars can form in them.

 \item
 \textcolor{black}{Our observed bar fraction is consistent with the bar fraction from $z \sim$ 0.5--4 predicted by TNG50 simulations for bars with a semi-major axis $a_{\rm bar}$ $> 1.5$ kpc.  However, TNG50 simulations predict a large population of small bars with $a_{\rm bar}$ $< 1.5$ kpc that our data cannot robustly detect.  If such a population of small bars exists, the true bar fraction at $z \sim$ 2--4 may be significantly higher than our results. Our observed bar fraction is also broadly consistent with the bar fraction \textcolor{black}{from $z \sim$ 0.5--1.5} predicted by the Auriga simulations using a sample of 39 Milky-Way-mass halos.}

\item 
Bars can form through spontaneous $m=2$ instabilities in isolated galaxies or {be induced} via tidal interactions. We present preliminary results suggesting that at $ z \geq 1.5 $, our sample of barred galaxies shows a larger frequency of nearby neighbors or/and signs of tidal interactions than a control sample of unbarred galaxies. These preliminary results raise the tantalizing possibility that at early epochs ($ z \geq 1.5 $), bars may preferentially form {via tidal interaction}. However, future detailed work will be needed to confirm this important result.

\item  
As {cosmic} time progresses from $z \sim 4$ to $z \sim 0.5$, we see a rise in bar abundance, bar length, and bar strength. The average projected bar length increases from  $\sim$ 1.7--2.5 kpc at $z \sim$ 2.5 to {$\sim$ 2.6--3.8 kpc at $z \sim$ 0.75} and the tail of strong bars with high $ e_{\rm bar} \sim 0.6 $ to 0.8 becomes more prominent at later times, showing the growth of bars over time.

\item 
{Bars are important in shaping the secular evolution of disk galaxies. As the bar fraction appears to rise from $z \sim 4$ to today, we conclude that bar-driven secular evolution will become progressively more important for driving galaxy evolution towards the present day.} 
{At earlier times, when the Universe was less than 4 billion years old ($z \gtrsim 2$), alternative mechanisms, such as gas accretion 
galaxy mergers and tidal interactions, 
likely played a dominant role in the growth and morphological transformation of galaxies.}
\end{enumerate}

\section*{Acknowledgements}
We acknowledge that the location where this work took place, the University of Texas at Austin, sits on indigenous land. The Tonkawa lived in central Texas, and the Comanche and Apache moved through this area. We pay our respects to all the American Indian and Indigenous Peoples and communities who have been or have become a part of these lands and territories in Texas on this piece of Turtle Island. 

YG and SJ acknowledge support from the Roland K. Blumberg Endowment in Astronomy, NSF grant AST 2244278, and the Heising-Simons Foundation grant 2017-0464. We acknowledge support from NASA through STScI ERS award JWST-ERS-1345. 

\textcolor{black}{The authors thank Yetli Rosas-Guevara and Francesca Fragkoudi for sharing information on the TNG50 and Auriga simulations, respectively.} The authors acknowledge the Texas Advanced Computing Center (TACC) at the University of Texas at Austin for providing HPC and visualization resources that have contributed to the research results reported within this paper. URL: http://www.tacc.utexas.edu







\appendix 

\section{Uncertainty in bar classifications}\label{APPEN-bar-classification-uncertainty}
\par
{In \S~\ref{sec:results} and in Figure \ref{fig:fbar-all}, we presented the overall observed bar fraction results with statistical errors as the error bars. For each method, the statistical errors are calculated using the sample size of robustly classified bars and the sample size of the moderately inclined disks. In this section, we explored a different potential error that relates to the uncertainty in bar classifications via visual classification and ellipse fits, respectively, and present the extent of the uncertainties in Figure \ref{fig: Appen-B}.}

\begin{figure}[h!]
    \centering
    \includegraphics[width=0.67\textwidth]{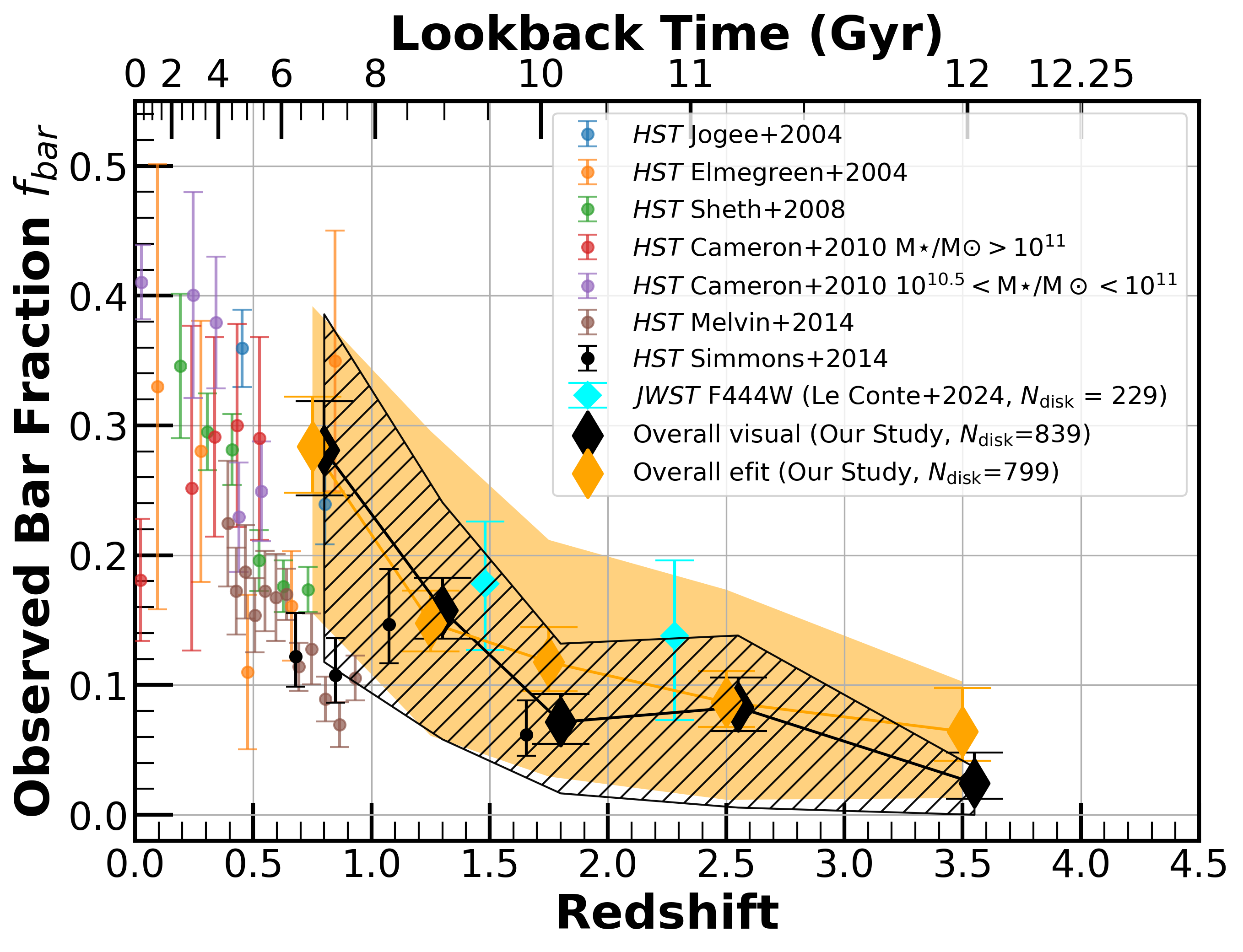}

\caption{{Lower and upper limits of the observed bar fraction, accounting for uncertainties in the classification of bars. The figure is adapted from Figure \ref{fig:fbar-all}. The black-hatched (orange-shaded) region represents the uncertainty in bar classifications via visual classification (ellipse fits). Including the uncertainties in classification, we still observe an early bar emergence and an increasing bar fraction from $z \sim$ 4 to  $z \sim$ 0.5. The large uncertainties highlight the complexities of identifying bars at higher redshifts. }}
\label{fig: Appen-B}
\end{figure}

\hspace{3mm}
\par
\textbf{Bar identification via Visual Classification:} {We explained the difficulties of classifying bars via visual classification in \S~\ref{sec:bar-vis-criteria}. In Figure \ref{fig:vis-bar-frac} and Figure \ref{fig:fbar-all}, we only reported robustly identified bars that have an average score $\ge $ 1.33. In this section, we try to estimate the uncertainties by gauging how the results would change if we used different criteria. Specifically, we set lower and upper limits to the observed bar fraction by considering bars identified using less stringent and more stringent criteria, respectively:
\begin{itemize}
    \item We define a lower limit of observed bar fraction (via visual classification) when we only count bars with an average score $\ge $ 2, corresponding to all classifiers classifying the source as likely barred or clearly barred in either band.
    \item We define an upper limit of observed bar fraction (via visual classification) when we count bars with an average score $\ge $ 1, which corresponds to at least 1 out of 3 classifiers classifying the source as clearly barred or all three classifiers classifying the source as likely unbarred, in either band.
\end{itemize} 

\par 
The lower and upper limits of the observed bar fraction via visual classification are presented as the lower and upper edge of the black-hatched region in Figure \ref{fig: Appen-B}. These limits illustrate the uncertainties in our visual bar fraction. It is notable that these uncertainties encompass our results and those of LC24, and that despite the uncertainties, the trend of a declining bar fraction with redshift remains.
\par

\hspace{3mm}
\par 
\textbf{Bar identification via Ellipse fits:} {When identifying bars with ellipse fits, we visually inspected the radial profiles of $e$ and PA, and the fitted ellipses overlaid on the original image to claim the sources are robustly identified bars. In this process, we assigned an internal confidence score to each set of ellipse fits. The scores are defined as 3 (clearly barred),  2 (likely barred),  1 (likely unbarred), and 0 (clearly unbarred): }
\begin{itemize}
     \item {A score of 3 (clearly barred) is assigned if the ellipse fits are good and show a clear bar signature that strictly follows the criteria in \S~\ref{sec:bar-efit-criteria} (e.g, $e_{\rm bar}>0.25$, $20^{\circ}< \Delta \theta_1$, \textcolor{black}{$\Delta \theta_2 > 10^{\circ}$}), and the ellipses overlaid on the image suggest that the bar signature is driven by the bar itself.}
     \item {A score of 2 (likely barred) rather than a score of 3 (clearly barred) is assigned if the fits show a bar signature that roughly meets the above criteria (e.g., $20^{\circ}< \Delta \theta_1 <30^{\circ}$, $\Delta \theta_2 <10^{\circ}$ but \textcolor{black}{an almost face-on disk} is present) or the quality of the fits is affected by the complicated morphology inside the disk (e.g., spirals, clumps, rings), and the ellipses overlaid on the image suggest that the bar signature is likely driven by the bar itself.}
     \item {A score of 1 (likely unbarred) rather than a score of 2 (likely barred) or a score of 0 (clearly unbarred) is assigned if the fits show a possible bar signature in its profiles, but the ellipses overlaid on the image suggest the signature is driven by other structures, or the quality of the fits is poor, or shows other similar issues.} 
     \item {A score of 0 (clearly unbarred) is assigned if the ellipse fits and radial profiles of $e$ and PA do not show any evidence of a bar.}

 \end{itemize}
{In Figure \ref{fig:observed-fbar-efit} and  Figure \ref{fig:fbar-all}, 
we only reported robustly identified bars that have a confidence score $\ge $ 2. In this section, we also define a lower limit of observed bar fraction (via ellipse fits) when we only count bars that have a confidence score of 3 (in either band) and an upper limit of observed bar fraction (via ellipse fits) when we count bars that have a confidence score of at least 1 (in either band). Similarly, the lower and upper limits are presented as the lower and upper edge of the orange shaded region in Figure \ref{fig: Appen-B} for the observed bar fraction derived from ellipse fits.}
\par
{At lower redshifts, the uncertainty in the observed bar fraction is consistent with the wide range of \textit{HST}--based results. At $z>1$, the upper limit of the observed bar fraction is roughly $\lesssim 5-15$ percentage \textcolor{black}{points} higher than our reported bar fraction for robustly identified bars. The large uncertainties highlight the complexities of identifying bars at higher redshifts. Overall, even considering the uncertainties in classifications, we still observe an early bar emergence and an increasing bar fraction from $z \sim$ 4 to  $z \sim$ 0.5. }

\section{Mass Dependence of the Observed Bar Fraction}\label{APPEN-mass}

\begin{figure}[t!]
    \centering
    \includegraphics[width=0.6\textwidth]{Fig7-all-bar-August.png}
    \includegraphics[width=0.47\textwidth]{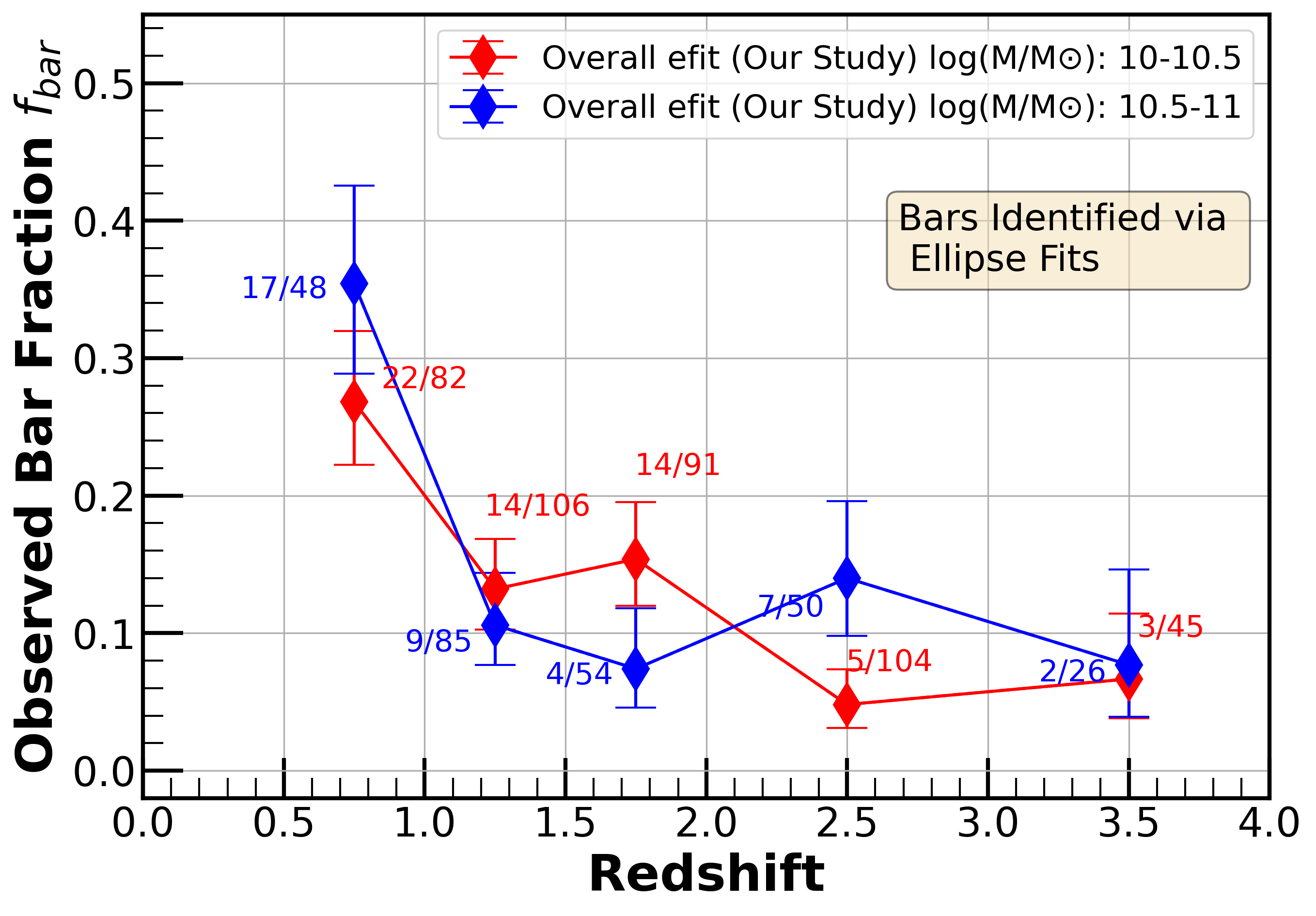}
    \includegraphics[width=0.47\textwidth]{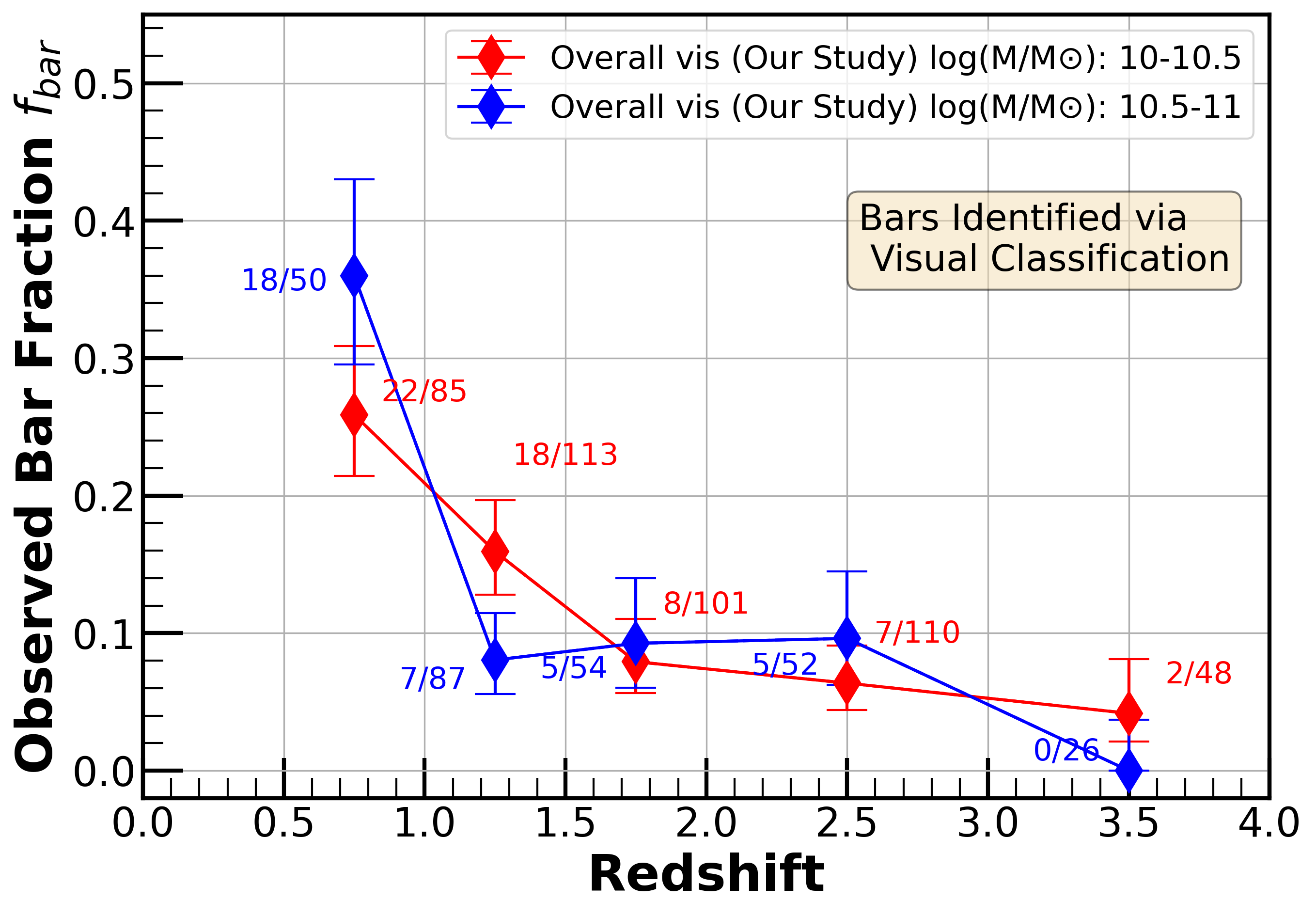}
\caption{{a) The top panel reproduces Figure \ref{fig:fbar-all} from \S~\ref{sec:HST+JWST}, which highlights our \textit{JWST}-based observed bar fraction across $z \sim$ 0.5--4 for a mass-complete sample with stellar mass $M_{\star} > 10^{10} M_{\odot}$. The lower panels explore how the results would change for different stellar mass ranges.
 b) Lower left and right panels:  The observed bar fraction based on ellipse fit (lower left panel) and visual classification (lower right panel) is shown for two stellar mass bins: $M_{\star} \sim 10^{10}-10^{10.5} M_{\odot}$ (red diamonds) and $M_{\star} \sim 10^{10.5}-10^{11} M_{\odot}$ (blue diamonds). For each data point, the number of bars identified and the size of the moderately inclined disk sample are annotated near the data point using the same color. The error bars are statistical errors. 
 Both panels show that the evolution of the observed bar fraction with redshift is similar for both stellar mass ranges.} }\label{fig: Appen-D}
   
\end{figure}

{Previous studies have shown that the evolution of the observed bar fraction with redshift at $z<1$ depends on the stellar mass range selected (e.g., \citealt{Cameron-etal-2010, Melvin-etal-2014, Erwin2018, Erwin-2023, Mendez-Abreu-etal-2023}). In Figure \ref{fig:fbar-all} and \S~\ref{sec:HST+JWST}, we showed our \textit{JWST}-based observed bar fraction across $z \sim$ 0.5--4 for a mass-complete sample with stellar mass $M_{\star} > 10^{10} M_{\odot}$. In this appendix, we explore how the evolution of the observed bar fraction with redshift shown in Figure \ref{fig:fbar-all} depends on the stellar mass range.

\textcolor{black}{At $z<1$, our results are consistent with previous studies (e.g., \citealt{Cameron-etal-2010, Melvin-etal-2014}) that find observed bar fraction is higher in more massive ($M_{\star}\gtrsim 10^{10.5} M_{\odot}$) disk galaxies. }In terms of the evolution of observed bar fraction at $z \sim 0.5-4$, as shown in Figure \ref{fig: Appen-D}, we find no significant difference in the bar fraction for galaxies in two stellar mass ranges:  $M_{\star} \sim 10^{10}-10^{10.5} M_{\odot}$ and $M_{\star} \sim 10^{10.5}-10^{11} M_{\odot}$. There are too few barred galaxies with $M_{\star} > 10^{11} M_{\odot}$ in our sample, so we do not discuss this mass range. }

\section{Comparison With \textit{HST}-based Studies}\label{APPEN-a-vs-HST-tests}
\par
{In \S~\ref{sec:HST+JWST} and Figure \ref{fig:fbar-all}, we noted that at $z \sim$ 0.5–1.5 where both \textit{HST} and $\textit{JWST}$ results exist, the $\textit{JWST}$-based observed bar fraction is up to two times higher than that from most \textit{HST} studies. We suggested in \S~\ref{sec:HST+JWST} that this difference likely arises because the higher spatial resolution and longer rest-frame NIR wavelengths of $\textit{JWST}$ NIRCam images can unveil a population of bars that were previously missed in $\textit{HST}$ data. In this Appendix, we further substantiate this suggestion through several tests.}

\par
{We can robustly Identify bars with size $a_{\rm bar} > 2$ times the PSF FWHM (see \S~\ref{sec: levarage-200and444} for details) during ellipse fits. This translates to bars with a size $a_{\rm bar}$$\gtrsim 1.5$ kpc when using $\textit{JWST}$ F200W images (PSF FWHM $\sim$ 0\farcs08, corresponding to $\sim$ 0.6--0.7 kpc across $z ~\sim$ 0.5--1) and bars with size $a_{\rm bar}\gtrsim 2.2$ kpc with \textit{HST} WFC3 F160W images (PSF $\sim$ 0\farcs13, corresponding to $\sim$ 1--1.2 kpc across $z\sim$ 0.5--1.5).  In the upper right panel of Figure \ref{fig: Appen-A}, we remove bars with $a_{\rm bar}\le 2.2$ kpc that would be robustly identified with $\textit{JWST}$ F200W images, but not with \textit{HST} WFC3 images. This leads to the removal of 16 bars in the redshift bin $z \sim$ 0.5--1, and lowers the overall $\textit{JWST}$ bar fraction based on ellipse fits from $\sim$ 28\% to $\sim$ 18\%, which is closer to the \textit{HST} bar fraction of $\sim$ 7--18\%.}

\begin{figure}[t!]
    \centering
    \includegraphics[width=1\textwidth]{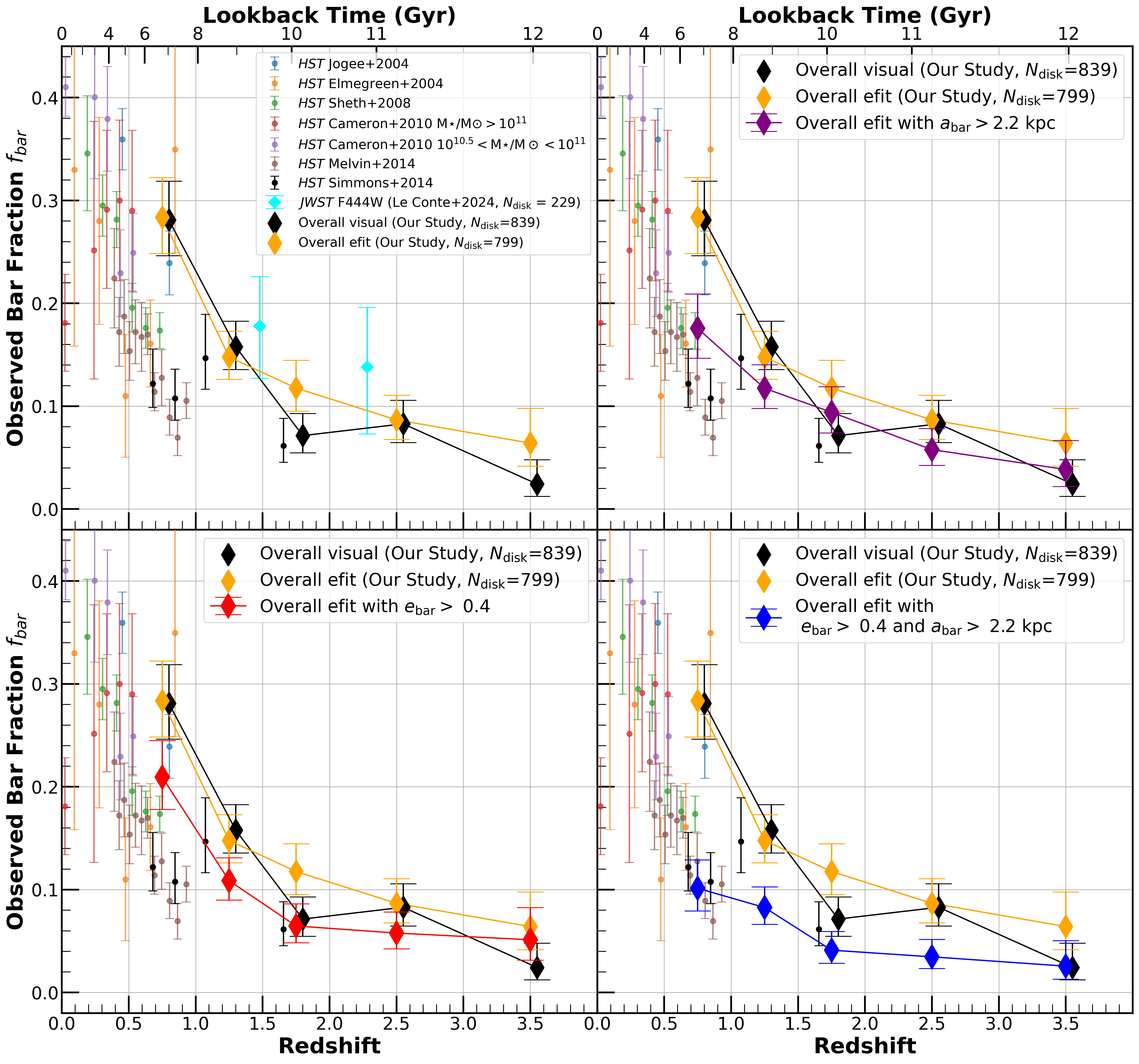}

\caption{{{(a) The upper \textcolor{black}{left} panel reproduces Figure \ref{fig:fbar-all} from \S~\ref{sec:HST+JWST}. The observed bar fraction at $z \sim$ \textcolor{black}{0.5--1} from our \textit{JWST}-based studies is $\sim$ \textcolor{black}{28}\%, which is approximately up to two times higher than the bar fraction reported by most \textit{HST} studies. In the other three panels, we explore potential reasons for this difference tied to the sharper PSF and longer rest-frame wavelength of \textit{JWST} data.
(b)Upper right panel: we remove bars with $a_{\rm bar}\le 2.2$ kpc that would be robustly identified with $\textit{JWST}$ F200W images, but not with \textit{HST} WFC3 images. This lowers the overall $\textit{JWST}$ bar fraction based on ellipse fits from $\sim$ 28\% to $\sim$ 18\%.
(c) Lower left panel: we remove bars with $e_{\rm bar} \le 0.4$ from the \textit{JWST} bar fraction. This lowers the overall \textit{JWST} bar fraction from $\sim$ 28\% to $\sim$ 21\%.
(d) Lower right panel: This shows the effect of removing both bars ($a_{\rm bar}\le 2.2 $ kpc) and weak bars ($e_{\rm bar} \le 0.4$) from the \textit{JWST}  bar fraction.
This lowers the overall \textit{JWST} bar fraction at $z \sim$ 0.5--1 from $\sim$ 28\% to $\sim$ 10\%, which is consistent with most of the \textit{HST} results.}}}

\label{fig: Appen-A}
\end{figure}

\par
{Another advantage of \textit{JWST} images over \textit{HST} WFC3 images is that they can more effectively identify {dust-obscured} bars due to their longer wavelength. \textit{JWST} F444W images trace longer rest-frame wavelength ($\lambda_{rest} \gtrsim 1 \mu m$) than \textit{HST} ACS or WFC3 images and can more effectively reveal dust-obscured bars. These obscured bars tend to preferentially be weak bars for the following reasons. A strong bar is less likely to be obscured by dust and recent SF because the large gas inflows and strong shear along the bar suppress the accumulation of gas and dust and the triggering of SF along the bar (e.g.,  \citealt{Jogee-Scoville-Kenney2005, James-etal-2009, Diaz-Garcia-etal-2020}).  Conversely, weak bars are more prone to have gas, dust, and SF along the bar and to be obscured by dust. To explore this effect, we remove bars with $e_{\rm bar} \le 0.4$ {measured in F444W images} from the \textit{JWST} bar fraction. This leads to the removal of 11 bars and lowers the overall \textit{JWST} bar fraction from $\sim$ 28\% to $\sim$ 21\%, which is closer to the \textit{HST} bar fraction of $\sim$ 7--18\% (lower left panel of Figure \ref{fig: Appen-A})
}

\par
{To assess the combined impact of \textit{JWST} data being able to more robustly identify shorter and weaker bars than \textit{HST} data, we removed both bars ($\le 2.2 $ kpc) and weak bars ($e_{\rm bar} \le 0.4$) from the \textit{JWST}  bar fraction (see lower right panel of Figure \ref{fig: Appen-A}).  This lowers the overall \textit{JWST} bar fraction at $z \sim$ 0.5--1 from $\sim$ 28\% to $\sim$ 10\%, which is consistent with most of the \textit{HST} results.} {The higher bar fraction measured in our study at $z \sim$ 0.5--1 highlight the power of \textit{JWST} NIRCam images.}



\newpage
\bibliographystyle{aasjournal}
\bibliography{bar-fraction} 

\providecommand{\noopsort}[1]{}
\begin{thebibliography}{}
\expandafter\ifx\csname natexlab\endcsname\relax\def\natexlab#1{#1}\fi
\providecommand{\url}[1]{\href{#1}{#1}}
\providecommand{\dodoi}[1]{doi:~\href{http://doi.org/#1}{\nolinkurl{#1}}}
\providecommand{\doeprint}[1]{\href{http://ascl.net/#1}{\nolinkurl{http://ascl.net/#1}}}
\providecommand{\doarXiv}[1]{\href{https://arxiv.org/abs/#1}{\nolinkurl{https://arxiv.org/abs/#1}}}

\bibitem[{{Abraham} {et~al.}(1999){Abraham}, {Merrifield}, {Ellis}, {Tanvir},
  \& {Brinchmann}}]{Abraham-etal-1999}
{Abraham}, R.~G., {Merrifield}, M.~R., {Ellis}, R.~S., {Tanvir}, N.~R., \&
  {Brinchmann}, J. 1999, \mnras, 308, 569,
  \dodoi{10.1046/j.1365-8711.1999.02766.x}

\bibitem[{{Algorry} {et~al.}(2017){Algorry}, {Navarro}, {Abadi}, {Sales},
  {Bower}, {Crain}, {Dalla Vecchia}, {Frenk}, {Schaller}, {Schaye}, \&
  {Theuns}}]{Algorry-etal-2017}
{Algorry}, D.~G., {Navarro}, J.~F., {Abadi}, M.~G., {et~al.} 2017, \mnras, 469,
  1054, \dodoi{10.1093/mnras/stx1008}

\bibitem[{{Amvrosiadis} {et~al.}(2024){Amvrosiadis}, {Lange}, {Nightingale},
  {He}, {Frenk}, {Oman}, {Smail}, {Swinbank}, {Fragkoudi}, {Gadotti}, {Cole},
  {Borsato}, {Robertson}, {Massey}, {Cao}, \& {Li}}]{Amvrosiadis-etal-2024}
{Amvrosiadis}, A., {Lange}, S., {Nightingale}, J., {et~al.} 2024, arXiv
  e-prints, arXiv:2404.01918, \dodoi{10.48550/arXiv.2404.01918}

\bibitem[{{Ansar} {et~al.}(2023){Ansar}, {Pearson}, {Sanderson}, {Arora},
  {Hopkins}, {Wetzel}, {Cunningham}, \& {Quinn}}]{Ansar-etal-2023}
{Ansar}, S., {Pearson}, S., {Sanderson}, R.~E., {et~al.} 2023, arXiv e-prints,
  arXiv:2309.16811, \dodoi{10.48550/arXiv.2309.16811}

\bibitem[{{Arrabal Haro} {et~al.}(2023){Arrabal Haro}, {Dickinson},
  {Finkelstein}, {Fujimoto}, {Fern{\'a}ndez}, {Kartaltepe}, {Jung}, {Cole},
  {Burgarella}, {Chworowsky}, {Hutchison}, {Morales}, {Papovich}, {Simons},
  {Amor{\'\i}n}, {Backhaus}, {Bagley}, {Bisigello}, {Calabr{\`o}},
  {Castellano}, {Cleri}, {Dav{\'e}}, {Dekel}, {Ferguson}, {Fontana}, {Gawiser},
  {Giavalisco}, {Harish}, {Hathi}, {Hirschmann}, {Holwerda}, {Huertas-Company},
  {Koekemoer}, {Larson}, {Lucas}, {Mobasher}, {P{\'e}rez-Gonz{\'a}lez},
  {Pirzkal}, {Rose}, {Santini}, {Trump}, {de la Vega}, {Wang}, {Weiner},
  {Wilkins}, {Yang}, {Yung}, \& {Zavala}}]{Arrabal-Haro-etal-2023}
{Arrabal Haro}, P., {Dickinson}, M., {Finkelstein}, S.~L., {et~al.} 2023,
  \apjl, 951, L22, \dodoi{10.3847/2041-8213/acdd54}

\bibitem[{{Athanassoula}(2002)}]{Athanassoula2002}
{Athanassoula}, E. 2002, \apjl, 569, L83, \dodoi{10.1086/340784}

\bibitem[{{Athanassoula}(2003)}]{Athanassoula2003}
---. 2003, \mnras, 341, 1179, \dodoi{10.1046/j.1365-8711.2003.06473.x}

\bibitem[{{Athanassoula} {et~al.}(2005){Athanassoula}, {Lambert}, \&
  {Dehnen}}]{Athanassoula-Lambert-Dehnen2005}
{Athanassoula}, E., {Lambert}, J.~C., \& {Dehnen}, W. 2005, \mnras, 363, 496,
  \dodoi{10.1111/j.1365-2966.2005.09445.x}

\bibitem[{{Athanassoula} \& {Misiriotis}(2002)}]{Athanassoula-Misiriotis2002}
{Athanassoula}, E., \& {Misiriotis}, A. 2002, \mnras, 330, 35,
  \dodoi{10.1046/j.1365-8711.2002.05028.x}

\bibitem[{{Athanassoula} {et~al.}(1990){Athanassoula}, {Morin}, {Wozniak},
  {Puy}, {Pierce}, {Lombard}, \& {Bosma}}]{Athanassoula-etal-1990}
{Athanassoula}, E., {Morin}, S., {Wozniak}, H., {et~al.} 1990, \mnras, 245, 130

\bibitem[{{Aumer} \& {Binney}(2017)}]{Aumer-etal-2017}
{Aumer}, M., \& {Binney}, J. 2017, \mnras, 470, 2113,
  \dodoi{10.1093/mnras/stx1300}

\bibitem[{{Bagley} {et~al.}(2023){Bagley}, {Finkelstein}, {Koekemoer},
  {Ferguson}, {Arrabal Haro}, {Dickinson}, {Kartaltepe}, {Papovich},
  {P{\'e}rez-Gonz{\'a}lez}, {Pirzkal}, {Somerville}, {Willmer}, {Yang}, {Yung},
  {Fontana}, {Grazian}, {Grogin}, {Hirschmann}, {Kewley}, {Kirkpatrick},
  {Kocevski}, {Lotz}, {Medrano}, {Morales}, {Pentericci}, {Ravindranath},
  {Trump}, {Wilkins}, {Calabr{\`o}}, {Cooper}, {Costantin}, {de la Vega},
  {Hilbert}, {Hutchison}, {Larson}, {Lucas}, {McGrath}, {Ryan}, {Wang}, \&
  {Wuyts}}]{Bagley-etal-2023}
{Bagley}, M.~B., {Finkelstein}, S.~L., {Koekemoer}, A.~M., {et~al.} 2023,
  \apjl, 946, L12, \dodoi{10.3847/2041-8213/acbb08}

\bibitem[{{Bertin} \& {Arnouts}(1996)}]{Bertin-Arnouts96}
{Bertin}, E., \& {Arnouts}, S. 1996, \aaps, 117, 393

\bibitem[{{Bi} {et~al.}(2022){Bi}, {Shlosman}, \&
  {Romano-D{\'\i}az}}]{Bi-Shlosman-Romano-Diaz2022}
{Bi}, D., {Shlosman}, I., \& {Romano-D{\'\i}az}, E. 2022, \apj, 934, 52,
  \dodoi{10.3847/1538-4357/ac779b}

\bibitem[{{Birkin} {et~al.}(2023){Birkin}, {Smail}, {Swinbank}, {An},
  {Chapman}, {Chen}, {Conselice}, {Dudzevi{\v{c}}i{\={u}}t{\.{e}}}, {Farrah},
  {Gullberg}, {Matsuda}, {Puglisi}, {Schinnerer}, {Scott}, {Wardlow}, \& {van
  der Werf}}]{Birkin-etal-2023}
{Birkin}, J.~E., {Smail}, I., {Swinbank}, A.~M., {et~al.} 2023, arXiv e-prints,
  arXiv:2301.05720, \dodoi{10.48550/arXiv.2301.05720}

\bibitem[{{Bland-Hawthorn} {et~al.}(2024){Bland-Hawthorn}, {Tepper-Garcia},
  {Agertz}, \& {Federrath}}]{Bland-Hawthorn-etal-2024}
{Bland-Hawthorn}, J., {Tepper-Garcia}, T., {Agertz}, O., \& {Federrath}, C.
  2024, arXiv e-prints, arXiv:2402.06060, \dodoi{10.48550/arXiv.2402.06060}

\bibitem[{{Bland-Hawthorn} {et~al.}(2023){Bland-Hawthorn}, {Tepper-Garcia},
  {Agertz}, \& {Freeman}}]{Bland-Hawthorn-etal-2023}
{Bland-Hawthorn}, J., {Tepper-Garcia}, T., {Agertz}, O., \& {Freeman}, K. 2023,
  \apj, 947, 80, \dodoi{10.3847/1538-4357/acc469}

\bibitem[{{Bonoli} {et~al.}(2016){Bonoli}, {Mayer}, {Kazantzidis}, {Madau},
  {Bellovary}, \& {Governato}}]{Bonoli-etal-2016}
{Bonoli}, S., {Mayer}, L., {Kazantzidis}, S., {et~al.} 2016, \mnras, 459, 2603,
  \dodoi{10.1093/mnras/stw694}

\bibitem[{{Bournaud} \& {Combes}(2002)}]{Bournaud-Combes2002}
{Bournaud}, F., \& {Combes}, F. 2002, \aap, 392, 83,
  \dodoi{10.1051/0004-6361:20020920}

\bibitem[{{Bradley} {et~al.}(2020){Bradley}, {Sip{\H{o}}cz}, {Robitaille},
  {Tollerud}, {Vin{\'\i}cius}, {Deil}, {Barbary}, {Wilson}, {Busko},
  {G{\"u}nther}, {Cara}, {Conseil}, {Bostroem}, {Droettboom}, {Bray}, {Andersen
  Bratholm}, {Lim}, {Barentsen}, {Craig}, {Pascual}, {Perren}, {Greco},
  {Donath}, {de Val-Borro}, {Kerzendorf}, {Bach}, {Weaver}, {D'Eugenio},
  {Souchereau}, \& {Ferreira}}]{photutils}
{Bradley}, L., {Sip{\H{o}}cz}, B., {Robitaille}, T., {et~al.} 2020,
  {astropy/photutils: 1.0.0}, 1.0.0, Zenodo,  Zenodo,
  \dodoi{10.5281/zenodo.4044744}

\bibitem[{{Brammer} {et~al.}(2008){Brammer}, {van Dokkum}, \&
  {Coppi}}]{Brammer-et-al-2008}
{Brammer}, G.~B., {van Dokkum}, P.~G., \& {Coppi}, P. 2008, \apj, 686, 1503,
  \dodoi{10.1086/591786}

\bibitem[{{Buta} \& {Combes}(1996)}]{Buta-Combes1996}
{Buta}, R., \& {Combes}, F. 1996, \fcp, 17, 95

\bibitem[{{Buta} {et~al.}(2015){Buta}, {Sheth}, {Athanassoula}, {Bosma},
  {Knapen}, {Laurikainen}, {Salo}, {Elmegreen}, {Ho}, {Zaritsky}, {Courtois},
  {Hinz}, {Mu{\~n}oz-Mateos}, {Kim}, {Regan}, {Gadotti}, {Gil de Paz}, {Laine},
  {Men{\'e}ndez-Delmestre}, {Comer{\'o}n}, {Erroz Ferrer}, {Seibert},
  {Mizusawa}, {Holwerda}, \& {Madore}}]{Buta-etal-2015}
{Buta}, R.~J., {Sheth}, K., {Athanassoula}, E., {et~al.} 2015, \apjs, 217, 32,
  \dodoi{10.1088/0067-0049/217/2/32}

\bibitem[{{Calzetti} {et~al.}(2000){Calzetti}, {Armus}, {Bohlin}, {Kinney},
  {Koornneef}, \& {Storchi-Bergmann}}]{Calzetti-etal-2001}
{Calzetti}, D., {Armus}, L., {Bohlin}, R.~C., {et~al.} 2000, \apj, 533, 682,
  \dodoi{10.1086/308692}

\bibitem[{{Cameron} {et~al.}(2010){Cameron}, {Carollo}, {Oesch}, {Aller},
  {Bschorr}, {Cerulo}, {Aussel}, {Capak}, {Le Floc'h}, {Ilbert}, {Kneib},
  {Koekemoer}, {Leauthaud}, {Lilly}, {Massey}, {McCracken}, {Rhodes},
  {Salvato}, {Sanders}, {Scoville}, {Sheth}, {Taniguchi}, \&
  {Thompson}}]{Cameron-etal-2010}
{Cameron}, E., {Carollo}, C.~M., {Oesch}, P., {et~al.} 2010, \mnras, 409, 346,
  \dodoi{10.1111/j.1365-2966.2010.17314.x}

\bibitem[{{Casey} {et~al.}(2023){Casey}, {Kartaltepe}, {Drakos}, {Franco},
  {Harish}, {Paquereau}, {Ilbert}, {Rose}, {Cox}, {Nightingale}, {Robertson},
  {Silverman}, {Koekemoer}, {Massey}, {McCracken}, {Rhodes}, {Akins}, {Allen},
  {Amvrosiadis}, {Arango-Toro}, {Bagley}, {Bongiorno}, {Capak}, {Champagne},
  {Chartab}, {Ch{\'a}vez Ortiz}, {Chworowsky}, {Cooke}, {Cooper}, {Darvish},
  {Ding}, {Faisst}, {Finkelstein}, {Fujimoto}, {Gentile}, {Gillman}, {Gould},
  {Gozaliasl}, {Hayward}, {He}, {Hemmati}, {Hirschmann}, {Jahnke}, {Jin},
  {Khostovan}, {Kokorev}, {Lambrides}, {Laigle}, {Larson}, {Leung}, {Liu},
  {Liaudat}, {Long}, {Magdis}, {Mahler}, {Mainieri}, {Manning}, {Maraston},
  {Martin}, {McCleary}, {McKinney}, {McPartland}, {Mobasher}, {Pattnaik},
  {Renzini}, {Rich}, {Sanders}, {Sattari}, {Scognamiglio}, {Scoville}, {Sheth},
  {Shuntov}, {Sparre}, {Suzuki}, {Talia}, {Toft}, {Trakhtenbrot}, {Urry},
  {Valentino}, {Vanderhoof}, {Vardoulaki}, {Weaver}, {Whitaker}, {Wilkins},
  {Yang}, \& {Zavala}}]{Casey-etal-2023}
{Casey}, C.~M., {Kartaltepe}, J.~S., {Drakos}, N.~E., {et~al.} 2023, \apj, 954,
  31, \dodoi{10.3847/1538-4357/acc2bc}

\bibitem[{{Ceverino} {et~al.}(2023){Ceverino}, {Mandelker}, {Snyder},
  {Lapiner}, {Dekel}, {Primack}, {Ginzburg}, \& {Larkin}}]{Ceverino-etal-2023}
{Ceverino}, D., {Mandelker}, N., {Snyder}, G.~F., {et~al.} 2023, \mnras, 522,
  3912, \dodoi{10.1093/mnras/stad1255}

\bibitem[{{Chabrier}(2003)}]{Chabrier-2003}
{Chabrier}, G. 2003, \pasp, 115, 763, \dodoi{10.1086/376392}

\bibitem[{{Cheung} {et~al.}(2013){Cheung}, {Athanassoula}, {Masters}, {Nichol},
  {Bosma}, {Bell}, {Faber}, {Koo}, {Lintott}, {Melvin}, {Schawinski}, {Skibba},
  \& {Willett}}]{Cheung-etal-2013}
{Cheung}, E., {Athanassoula}, E., {Masters}, K.~L., {et~al.} 2013, \apj, 779,
  162, \dodoi{10.1088/0004-637X/779/2/162}

\bibitem[{{Cheung} {et~al.}(2015){Cheung}, {Trump}, {Athanassoula}, {Bamford},
  {Bell}, {Bosma}, {Cardamone}, {Casteels}, {Faber}, {Fang}, {Fortson},
  {Kocevski}, {Koo}, {Laine}, {Lintott}, {Masters}, {Melvin}, {Nichol},
  {Schawinski}, {Simmons}, {Smethurst}, \& {Willett}}]{Cheung-etal-2015}
{Cheung}, E., {Trump}, J.~R., {Athanassoula}, E., {et~al.} 2015, \mnras, 447,
  506, \dodoi{10.1093/mnras/stu2462}

\bibitem[{{Coelho} \& {Gadotti}(2011)}]{Coelho-Gadotti-2011}
{Coelho}, P., \& {Gadotti}, D.~A. 2011, \apjl, 743, L13,
  \dodoi{10.1088/2041-8205/743/1/L13}

\bibitem[{{Collier} {et~al.}(2018){Collier}, {Shlosman}, \&
  {Heller}}]{Collier-Shlosman-Heller2018}
{Collier}, A., {Shlosman}, I., \& {Heller}, C. 2018, \mnras, 476, 1331,
  \dodoi{10.1093/mnras/sty270}

\bibitem[{{Collier} {et~al.}(2019){Collier}, {Shlosman}, \&
  {Heller}}]{Collier-etal-2019}
---. 2019, \mnras, 488, 5788, \dodoi{10.1093/mnras/stz2144}

\bibitem[{{Conselice} {et~al.}(2003){Conselice}, {Bershady}, {Dickinson}, \&
  {Papovich}}]{Conselice-etal-2003}
{Conselice}, C.~J., {Bershady}, M.~A., {Dickinson}, M., \& {Papovich}, C. 2003,
  \aj, 126, 1183, \dodoi{10.1086/377318}

\bibitem[{{Costantin} {et~al.}(2023){Costantin}, {P{\'e}rez-Gonz{\'a}lez},
  {Guo}, {Buttitta}, {Jogee}, {Bagley}, {Barro}, {Kartaltepe}, {Koekemoer},
  {Cabello}, {Corsini}, {M{\'e}ndez-Abreu}, {de la Vega}, {Iyer}, {Bisigello},
  {Cheng}, {Morelli}, {Arrabal Haro}, {Buitrago}, {Cooper}, {Dekel},
  {Dickinson}, {Finkelstein}, {Giavalisco}, {Holwerda}, {Huertas-Company},
  {Lucas}, {Papovich}, {Pirzkal}, {Seill{\'e}}, {Vega-Ferrero}, {Wuyts}, \&
  {Yung}}]{Costantin-etal-2023}
{Costantin}, L., {P{\'e}rez-Gonz{\'a}lez}, P.~G., {Guo}, Y., {et~al.} 2023,
  \nat, 623, 499, \dodoi{10.1038/s41586-023-06636-x}

\bibitem[{{de Lorenzo-C{\'a}ceres} {et~al.}(2019){de Lorenzo-C{\'a}ceres},
  {S{\'a}nchez-Bl{\'a}zquez}, {M{\'e}ndez-Abreu}, {Gadotti},
  {Falc{\'o}n-Barroso}, {Mart{\'\i}nez-Valpuesta}, {Coelho}, {Fragkoudi},
  {Husemann}, {Leaman}, {P{\'e}rez}, {Querejeta}, {Seidel}, \& {van de
  Ven}}]{deLorenzo-Caceres-etal-2019}
{de Lorenzo-C{\'a}ceres}, A., {S{\'a}nchez-Bl{\'a}zquez}, P.,
  {M{\'e}ndez-Abreu}, J., {et~al.} 2019, \mnras, 484, 5296,
  \dodoi{10.1093/mnras/stz221}

\bibitem[{{de Vaucouleurs} {et~al.}(1991){de Vaucouleurs}, {de Vaucouleurs},
  {Corwin}, {Buta}, {Paturel}, \& {Fouque}}]{RC3}
{de Vaucouleurs}, G., {de Vaucouleurs}, A., {Corwin}, Herold~G., J., {et~al.}
  1991, {Third Reference Catalogue of Bright Galaxies}

\bibitem[{{Dekel} \& {Birnboim}(2006)}]{Dekel-Birnboim2006}
{Dekel}, A., \& {Birnboim}, Y. 2006, \mnras, 368, 2,
  \dodoi{10.1111/j.1365-2966.2006.10145.x}

\bibitem[{{D{\'\i}az-Garc{\'\i}a} {et~al.}(2020){D{\'\i}az-Garc{\'\i}a},
  {Moyano}, {Comer{\'o}n}, {Knapen}, {Salo}, \&
  {Bouquin}}]{Diaz-Garcia-etal-2020}
{D{\'\i}az-Garc{\'\i}a}, S., {Moyano}, F.~D., {Comer{\'o}n}, S., {et~al.} 2020,
  \aap, 644, A38, \dodoi{10.1051/0004-6361/202039162}

\bibitem[{{D{\'\i}az-Garc{\'\i}a} {et~al.}(2016){D{\'\i}az-Garc{\'\i}a},
  {Salo}, {Laurikainen}, \& {Herrera-Endoqui}}]{Diaz-Garcia-etal-2016}
{D{\'\i}az-Garc{\'\i}a}, S., {Salo}, H., {Laurikainen}, E., \&
  {Herrera-Endoqui}, M. 2016, \aap, 587, A160,
  \dodoi{10.1051/0004-6361/201526161}

\bibitem[{{Donohoe-Keyes} {et~al.}(2019){Donohoe-Keyes}, {Martig}, {James}, \&
  {Kraljic}}]{Donohoe-Keyes-etal-2019}
{Donohoe-Keyes}, C.~E., {Martig}, M., {James}, P.~A., \& {Kraljic}, K. 2019,
  \mnras, 489, 4992, \dodoi{10.1093/mnras/stz2474}

\bibitem[{{Efstathiou} {et~al.}(1982){Efstathiou}, {Lake}, \&
  {Negroponte}}]{Efstathiou-etal-1982}
{Efstathiou}, G., {Lake}, G., \& {Negroponte}, J. 1982, \mnras, 199, 1069,
  \dodoi{10.1093/mnras/199.4.1069}

\bibitem[{{Elmegreen} {et~al.}(2004){Elmegreen}, {Elmegreen}, \&
  {Hirst}}]{Elmegreen-Elmegreen-Hirst2004}
{Elmegreen}, B.~G., {Elmegreen}, D.~M., \& {Hirst}, A.~C. 2004, \apj, 612, 191,
  \dodoi{10.1086/422407}

\bibitem[{{Erwin}(2005)}]{Erwin2005}
{Erwin}, P. 2005, \mnras, 364, 283, \dodoi{10.1111/j.1365-2966.2005.09560.x}

\bibitem[{{Erwin}(2018)}]{Erwin2018}
---. 2018, \mnras, 474, 5372, \dodoi{10.1093/mnras/stx3117}

\bibitem[{{Erwin}(2023)}]{Erwin-2023}
---. 2023, \mnras, \dodoi{10.1093/mnras/stad3944}

\bibitem[{{Eskridge} {et~al.}(2000){Eskridge}, {Frogel}, {Pogge}, {Quillen},
  {Davies}, {DePoy}, {Houdashelt}, {Kuchinski}, {Ram{\'\i}rez}, {Sellgren},
  {Terndrup}, \& {Tiede}}]{Eskridge-etal-2000}
{Eskridge}, P.~B., {Frogel}, J.~A., {Pogge}, R.~W., {et~al.} 2000, \aj, 119,
  536, \dodoi{10.1086/301203}

\bibitem[{{Faucher-Gigu{\`e}re} \&
  {Kere{\v{s}}}(2011)}]{Faucher-Giguere-Keres2011}
{Faucher-Gigu{\`e}re}, C.-A., \& {Kere{\v{s}}}, D. 2011, \mnras, 412, L118,
  \dodoi{10.1111/j.1745-3933.2011.01018.x}

\bibitem[{{Ferreira} {et~al.}(2022{\natexlab{a}}){Ferreira}, {Adams},
  {Conselice}, {Sazonova}, {Austin}, {Caruana}, {Ferrari}, {Verma}, {Trussler},
  {Broadhurst}, {Diego}, {Frye}, {Pascale}, {Wilkins}, {Windhorst}, \&
  {Zitrin}}]{Ferreira-etal-2022a}
{Ferreira}, L., {Adams}, N., {Conselice}, C.~J., {et~al.} 2022{\natexlab{a}},
  \apjl, 938, L2, \dodoi{10.3847/2041-8213/ac947c}

\bibitem[{{Ferreira} {et~al.}(2022{\natexlab{b}}){Ferreira}, {Conselice},
  {Sazonova}, {Ferrari}, {Caruana}, {Tohill}, {Lucatelli}, {Adams}, {Irodotou},
  {Marshall}, {Roper}, {Lovell}, {Verma}, {Austin}, {Trussler}, \&
  {Wilkins}}]{Ferreira-etal-2022b}
{Ferreira}, L., {Conselice}, C.~J., {Sazonova}, E., {et~al.}
  2022{\natexlab{b}}, arXiv e-prints, arXiv:2210.01110.
\newblock \doarXiv{2210.01110}

\bibitem[{{Finkelstein} {et~al.}(2022){Finkelstein}, {Bagley}, {Haro},
  {Dickinson}, {Ferguson}, {Kartaltepe}, {Papovich}, {Burgarella}, {Kocevski},
  {Huertas-Company}, {Iyer}, {Koekemoer}, {Larson}, {P{\'e}rez-Gonz{\'a}lez},
  {Rose}, {Tacchella}, {Wilkins}, {Chworowsky}, {Medrano}, {Morales},
  {Somerville}, {Yung}, {Fontana}, {Giavalisco}, {Grazian}, {Grogin}, {Kewley},
  {Kirkpatrick}, {Kurczynski}, {Lotz}, {Pentericci}, {Pirzkal}, {Ravindranath},
  {Ryan}, {Trump}, {Yang}, {Almaini}, {Amor{\'\i}n}, {Annunziatella},
  {Backhaus}, {Barro}, {Behroozi}, {Bell}, {Bhatawdekar}, {Bisigello}, {Bromm},
  {Buat}, {Buitrago}, {Calabr{\`o}}, {Casey}, {Castellano}, {Ch{\'a}vez Ortiz},
  {Ciesla}, {Cleri}, {Cohen}, {Cole}, {Cooke}, {Cooper}, {Cooray}, {Costantin},
  {Cox}, {Croton}, {Daddi}, {Dav{\'e}}, {de La Vega}, {Dekel}, {Elbaz},
  {Estrada-Carpenter}, {Faber}, {Fern{\'a}ndez}, {Finkelstein}, {Freundlich},
  {Fujimoto}, {Garc{\'\i}a-Argum{\'a}nez}, {Gardner}, {Gawiser},
  {G{\'o}mez-Guijarro}, {Guo}, {Hamblin}, {Hamilton}, {Hathi}, {Holwerda},
  {Hirschmann}, {Hutchison}, {Jaskot}, {Jha}, {Jogee}, {Juneau}, {Jung},
  {Kassin}, {Le Bail}, {Leung}, {Lucas}, {Magnelli}, {Mantha}, {Matharu},
  {McGrath}, {McIntosh}, {Merlin}, {Mobasher}, {Newman}, {Nicholls}, {Pandya},
  {Rafelski}, {Ronayne}, {Santini}, {Seill{\'e}}, {Shah}, {Shen}, {Simons},
  {Snyder}, {Stanway}, {Straughn}, {Teplitz}, {Vanderhoof}, {Vega-Ferrero},
  {Wang}, {Weiner}, {Willmer}, {Wuyts}, {Zavala}, \& {CEERS
  Team}}]{Finkelstein-etal-2022}
{Finkelstein}, S.~L., {Bagley}, M.~B., {Haro}, P.~A., {et~al.} 2022, \apjl,
  940, L55, \dodoi{10.3847/2041-8213/ac966e}

\bibitem[{{Finkelstein} {et~al.}(2023){Finkelstein}, {Bagley}, {Ferguson},
  {Wilkins}, {Kartaltepe}, {Papovich}, {Yung}, {Haro}, {Behroozi}, {Dickinson},
  {Kocevski}, {Koekemoer}, {Larson}, {Le Bail}, {Morales},
  {P{\'e}rez-Gonz{\'a}lez}, {Burgarella}, {Dav{\'e}}, {Hirschmann},
  {Somerville}, {Wuyts}, {Bromm}, {Casey}, {Fontana}, {Fujimoto}, {Gardner},
  {Giavalisco}, {Grazian}, {Grogin}, {Hathi}, {Hutchison}, {Jha}, {Jogee},
  {Kewley}, {Kirkpatrick}, {Long}, {Lotz}, {Pentericci}, {Pierel}, {Pirzkal},
  {Ravindranath}, {Ryan}, {Trump}, {Yang}, {Bhatawdekar}, {Bisigello}, {Buat},
  {Calabr{\`o}}, {Castellano}, {Cleri}, {Cooper}, {Croton}, {Daddi}, {Dekel},
  {Elbaz}, {Franco}, {Gawiser}, {Holwerda}, {Huertas-Company}, {Jaskot},
  {Leung}, {Lucas}, {Mobasher}, {Pandya}, {Tacchella}, {Weiner}, \&
  {Zavala}}]{Finkelstein-et-al-2023}
{Finkelstein}, S.~L., {Bagley}, M.~B., {Ferguson}, H.~C., {et~al.} 2023, \apjl,
  946, L13, \dodoi{10.3847/2041-8213/acade4}

\bibitem[{{Finkelstein} {et~al.}(2024){Finkelstein}, {Leung}, {Bagley},
  {Dickinson}, {Ferguson}, {Papovich}, {Akins}, {Arrabal Haro}, {Dav{\'e}},
  {Dekel}, {Kartaltepe}, {Kocevski}, {Koekemoer}, {Pirzkal}, {Somerville},
  {Yung}, {Amor{\'\i}n}, {Backhaus}, {Behroozi}, {Bisigello}, {Bromm}, {Casey},
  {Ch{\'a}vez Ortiz}, {Cheng}, {Chworowsky}, {Cleri}, {Cooper}, {Davis}, {de la
  Vega}, {Elbaz}, {Franco}, {Fontana}, {Fujimoto}, {Giavalisco}, {Grogin},
  {Holwerda}, {Huertas-Company}, {Hirschmann}, {Iyer}, {Jogee}, {Jung},
  {Larson}, {Lucas}, {Mobasher}, {Morales}, {Morley}, {Mukherjee},
  {P{\'e}rez-Gonz{\'a}lez}, {Ravindranath}, {Rodighiero}, {Rowland},
  {Tacchella}, {Taylor}, {Trump}, \&
  {Wilkins}}]{Finkelstein-etal-2024-internal-catalog}
{Finkelstein}, S.~L., {Leung}, G. C.~K., {Bagley}, M.~B., {et~al.} 2024, \apjl,
  969, L2, \dodoi{10.3847/2041-8213/ad4495}

\bibitem[{{Fragkoudi} {et~al.}(2025){Fragkoudi}, {Grand}, {Pakmor},
  {G{\'o}mez}, {Marinacci}, \& {Springel}}]{Fragkoudi-etal-2025}
{Fragkoudi}, F., {Grand}, R. J.~J., {Pakmor}, R., {et~al.} 2025, \mnras, 538,
  1587, \dodoi{10.1093/mnras/staf389}

\bibitem[{{Fragkoudi} {et~al.}(2021){Fragkoudi}, {Grand}, {Pakmor}, {Springel},
  {White}, {Marinacci}, {Gomez}, \& {Navarro}}]{Fragkoudi-etal-2021}
{Fragkoudi}, F., {Grand}, R.~J.~J., {Pakmor}, R., {et~al.} 2021, \aap, 650,
  L16, \dodoi{10.1051/0004-6361/202140320}

\bibitem[{Frogel {et~al.}(1996)Frogel, Quillen, \&
  Pogge}]{Frogel-Quillen-Pogge1996}
Frogel, J.~A., Quillen, A.~C., \& Pogge, R.~W. 1996, in New Extragalactic
  Perspectives in the New South Africa, ed. D.~L. Block \& J.~M. Greenberg
  (Dordrecht: Springer Netherlands), 65--83

\bibitem[{{Gadotti}(2011)}]{Gadotti-2011}
{Gadotti}, D.~A. 2011, \mnras, 415, 3308,
  \dodoi{10.1111/j.1365-2966.2011.18945.x}

\bibitem[{{Gadotti} {et~al.}(2015){Gadotti}, {Seidel},
  {S{\'a}nchez-Bl{\'a}zquez}, {Falc{\'o}n-Barroso}, {Husemann}, {Coelho}, \&
  {P{\'e}rez}}]{Gadotti-etal-2015}
{Gadotti}, D.~A., {Seidel}, M.~K., {S{\'a}nchez-Bl{\'a}zquez}, P., {et~al.}
  2015, \aap, 584, A90, \dodoi{10.1051/0004-6361/201526677}

\bibitem[{{Galloway} {et~al.}(2015){Galloway}, {Willett}, {Fortson},
  {Cardamone}, {Schawinski}, {Cheung}, {Lintott}, {Masters}, {Melvin}, \&
  {Simmons}}]{Galloway-etal-2015}
{Galloway}, M.~A., {Willett}, K.~W., {Fortson}, L.~F., {et~al.} 2015, \mnras,
  448, 3442, \dodoi{10.1093/mnras/stv235}

\bibitem[{{Gardner} {et~al.}(2023){Gardner}, {Mather}, {Abbott}, {Abell},
  {Abernathy}, {Abney}, {Abraham}, {Abraham}, {Abul-Huda}, {Acton}, {Adams},
  {Adams}, {Adler}, {Adriaensen}, {Aguilar}, {Ahmed}, {Ahmed}, {Ahmed},
  {Albat}, {Albert}, {Alberts}, {Aldridge}, {Allen}, {Allen}, {Altenburg},
  {Altunc}, {Alvarez}, {{\'A}lvarez-M{\'a}rquez}, {Alves de Oliveira},
  {Ambrose}, {Anandakrishnan}, {Andersen}, {Anderson}, {Anderson}, {Anderson},
  {Anderson}, {Aprea}, {Archer}, {Arenberg}, {Argyriou}, {Arribas}, {Artigau},
  {Arvai}, {Atcheson}, {Atkinson}, {Averbukh}, {Aymergen}, {Bacinski},
  {Baggett}, {Bagnasco}, {Baker}, {Balzano}, {Banks}, {Baran}, {Barker},
  {Barrett}, {Barringer}, {Barto}, {Bast}, {Baudoz}, {Baum}, {Beatty},
  {Beaulieu}, {Bechtold}, {Beck}, {Beddard}, {Beichman}, {Bellagama}, {Bely},
  {Berger}, {Bergeron}, {Bernier}, {Bertch}, {Beskow}, {Betz}, {Biagetti},
  {Birkmann}, {Bjorklund}, {Blackwood}, {Blazek}, {Blossfeld}, {Bluth},
  {Boccaletti}, {Boegner}, {Bohlin}, {Boia}, {B{\"o}ker}, {Bonaventura},
  {Bond}, {Bosley}, {Boucarut}, {Bouchet}, {Bouwman}, {Bower}, {Bowers},
  {Bowers}, {Boyce}, {Boyer}, {Boyer}, {Boyer}, {Boyer}, {Bradley}, {Brady},
  {Brandl}, {Brannen}, {Breda}, {Bremmer}, {Brennan}, {Bresnahan}, {Bright},
  {Broiles}, {Bromenschenkel}, {Brooks}, {Brooks}, {Brown}, {Brown}, {Brown},
  {Bruce}, {Bryson}, {Bujanda}, {Bullock}, {Bunker}, {Bureo}, {Burt}, {Bush},
  {Bushouse}, {Bussman}, {Cabaud}, {Cale}, {Calhoon}, {Calvani}, {Canipe},
  {Caputo}, {Cara}, {Carey}, {Case}, {Cesari}, {Cetorelli}, {Chance},
  {Chandler}, {Chaney}, {Chapman}, {Charlot}, {Chayer}, {Cheezum}, {Chen},
  {Chen}, {Cherinka}, {Chichester}, {Chilton}, {Chittiraibalan}, {Clampin},
  {Clark}, {Clark}, {Clark}, {Claybrooks}, {Cleveland}, {Cohen}, {Cohen},
  {Col{\'o}n}, {Coleman}, {Colina}, {Comber}, {Comeau}, {Comer}, {Conde Reis},
  {Connolly}, {Conroy}, {Contos}, {Contreras}, {Cook}, {Cooper}, {Cooper},
  {Correia}, {Correnti}, {Cossou}, {Costanza}, {Coulais}, {Cox}, {Coyle},
  {Cracraft}, {Crew}, {Curtis}, {Cusveller}, {Da Costa Maciel}, {Dailey},
  {Daugeron}, {Davidson}, {Davies}, {Davis}, {Davis}, {Day}, {de Chambure}, {de
  Jong}, {De Marchi}, {Dean}, {Decker}, {Delisa}, {Dell}, {Dellagatta},
  {Dembinska}, {Demosthenes}, {Dencheva}, {Deneu}, {DePriest}, {Deschenes},
  {Dethienne}, {Detre}, {Diaz}, {Dicken}, {DiFelice}, {Dillman}, {Disharoon},
  {Dixon}, {Doggett}, {Dominguez}, {Donaldson}, {Doria-Warner}, {Santos},
  {Doty}, {Douglas}, {Doyon}, {Dressler}, {Driggers}, {Driggers}, {Dunn},
  {DuPrie}, {Dupuis}, {Durning}, {Dutta}, {Earl}, {Eccleston}, {Ecobichon},
  {Egami}, {Ehrenwinkler}, {Eisenhamer}, {Eisenhower}, {Eisenstein}, {El
  Hamel}, {Elie}, {Elliott}, {Elliott}, {Engesser}, {Espinoza}, {Etienne},
  {Etxaluze}, {Evans}, {Fabreguettes}, {Falcolini}, {Falini}, {Fatig},
  {Feeney}, {Feinberg}, {Fels}, {Ferdous}, {Ferguson}, {Ferrarese}, {Ferreira},
  {Ferruit}, {Ferry}, {Filippazzo}, {Firre}, {Fix}, {Flagey}, {Flanagan},
  {Fleming}, {Florian}, {Flynn}, {Foiadelli}, {Fontaine}, {Fontanella},
  {Forshay}, {Fortner}, {Fox}, {Framarini}, {Francisco}, {Franck}, {Franx},
  {Franz}, {Friedman}, {Friend}, {Frost}, {Fu}, {Fullerton}, {Gaillard},
  {Galkin}, {Gallagher}, {Galyer}, {Garc{\'\i}a Mar{\'\i}n}, {Gardner},
  {Garland}, {Garrett}, {Gasman}, {G{\'a}sp{\'a}r}, {Gastaud}, {Gaudreau},
  {Gauthier}, {Geers}, {Geithner}, {Gennaro}, {Gerber}, {Gereau}, {Giampaoli},
  {Giardino}, {Gibbons}, {Gilbert}, {Gilman}, {Girard}, {Giuliano}, {Gkountis},
  {Glasse}, {Glassmire}, {Glauser}, {Glazer}, {Goldberg}, {Golimowski},
  {Gonzaga}, {Gordon}, {Gordon}, {Goudfrooij}, {Gough}, {Graham}, {Grau},
  {Green}, {Greene}, {Greene}, {Greenfield}, {Greenhouse}, {Greve}, {Greville},
  {Grimaldi}, {Groe}, {Groebner}, {Grumm}, {Grundy}, {G{\"u}del}, {Guillard},
  {Guldalian}, {Gunn}, {Gurule}, {Gutman}, {Guy}, {Guyot}, {Hack}, {Haderlein},
  {Hagan}, {Hagedorn}, {Hainline}, {Haley}, {Hami}, {Hamilton}, {Hammann},
  {Hammel}, {Hanley}, {Hansen}, {Hardy}, {Harnisch}, {Harr}, {Harris}, {Hart},
  {Hartig}, {Hasan}, {Hashim}, {Hashimoto}, {Haskins}, {Hawkins}, {Hayden},
  {Hayden}, {Healy}, {Hecht}, {Heeg}, {Hejal}, {Helm}, {Hengemihle}, {Henning},
  {Henry}, {Henry}, {Henshaw}, {Hernandez}, {Herrington}, {Heske}, {Hesman},
  {Hickey}, {Hilbert}, {Hines}, {Hinz}, {Hirsch}, {Hitcho}, {Hodapp}, {Hodge},
  {Hoffman}, {Holfeltz}, {Holler}, {Hoppa}, {Horner}, {Howard}, {Howard},
  {Huber}, {Hunkeler}, {Hunter}, {Hunter}, {Hurd}, {Hurst}, {Hutchings},
  {Hylan}, {Ignat}, {Illingworth}, {Irish}, {Isaacs}, {Jackson}, {Jaffe},
  {Jahic}, {Jahromi}, {Jakobsen}, {James}, {James}, {James}, {Jamieson},
  {Jandra}, {Jayawardhana}, {Jedrzejewski}, {Jeffers}, {Jensen}, {Joanne},
  {Johns}, {Johnson}, {Johnson}, {Johnson}, {Johnson}, {Johnson}, {Johnson},
  {Johnstone}, {Jollet}, {Jones}, {Jones}, {Jones}, {Jones}, {Jones}, {Jordan},
  {Jordan}, {Jue}, {Jurkowski}, {Justis}, {Justtanont}, {Kaleida}, {Kalirai},
  {Kalmanson}, {Kaltenegger}, {Kammerer}, {Kan}, {Kanarek}, {Kao}, {Karakla},
  {Karl}, {Kassin}, {Kauffman}, {Kavanagh}, {Kelley}, {Kelly}, {Kendrew},
  {Kennedy}, {Kenny}, {Keski-Kuha}, {Keyes}, {Khan}, {Kidwell}, {Kimble},
  {King}, {King}, {Kinzel}, {Kirk}, {Kirkpatrick}, {Klaassen}, {Klingemann},
  {Klintworth}, {Knapp}, {Knight}, {Knollenberg}, {Knutsen}, {Koehler},
  {Koekemoer}, {Kofler}, {Kontson}, {Kovacs}, {Kozhurina-Platais}, {Krause},
  {Kriss}, {Krist}, {Kristoffersen}, {Krogel}, {Krueger}, {Kulp}, {Kumari},
  {Kwan}, {Kyprianou}, {Labador}, {Labiano}, {Lafreni{\`e}re}, {Lagage},
  {Laidler}, {Laine}, {Laird}, {Lajoie}, {Lallo}, {Lam}, {LaMassa}, {Lambros},
  {Lampenfield}, {Lander}, {Langston}, {Larson}, {Larson}, {LaVerghetta},
  {Law}, {Lawrence}, {Lee}, {Lee}, {Lee}, {Leisenring}, {Leveille}, {Levenson},
  {Levi}, {Levine}, {Lewis}, {Lewis}, {Lewis}, {Libralato}, {Lidon},
  {Liebrecht}, {Lightsey}, {Lilly}, {Lim}, {Lim}, {Ling}, {Link}, {Link},
  {Lipinski}, {Liu}, {Lo}, {Lobmeyer}, {Logue}, {Long}, {Long}, {Long}, {Long},
  {L{\'o}pez-Caniego}, {Lotz}, {Love-Pruitt}, {Lubskiy}, {Luers}, {Luetgens},
  {Luevano}, {Lui}, {Lund}, {Lundquist}, {Lunine}, {L{\"u}tzgendorf}, {Lynch},
  {MacDonald}, {MacDonald}, {Macias}, {Macklis}, {Maghami}, {Maharaja},
  {Maiolino}, {Makrygiannis}, {Malla}, {Malumuth}, {Manjavacas}, {Marini},
  {Marrione}, {Marston}, {Martel}, {Martin}, {Martin}, {Martinez}, {Maschmann},
  {Masci}, {Masetti}, {Maszkiewicz}, {Matthews}, {Matuskey}, {McBrayer},
  {McCarthy}, {McCaughrean}, {McClare}, {McClare}, {McCloskey}, {McClurg},
  {McCoy}, {McElwain}, {McGregor}, {McGuffey}, {McKay}, {McKenzie}, {McLean},
  {McMaster}, {McNeil}, {De Meester}, {Mehalick}, {Meixner}, {Mel{\'e}ndez},
  {Menzel}, {Menzel}, {Merz}, {Mesterharm}, {Meyer}, {Meyett}, {Meza},
  {Midwinter}, {Milam}, {Miller}, {Miller}, {Miskey}, {Misselt}, {Mitchell},
  {Mohan}, {Montoya}, {Moran}, {Morishita}, {Moro-Mart{\'\i}n}, {Morrison},
  {Morrison}, {Morse}, {Moschos}, {Moseley}, {Mosier}, {Mosner}, {Mountain},
  {Muckenthaler}, {Mueller}, {Mueller}, {Muhiem}, {M{\"u}hlmann}, {Mullally},
  {Mullen}, {Munger}, {Murphy}, {Murray}, {Muzerolle}, {Mycroft}, {Myers},
  {Myers}, {Myers}, {Myers}, {Myrick}, {Nagle}, {Nayak}, {Naylor}, {Neff},
  {Nelan}, {Nella}, {Nguyen}, {Nguyen}, {Nickson}, {Nidhiry}, {Niedner},
  {Nieto-Santisteban}, {Nikolov}, {Nishisaka}, {Noriega-Crespo}, {Nota},
  {O'Mara}, {Oboryshko}, {O'Brien}, {Ochs}, {Offenberg}, {Ogle}, {Ohl},
  {Olmsted}, {Osborne}, {O'Shaughnessy}, {{\"O}stlin}, {O'Sullivan}, {Otor},
  {Ottens}, {Ouellette}, {Outlaw}, {Owens}, {Pacifici}, {Page}, {Paranilam},
  {Park}, {Parrish}, {Paschal}, {Patapis}, {Patel}, {Patrick}, {Pattishall},
  {Paul}, {Paul}, {Pauly}, {Pavlovsky}, {Pe{\~n}a-Guerrero}, {Pedder}, {Peek},
  {Pelham}, {Penanen}, {Perriello}, {Perrin}, {Perrine}, {Perrygo}, {Peslier},
  {Petach}, {Peterson}, {Pfarr}, {Pierson}, {Pietraszkiewicz}, {Pilchen},
  {Pipher}, {Pirzkal}, {Pitman}, {Player}, {Plesha}, {Plitzke}, {Pohner},
  {Poletis}, {Pollizzi}, {Polster}, {Pontius}, {Pontoppidan}, {Porges},
  {Potter}, {Prescott}, {Proffitt}, {Pueyo}, {Quispe Neira}, {Radich}, {Rager},
  {Rameau}, {Ramey}, {Ramos Alarcon}, {Rampini}, {Rapp}, {Rashford},
  {Rauscher}, {Ravindranath}, {Rawle}, {Rawlings}, {Ray}, {Regan}, {Rehm},
  {Rehm}, {Reid}, {Reis}, {Renk}, {Reoch}, {Ressler}, {Rest}, {Reynolds},
  {Richon}, {Richon}, {Ridgaway}, {Riedel}, {Rieke}, {Rieke}, {Rifelli},
  {Rigby}, {Riggs}, {Ringel}, {Ritchie}, {Rix}, {Robberto}, {Robinson},
  {Robinson}, {Robinson}, {Rock}, {Rodriguez}, {Rodr{\'\i}guez del Pino},
  {Roellig}, {Rohrbach}, {Roman}, {Romelfanger}, {Romo}, {Rosales}, {Rose},
  {Roteliuk}, {Roth}, {Rothwell}, {Rouzaud}, {Rowe}, {Rowlands}, {Roy},
  {Royer}, {Rui}, {Rumler}, {Rumpl}, {Russ}, {Ryan}, {Ryan}, {Saad}, {Sabata},
  {Sabatino}, {Sabbi}, {Sabelhaus}, {Sabia}, {Sahu}, {Saif}, {Salvignol},
  {Samara-Ratna}, {Samuelson}, {Sanders}, {Sappington}, {Sargent}, {Sauer},
  {Savadkin}, {Sawicki}, {Schappell}, {Scheffer}, {Scheithauer}, {Scherer},
  {Schiff}, {Schlawin}, {Schmeitzky}, {Schmitz}, {Schmude}, {Schneider},
  {Schreiber}, {Schroeven-Deceuninck}, {Schultz}, {Schwab}, {Schwartz},
  {Scoccimarro}, {Scott}, {Scott}, {Seaton}, {Seely}, {Seery}, {Seidleck},
  {Sembach}, {Shanahan}, {Shaughnessy}, {Shaw}, {Shay}, {Sheehan}, {Sheth},
  {Shih}, {Shivaei}, {Siegel}, {Sienkiewicz}, {Simmons}, {Simon}, {Sirianni},
  {Sivaramakrishnan}, {Slade}, {Sloan}, {Slocum}, {Slowinski}, {Smith},
  {Smith}, {Smith}, {Smith}, {Smith}, {Smith}, {Smolik}, {Soderblom}, {Sohn},
  {Sokol}, {Sonneborn}, {Sontag}, {Sooy}, {Soummer}, {Southwood}, {Spain},
  {Sparmo}, {Speer}, {Spencer}, {Sprofera}, {Stallcup}, {Stanley},
  {Stansberry}, {Stark}, {Starr}, {Stassi}, {Steck}, {Steeley}, {Stephens},
  {Stephenson}, {Stewart}, {Stiavelli}, {}, {Strada}, {Straughn}, {Streetman},
  {Strickland}, {Strobele}, {Stuhlinger}, {Stys}, {Such}, {Sukhatme},
  {Sullivan}, {Sullivan}, {Sumner}, {Sun}, {Sunnquist}, {Swade}, {Swam},
  {Swenton}, {Swoish}, {Tam Litten}, {Tamas}, {Tao}, {Taylor}, {Taylor}, {te
  Plate}, {Van Tea}, {Teague}, {Telfer}, {Temim}, {Texter}, {Thatte},
  {Thompson}, {Thompson}, {Thomson}, {Thronson}, {Tierney}, {Tikkanen},
  {Tinnin}, {Tippet}, {Todd}, {Tran}, {Trauger}, {Trejo}, {Vinh Truong},
  {Tsukamoto}, {Tufail}, {Tumlinson}, {Tustain}, {Tyra}, {Ubeda}, {Underwood},
  {Uzzo}, {Vaclavik}, {Valenduc}, {Valenti}, {Van Campen}, {van de Wetering},
  {Van Der Marel}, {van Haarlem}, {Vandenbussche}, {van Dishoeck},
  {Vanterpool}, {Vernoy}, {Vila Costas}, {Volk}, {Voorzaat}, {Voyton}, {Vydra},
  {Waddy}, {Waelkens}, {Wahlgren}, {Walker}, {Wander}, {Warfield}, {Warner},
  {Wasiak}, {Wasiak}, {Wehner}, {Weiler}, {Weilert}, {Weiss}, {Wells}, {Welty},
  {Wheate}, {Wheeler}, {White}, {Whitehouse}, {Whiteleather}, {Whitman},
  {Williams}, {Willmer}, {Willott}, {Willoughby}, {Wilson}, {Wilson}, {Wilson},
  {Windhorst}, {Wislowski}, {Wolfe}, {Wolfe}, {Wolff}, {Wondel}, {Woo},
  {Woods}, {Worden}, {Workman}, {Wright}, {Wu}, {Wu}, {Wun}, {Wymer},
  {Yadetie}, {Yan}, {Yang}, {Yates}, {Yeager}, {Yerger}, {Young}, {Young},
  {Yu}, {Yu}, {Zak}, {Zeidler}, {Zepp}, {Zhou}, {Zincke}, {Zonak}, \&
  {Zondag}}]{Gardner-etal-2023}
{Gardner}, J.~P., {Mather}, J.~C., {Abbott}, R., {et~al.} 2023, \pasp, 135,
  068001, \dodoi{10.1088/1538-3873/acd1b5}

\bibitem[{{Garland} {et~al.}(2023){Garland}, {Fahey}, {Simmons}, {Smethurst},
  {Lintott}, {Shanahan}, {Silcock}, {Smith}, {Keel}, {Coil}, {G{\'e}ron},
  {Kruk}, {Masters}, {O'Ryan}, {Thorne}, \& {Wiersema}}]{Garland-etal-2023}
{Garland}, I.~L., {Fahey}, M.~J., {Simmons}, B.~D., {et~al.} 2023, \mnras, 522,
  211, \dodoi{10.1093/mnras/stad966}

\bibitem[{{Genzel} {et~al.}(2011){Genzel}, {Newman}, {Jones}, {F{\"o}rster
  Schreiber}, {Shapiro}, {Genel}, {Lilly}, {Renzini}, {Tacconi}, {Bouch{\'e}},
  {Burkert}, {Cresci}, {Buschkamp}, {Carollo}, {Ceverino}, {Davies}, {Dekel},
  {Eisenhauer}, {Hicks}, {Kurk}, {Lutz}, {Mancini}, {Naab}, {Peng},
  {Sternberg}, {Vergani}, \& {Zamorani}}]{Genzel-etal-2011}
{Genzel}, R., {Newman}, S., {Jones}, T., {et~al.} 2011, \apj, 733, 101,
  \dodoi{10.1088/0004-637X/733/2/101}

\bibitem[{{Ghosh} {et~al.}(2023){Ghosh}, {Fragkoudi}, {Di Matteo}, \&
  {Saha}}]{Ghosh-etal-2023}
{Ghosh}, S., {Fragkoudi}, F., {Di Matteo}, P., \& {Saha}, K. 2023, \aap, 674,
  A128, \dodoi{10.1051/0004-6361/202245275}

\bibitem[{{Ghosh} {et~al.}(2021){Ghosh}, {Saha}, {Di Matteo}, \&
  {Combes}}]{Ghosh-etal-2021}
{Ghosh}, S., {Saha}, K., {Di Matteo}, P., \& {Combes}, F. 2021, \mnras, 502,
  3085, \dodoi{10.1093/mnras/stab238}

\bibitem[{{Goulding} {et~al.}(2017){Goulding}, {Matthaey}, {Greene}, {Hickox},
  {Alexander}, {Forman}, {Jones}, {Lehmer}, {Griffis}, {Kanek}, \&
  {Oulmakki}}]{Goulding-etal-2017}
{Goulding}, A.~D., {Matthaey}, E., {Greene}, J.~E., {et~al.} 2017, \apj, 843,
  135, \dodoi{10.3847/1538-4357/aa755b}

\bibitem[{{Grand} {et~al.}(2024){Grand}, {Fragkoudi}, {G{\'o}mez}, {Jenkins},
  {Marinacci}, {Pakmor}, \& {Springel}}]{Grand-etal-2024}
{Grand}, R. J.~J., {Fragkoudi}, F., {G{\'o}mez}, F.~A., {et~al.} 2024, \mnras,
  532, 1814, \dodoi{10.1093/mnras/stae1598}

\bibitem[{{Grand} {et~al.}(2018){Grand}, {Bustamante}, {G{\'o}mez}, {Kawata},
  {Marinacci}, {Pakmor}, {Rix}, {Simpson}, {Sparre}, \&
  {Springel}}]{Grand-etal-2018}
{Grand}, R. J.~J., {Bustamante}, S., {G{\'o}mez}, F.~A., {et~al.} 2018, \mnras,
  474, 3629, \dodoi{10.1093/mnras/stx3025}

\bibitem[{{Grand} {et~al.}(2019){Grand}, {van de Voort}, {Zjupa}, {Fragkoudi},
  {G{\'o}mez}, {Kauffmann}, {Marinacci}, {Pakmor}, {Springel}, \&
  {White}}]{Grand-etal-2019}
{Grand}, R. J.~J., {van de Voort}, F., {Zjupa}, J., {et~al.} 2019, \mnras, 490,
  4786, \dodoi{10.1093/mnras/stz2928}

\bibitem[{{Grogin} {et~al.}(2011){Grogin}, {Kocevski}, {Faber}, {Ferguson},
  {Koekemoer}, {Riess}, {Acquaviva}, {Alexander}, {Almaini}, {Ashby}, {Barden},
  {Bell}, {Bournaud}, {Brown}, {Caputi}, {Casertano}, {Cassata}, {Castellano},
  {Challis}, {Chary}, {Cheung}, {Cirasuolo}, {Conselice}, {Roshan Cooray},
  {Croton}, {Daddi}, {Dahlen}, {Dav{\'e}}, {de Mello}, {Dekel}, {Dickinson},
  {Dolch}, {Donley}, {Dunlop}, {Dutton}, {Elbaz}, {Fazio}, {Filippenko},
  {Finkelstein}, {Fontana}, {Gardner}, {Garnavich}, {Gawiser}, {Giavalisco},
  {Grazian}, {Guo}, {Hathi}, {H{\"a}ussler}, {Hopkins}, {Huang}, {Huang},
  {Jha}, {Kartaltepe}, {Kirshner}, {Koo}, {Lai}, {Lee}, {Li}, {Lotz}, {Lucas},
  {Madau}, {McCarthy}, {McGrath}, {McIntosh}, {McLure}, {Mobasher},
  {Moustakas}, {Mozena}, {Nandra}, {Newman}, {Niemi}, {Noeske}, {Papovich},
  {Pentericci}, {Pope}, {Primack}, {Rajan}, {Ravindranath}, {Reddy}, {Renzini},
  {Rix}, {Robaina}, {Rodney}, {Rosario}, {Rosati}, {Salimbeni}, {Scarlata},
  {Siana}, {Simard}, {Smidt}, {Somerville}, {Spinrad}, {Straughn}, {Strolger},
  {Telford}, {Teplitz}, {Trump}, {van der Wel}, {Villforth}, {Wechsler},
  {Weiner}, {Wiklind}, {Wild}, {Wilson}, {Wuyts}, {Yan}, \&
  {Yun}}]{Grogin-etal-2011}
{Grogin}, N.~A., {Kocevski}, D.~D., {Faber}, S.~M., {et~al.} 2011, \apjs, 197,
  35, \dodoi{10.1088/0067-0049/197/2/35}

\bibitem[{{Guo} {et~al.}(2023){Guo}, {Jogee}, {Finkelstein}, {Chen}, {Wise},
  {Bagley}, {Barro}, {Wuyts}, {Kocevski}, {Kartaltepe}, {McGrath}, {Ferguson},
  {Mobasher}, {Giavalisco}, {Lucas}, {Zavala}, {Lotz}, {Grogin},
  {Huertas-Company}, {Vega-Ferrero}, {Hathi}, {Haro}, {Dickinson}, {Koekemoer},
  {Papovich}, {Pirzkal}, {Yung}, {Backhaus}, {Bell}, {Calabr{\`o}}, {Cleri},
  {Coogan}, {Cooper}, {Costantin}, {Croton}, {Davis}, {Dekel}, {Franco},
  {Gardner}, {Holwerda}, {Hutchison}, {Pandya}, {P{\'e}rez-Gonz{\'a}lez},
  {Ravindranath}, {Rose}, {Trump}, {de la Vega}, \& {Wang}}]{Guo-etal-2023}
{Guo}, Y., {Jogee}, S., {Finkelstein}, S.~L., {et~al.} 2023, \apjl, 945, L10,
  \dodoi{10.3847/2041-8213/acacfb}

\bibitem[{{Hamilton-Campos} {et~al.}(2023){Hamilton-Campos}, {Simons},
  {Peeples}, {Snyder}, \& {Heckman}}]{Hamilton-Campos-etal-2023}
{Hamilton-Campos}, K.~A., {Simons}, R.~C., {Peeples}, M.~S., {Snyder}, G.~F.,
  \& {Heckman}, T.~M. 2023, \apj, 956, 147, \dodoi{10.3847/1538-4357/acf211}

\bibitem[{{Hernquist} \& {Mihos}(1995)}]{Hernquist-Mihos1995}
{Hernquist}, L., \& {Mihos}, J.~C. 1995, \apj, 448, 41, \dodoi{10.1086/175940}

\bibitem[{{Ho} {et~al.}(1997){Ho}, {Filippenko}, \& {Sargent}}]{Ho-etal-1997}
{Ho}, L.~C., {Filippenko}, A.~V., \& {Sargent}, W. L.~W. 1997, \apj, 487, 591,
  \dodoi{10.1086/304643}

\bibitem[{{Hopkins} \& {Quataert}(2010)}]{Hopkins-etal-2010}
{Hopkins}, P.~F., \& {Quataert}, E. 2010, \mnras, 407, 1529,
  \dodoi{10.1111/j.1365-2966.2010.17064.x}

\bibitem[{{Huang} {et~al.}(2023){Huang}, {Kawabe}, {Kohno}, {Saito},
  {Mizukoshi}, {Iono}, {Michiyama}, {Tamura}, {Hayward}, \&
  {Umehata}}]{Huang-etal-2023}
{Huang}, S., {Kawabe}, R., {Kohno}, K., {et~al.} 2023, \apjl, 958, L26,
  \dodoi{10.3847/2041-8213/acff63}

\bibitem[{{Huertas-Company} {et~al.}(2024){Huertas-Company}, {Iyer},
  {Angeloudi}, {Bagley}, {Finkelstein}, {Kartaltepe}, {McGrath}, {Sarmiento},
  {Vega-Ferrero}, {Arrabal Haro}, {Behroozi}, {Buitrago}, {Cheng}, {Costantin},
  {Dekel}, {Dickinson}, {Elbaz}, {Grogin}, {Hathi}, {Holwerda}, {Koekemoer},
  {Lucas}, {Papovich}, {P{\'e}rez-Gonz{\'a}lez}, {Pirzkal}, {Seill{\'e}}, {de
  la Vega}, {Wuyts}, {Yang}, \& {Yung}}]{Huertas-Company-etal-2024}
{Huertas-Company}, M., {Iyer}, K.~G., {Angeloudi}, E., {et~al.} 2024, \aap,
  685, A48, \dodoi{10.1051/0004-6361/202346800}

\bibitem[{{Huertas-Company} {et~al.}(2025){Huertas-Company}, {Shuntov}, {Dong},
  {Walmsley}, {Ilbert}, {McCracken}, {Akins}, {Allen}, {Casey}, {Costantin},
  {Daddi}, {Dekel}, {Franco}, {Garland}, {G{\'e}ron}, {Gozaliasl},
  {Hirschmann}, {Kartaltepe}, {Koekemoer}, {Lintott}, {Liu}, {Lucas},
  {Masters}, {Pacucci}, {Paquereau}, {P'erez-Gonz'alez}, {Rhodes}, {Robertson},
  {Simmons}, {Smethurst}, {Toft}, \& {Yang}}]{Huertas-Company-etal-2025}
{Huertas-Company}, M., {Shuntov}, M., {Dong}, Y., {et~al.} 2025, arXiv
  e-prints, arXiv:2502.03532, \dodoi{10.48550/arXiv.2502.03532}

\bibitem[{{Iyer} \& {Gawiser}(2017)}]{Iyer-Gawiser-2017}
{Iyer}, K., \& {Gawiser}, E. 2017, \apj, 838, 127,
  \dodoi{10.3847/1538-4357/aa63f0}

\bibitem[{{Iyer} {et~al.}(2019){Iyer}, {Gawiser}, {Faber}, {Ferguson},
  {Kartaltepe}, {Koekemoer}, {Pacifici}, \& {Somerville}}]{Iyer-etal-2019}
{Iyer}, K.~G., {Gawiser}, E., {Faber}, S.~M., {et~al.} 2019, \apj, 879, 116,
  \dodoi{10.3847/1538-4357/ab2052}

\bibitem[{{Izquierdo-Villalba} {et~al.}(2022){Izquierdo-Villalba}, {Bonoli},
  {Rosas-Guevara}, {Springel}, {White}, {Zana}, {Dotti}, {Spinoso}, {Bonetti},
  \& {Lupi}}]{Izquierdo-Villalba-etal-2022}
{Izquierdo-Villalba}, D., {Bonoli}, S., {Rosas-Guevara}, Y., {et~al.} 2022,
  \mnras, 514, 1006, \dodoi{10.1093/mnras/stac1413}

\bibitem[{{Jacobs} {et~al.}(2022){Jacobs}, {Glazebrook}, {Calabr{\`o}}, {Treu},
  {Nanayakkara}, {Jones}, {Merlin}, {Abraham}, {Stevens}, {Vulcani}, {Yang},
  {Bonchi}, {Bradac}, {Castellano}, {Fontana}, {Mason}, {Morishita}, {Paris},
  {Trenti}, {Marchesini}, {Wang}, \& {Santini}}]{Jacobs-etal-2022}
{Jacobs}, C., {Glazebrook}, K., {Calabr{\`o}}, A., {et~al.} 2022, arXiv
  e-prints, arXiv:2208.06516.
\newblock \doarXiv{2208.06516}

\bibitem[{{James} {et~al.}(2009){James}, {Bretherton}, \&
  {Knapen}}]{James-etal-2009}
{James}, P.~A., {Bretherton}, C.~F., \& {Knapen}, J.~H. 2009, \aap, 501, 207,
  \dodoi{10.1051/0004-6361/200810715}

\bibitem[{{James} \& {Percival}(2016)}]{James-Percival-2016}
{James}, P.~A., \& {Percival}, S.~M. 2016, \mnras, 457, 917,
  \dodoi{10.1093/mnras/stv2978}

\bibitem[{{Jedrzejewski}(1987)}]{Jedrzejewski1987}
{Jedrzejewski}, R.~I. 1987, \mnras, 226, 747, \dodoi{10.1093/mnras/226.4.747}

\bibitem[{{Jogee}(2006)}]{Jogee2006}
{Jogee}, S. 2006, in Physics of Active Galactic Nuclei at all Scales, ed.
  D.~{Alloin}, Vol. 693, 143, \dodoi{10.1007/3-540-34621-X_6}

\bibitem[{{Jogee} {et~al.}(1999){Jogee}, {Kenney}, \&
  {Smith}}]{Jogee-Kenney-Smith1999}
{Jogee}, S., {Kenney}, J. D.~P., \& {Smith}, B.~J. 1999, \apj, 526, 665,
  \dodoi{10.1086/308021}

\bibitem[{{Jogee} {et~al.}(2002){Jogee}, {Knapen}, {Laine}, {Shlosman},
  {Scoville}, \& {Englmaier}}]{Jogee-etal-2002a}
{Jogee}, S., {Knapen}, J.~H., {Laine}, S., {et~al.} 2002, \apjl, 570, L55,
  \dodoi{10.1086/340974}

\bibitem[{{Jogee} {et~al.}(2005){Jogee}, {Scoville}, \&
  {Kenney}}]{Jogee-Scoville-Kenney2005}
{Jogee}, S., {Scoville}, N., \& {Kenney}, J. D.~P. 2005, \apj, 630, 837,
  \dodoi{10.1086/432106}

\bibitem[{{Jogee} {et~al.}(2004){Jogee}, {Barazza}, {Rix}, {Shlosman},
  {Barden}, {Wolf}, {Davies}, {Heyer}, {Beckwith}, {Bell}, {Borch}, {Caldwell},
  {Conselice}, {Dahlen}, {H{\"a}ussler}, {Heymans}, {Jahnke}, {Knapen},
  {Laine}, {Lubell}, {Mobasher}, {McIntosh}, {Meisenheimer}, {Peng},
  {Ravindranath}, {Sanchez}, {Somerville}, \& {Wisotzki}}]{Jogee-etal-2004}
{Jogee}, S., {Barazza}, F.~D., {Rix}, H.-W., {et~al.} 2004, \apjl, 615, L105,
  \dodoi{10.1086/426138}

\bibitem[{{Jogee} {et~al.}(2009){Jogee}, {Miller}, {Penner}, {Skelton},
  {Conselice}, {Somerville}, {Bell}, {Zheng}, {Rix}, {Robaina}, {Barazza},
  {Barden}, {Borch}, {Beckwith}, {Caldwell}, {Peng}, {Heymans}, {McIntosh},
  {H{\"a}u{\ss}ler}, {Jahnke}, {Meisenheimer}, {Sanchez}, {Wisotzki}, {Wolf},
  \& {Papovich}}]{Jogee-etal-2009}
{Jogee}, S., {Miller}, S.~H., {Penner}, K., {et~al.} 2009, \apj, 697, 1971,
  \dodoi{10.1088/0004-637X/697/2/1971}

\bibitem[{{Kartaltepe} {et~al.}(2007){Kartaltepe}, {Sanders}, {Scoville},
  {Calzetti}, {Capak}, {Koekemoer}, {Mobasher}, {Murayama}, {Salvato},
  {Sasaki}, \& {Taniguchi}}]{Kartaltepe-etal-2007}
{Kartaltepe}, J.~S., {Sanders}, D.~B., {Scoville}, N.~Z., {et~al.} 2007, \apjs,
  172, 320, \dodoi{10.1086/519953}

\bibitem[{{Kartaltepe} {et~al.}(2023){Kartaltepe}, {Rose}, {Vanderhoof},
  {McGrath}, {Costantin}, {Cox}, {Yung}, {Kocevski}, {Wuyts}, {Ferguson},
  {Bagley}, {Finkelstein}, {Amor{\'\i}n}, {Andrews}, {Arrabal Haro},
  {Backhaus}, {Behroozi}, {Bisigello}, {Calabr{\`o}}, {Casey}, {Coogan},
  {Cooper}, {Croton}, {de la Vega}, {Dickinson}, {Fontana}, {Franco},
  {Grazian}, {Grogin}, {Hathi}, {Holwerda}, {Huertas-Company}, {Iyer}, {Jogee},
  {Jung}, {Kewley}, {Kirkpatrick}, {Koekemoer}, {Liu}, {Lotz}, {Lucas},
  {Newman}, {Pacifici}, {Pandya}, {Papovich}, {Pentericci},
  {P{\'e}rez-Gonz{\'a}lez}, {Petersen}, {Pirzkal}, {Rafelski}, {Ravindranath},
  {Simons}, {Snyder}, {Somerville}, {Stanway}, {Straughn}, {Tacchella},
  {Trump}, {Vega-Ferrero}, {Wilkins}, {Yang}, \&
  {Zavala}}]{Kartaltepe-etal-2023}
{Kartaltepe}, J.~S., {Rose}, C., {Vanderhoof}, B.~N., {et~al.} 2023, \apjl,
  946, L15, \dodoi{10.3847/2041-8213/acad01}

\bibitem[{{Kataria} \& {Vivek}(2024)}]{Kataria-etal-2024}
{Kataria}, S.~K., \& {Vivek}, M. 2024, \mnras, 527, 3366,
  \dodoi{10.1093/mnras/stad3383}

\bibitem[{{Katz} {et~al.}(2003){Katz}, {Keres}, {Dave}, \&
  {Weinberg}}]{Katz-etal-2003}
{Katz}, N., {Keres}, D., {Dave}, R., \& {Weinberg}, D.~H. 2003, in Astrophysics
  and Space Science Library, Vol. 281, The IGM/Galaxy Connection. The
  Distribution of Baryons at z=0, ed. J.~L. {Rosenberg} \& M.~E. {Putman}, 185,
  \dodoi{10.1007/978-94-010-0115-1_34}

\bibitem[{{Kere{\v{s}}} {et~al.}(2005){Kere{\v{s}}}, {Katz}, {Weinberg}, \&
  {Dav{\'e}}}]{Keres-etal-2005}
{Kere{\v{s}}}, D., {Katz}, N., {Weinberg}, D.~H., \& {Dav{\'e}}, R. 2005,
  \mnras, 363, 2, \dodoi{10.1111/j.1365-2966.2005.09451.x}

\bibitem[{{Kere{\v{s}}} {et~al.}(2012){Kere{\v{s}}}, {Vogelsberger}, {Sijacki},
  {Springel}, \& {Hernquist}}]{Keres-etal-2012}
{Kere{\v{s}}}, D., {Vogelsberger}, M., {Sijacki}, D., {Springel}, V., \&
  {Hernquist}, L. 2012, \mnras, 425, 2027,
  \dodoi{10.1111/j.1365-2966.2012.21548.x}

\bibitem[{{Knapen} {et~al.}(1995){Knapen}, {Beckman}, {Heller}, {Shlosman}, \&
  {de Jong}}]{Knapen-etal-1995}
{Knapen}, J.~H., {Beckman}, J.~E., {Heller}, C.~H., {Shlosman}, I., \& {de
  Jong}, R.~S. 1995, \apj, 454, 623, \dodoi{10.1086/176516}

\bibitem[{{Knapen} {et~al.}(2000){Knapen}, {Shlosman}, \&
  {Peletier}}]{Knapen-Shlosman-Peletier2000}
{Knapen}, J.~H., {Shlosman}, I., \& {Peletier}, R.~F. 2000, \apj, 529, 93,
  \dodoi{10.1086/308266}

\bibitem[{{Koekemoer} {et~al.}(2011){Koekemoer}, {Faber}, {Ferguson}, {Grogin},
  {Kocevski}, {Koo}, {Lai}, {Lotz}, {Lucas}, {McGrath}, {Ogaz}, {Rajan},
  {Riess}, {Rodney}, {Strolger}, {Casertano}, {Castellano}, {Dahlen},
  {Dickinson}, {Dolch}, {Fontana}, {Giavalisco}, {Grazian}, {Guo}, {Hathi},
  {Huang}, {van der Wel}, {Yan}, {Acquaviva}, {Alexander}, {Almaini}, {Ashby},
  {Barden}, {Bell}, {Bournaud}, {Brown}, {Caputi}, {Cassata}, {Challis},
  {Chary}, {Cheung}, {Cirasuolo}, {Conselice}, {Roshan Cooray}, {Croton},
  {Daddi}, {Dav{\'e}}, {de Mello}, {de Ravel}, {Dekel}, {Donley}, {Dunlop},
  {Dutton}, {Elbaz}, {Fazio}, {Filippenko}, {Finkelstein}, {Frazer}, {Gardner},
  {Garnavich}, {Gawiser}, {Gruetzbauch}, {Hartley}, {H{\"a}ussler},
  {Herrington}, {Hopkins}, {Huang}, {Jha}, {Johnson}, {Kartaltepe},
  {Khostovan}, {Kirshner}, {Lani}, {Lee}, {Li}, {Madau}, {McCarthy},
  {McIntosh}, {McLure}, {McPartland}, {Mobasher}, {Moreira}, {Mortlock},
  {Moustakas}, {Mozena}, {Nandra}, {Newman}, {Nielsen}, {Niemi}, {Noeske},
  {Papovich}, {Pentericci}, {Pope}, {Primack}, {Ravindranath}, {Reddy},
  {Renzini}, {Rix}, {Robaina}, {Rosario}, {Rosati}, {Salimbeni}, {Scarlata},
  {Siana}, {Simard}, {Smidt}, {Snyder}, {Somerville}, {Spinrad}, {Straughn},
  {Telford}, {Teplitz}, {Trump}, {Vargas}, {Villforth}, {Wagner}, {Wandro},
  {Wechsler}, {Weiner}, {Wiklind}, {Wild}, {Wilson}, {Wuyts}, \&
  {Yun}}]{Koekemoer-etal-2011}
{Koekemoer}, A.~M., {Faber}, S.~M., {Ferguson}, H.~C., {et~al.} 2011, \apjs,
  197, 36, \dodoi{10.1088/0067-0049/197/2/36}

\bibitem[{{Kohandel} {et~al.}(2023){Kohandel}, {Pallottini}, {Ferrara},
  {Zanella}, {Rizzo}, \& {Carniani}}]{Kohandel-etal-2024}
{Kohandel}, M., {Pallottini}, A., {Ferrara}, A., {et~al.} 2023, arXiv e-prints,
  arXiv:2311.05832, \dodoi{10.48550/arXiv.2311.05832}

\bibitem[{{Kormendy} \&
  {Kennicutt}(2004{\natexlab{a}})}]{Kormendy&Kennicutt2004}
{Kormendy}, J., \& {Kennicutt}, Robert~C., J. 2004{\natexlab{a}}, \araa, 42,
  603, \dodoi{10.1146/annurev.astro.42.053102.134024}

\bibitem[{{Kormendy} \&
  {Kennicutt}(2004{\natexlab{b}})}]{Kormendy-Kennicutt2004}
---. 2004{\natexlab{b}}, \araa, 42, 603,
  \dodoi{10.1146/annurev.astro.42.053102.134024}

\bibitem[{{Kraljic} {et~al.}(2012){Kraljic}, {Bournaud}, \&
  {Martig}}]{Kraljic-Bournaud-Martig2012}
{Kraljic}, K., {Bournaud}, F., \& {Martig}, M. 2012, \apj, 757, 60,
  \dodoi{10.1088/0004-637X/757/1/60}

\bibitem[{{Larson} {et~al.}(2023){Larson}, {Finkelstein}, {Kocevski},
  {Hutchison}, {Trump}, {Haro}, {Bromm}, {Cleri}, {Dickinson}, {Fujimoto},
  {Kartaltepe}, {Koekemoer}, {Papovich}, {Pirzkal}, {Tacchella}, {Zavala},
  {Bagley}, {Behroozi}, {Champagne}, {Cole}, {Jung}, {Morales}, {Yang},
  {Zhang}, {Zitrin}, {Amor{\'\i}n}, {Burgarella}, {Casey}, {Ch{\'a}vez Ortiz},
  {Cox}, {Chworowsky}, {Fontana}, {Gawiser}, {Grazian}, {Grogin}, {Harish},
  {Hathi}, {Hirschmann}, {Holwerda}, {Juneau}, {Leung}, {Lucas}, {McGrath},
  {P{\'e}rez-Gonz{\'a}lez}, {Rigby}, {Seill{\'e}}, {Simons}, {de La Vega},
  {Weiner}, {Wilkins}, {Yung}, \& {Ceers Team}}]{Larson-et-al-2023}
{Larson}, R.~L., {Finkelstein}, S.~L., {Kocevski}, D.~D., {et~al.} 2023, \apjl,
  953, L29, \dodoi{10.3847/2041-8213/ace619}

\bibitem[{{Laurikainen} {et~al.}(2005){Laurikainen}, {Salo}, \&
  {Buta}}]{Laurikainen-Salo-Buta2005}
{Laurikainen}, E., {Salo}, H., \& {Buta}, R. 2005, \mnras, 362, 1319,
  \dodoi{10.1111/j.1365-2966.2005.09404.x}

\bibitem[{{Laurikainen} {et~al.}(2002){Laurikainen}, {Salo}, \&
  {Rautiainen}}]{Laurikainen-Salo-Rautianinen2002}
{Laurikainen}, E., {Salo}, H., \& {Rautiainen}, P. 2002, \mnras, 331, 880,
  \dodoi{10.1046/j.1365-8711.2002.05243.x}

\bibitem[{{Le Conte} {et~al.}(2024){Le Conte}, {Gadotti}, {Ferreira},
  {Conselice}, {de S{\'a}-Freitas}, {Kim}, {Neumann}, {Fragkoudi},
  {Athanassoula}, \& {Adams}}]{Le-Conte-etal-2024}
{Le Conte}, Z.~A., {Gadotti}, D.~A., {Ferreira}, L., {et~al.} 2024, \mnras,
  530, 1984, \dodoi{10.1093/mnras/stae921}

\bibitem[{{Lee} {et~al.}(2023){Lee}, {Park}, {Hwang}, \&
  {Kwon}}]{Lee-etal-2023}
{Lee}, J.~H., {Park}, C., {Hwang}, H.~S., \& {Kwon}, M. 2023, arXiv e-prints,
  arXiv:2312.04899, \dodoi{10.48550/arXiv.2312.04899}

\bibitem[{{Leja} {et~al.}(2019){Leja}, {Carnall}, {Johnson}, {Conroy}, \&
  {Speagle}}]{Leja-etal-2019}
{Leja}, J., {Carnall}, A.~C., {Johnson}, B.~D., {Conroy}, C., \& {Speagle},
  J.~S. 2019, \apj, 876, 3, \dodoi{10.3847/1538-4357/ab133c}

\bibitem[{{Leroy} {et~al.}(2021){Leroy}, {Schinnerer}, {Hughes}, {Rosolowsky},
  {Pety}, {Schruba}, {Usero}, {Blanc}, {Chevance}, {Emsellem}, {Faesi},
  {Herrera}, {Liu}, {Meidt}, {Querejeta}, {Saito}, {Sandstrom}, {Sun},
  {Williams}, {Anand}, {Barnes}, {Behrens}, {Belfiore}, {Benincasa},
  {Be{\v{s}}li{\'c}}, {Bigiel}, {Bolatto}, {den Brok}, {Cao}, {Chandar},
  {Chastenet}, {Chiang}, {Congiu}, {Dale}, {Deger}, {Eibensteiner}, {Egorov},
  {Garc{\'\i}a-Rodr{\'\i}guez}, {Glover}, {Grasha}, {Henshaw}, {Ho}, {Kepley},
  {Kim}, {Klessen}, {Kreckel}, {Koch}, {Kruijssen}, {Larson}, {Lee}, {Lopez},
  {Machado}, {Mayker}, {McElroy}, {Murphy}, {Ostriker}, {Pan}, {Pessa},
  {Puschnig}, {Razza}, {S{\'a}nchez-Bl{\'a}zquez}, {Santoro}, {Sardone},
  {Scheuermann}, {Sliwa}, {Sormani}, {Stuber}, {Thilker}, {Turner}, {Utomo},
  {Watkins}, \& {Whitmore}}]{Leroy-etal-2021}
{Leroy}, A.~K., {Schinnerer}, E., {Hughes}, A., {et~al.} 2021, \apjs, 257, 43,
  \dodoi{10.3847/1538-4365/ac17f3}

\bibitem[{{Liang} {et~al.}(2024){Liang}, {Yu}, {Fang}, \&
  {Ho}}]{Liang-etal-2023}
{Liang}, X., {Yu}, S.-Y., {Fang}, T., \& {Ho}, L.~C. 2024, \aap, 688, A158,
  \dodoi{10.1051/0004-6361/202348539}

\bibitem[{{Lin} {et~al.}(2020){Lin}, {Li}, {Du}, {Wang}, {Xiao}, {Bureau},
  {Fraser-McKelvie}, {Masters}, {Lin}, {Wake}, \& {Hao}}]{Lin-Li-Du2020}
{Lin}, L., {Li}, C., {Du}, C., {et~al.} 2020, \mnras, 499, 1406,
  \dodoi{10.1093/mnras/staa2913}

\bibitem[{{Lotz} {et~al.}(2010){Lotz}, {Jonsson}, {Cox}, \&
  {Primack}}]{Lotz-etal-2010}
{Lotz}, J.~M., {Jonsson}, P., {Cox}, T.~J., \& {Primack}, J.~R. 2010, \mnras,
  404, 590, \dodoi{10.1111/j.1365-2966.2010.16269.x}

\bibitem[{{Lower} {et~al.}(2020){Lower}, {Narayanan}, {Leja}, {Johnson},
  {Conroy}, \& {Dav{\'e}}}]{Lower-etal-2020}
{Lower}, S., {Narayanan}, D., {Leja}, J., {et~al.} 2020, \apj, 904, 33,
  \dodoi{10.3847/1538-4357/abbfa7}

\bibitem[{{Madau}(1995)}]{Madau-1995}
{Madau}, P. 1995, \apj, 441, 18, \dodoi{10.1086/175332}

\bibitem[{{Marinacci} {et~al.}(2018){Marinacci}, {Vogelsberger}, {Pakmor},
  {Torrey}, {Springel}, {Hernquist}, {Nelson}, {Weinberger}, {Pillepich},
  {Naiman}, \& {Genel}}]{Marinacci-etal-2018}
{Marinacci}, F., {Vogelsberger}, M., {Pakmor}, R., {et~al.} 2018, \mnras, 480,
  5113, \dodoi{10.1093/mnras/sty2206}

\bibitem[{{Marinova} \& {Jogee}(2007)}]{Marinova-Jogee2007}
{Marinova}, I., \& {Jogee}, S. 2007, \apj, 659, 1176, \dodoi{10.1086/512355}

\bibitem[{{Martinez-Valpuesta} {et~al.}(2006){Martinez-Valpuesta}, {Shlosman},
  \& {Heller}}]{Martinez-Valpuesta-Shlosman-Heller2006}
{Martinez-Valpuesta}, I., {Shlosman}, I., \& {Heller}, C. 2006, \apj, 637, 214,
  \dodoi{10.1086/498338}

\bibitem[{{Masters} {et~al.}(2010){Masters}, {Mosleh}, {Romer}, {Nichol},
  {Bamford}, {Schawinski}, {Lintott}, {Andreescu}, {Campbell}, {Crowcroft},
  {Doyle}, {Edmondson}, {Murray}, {Raddick}, {Slosar}, {Szalay}, \&
  {Vandenberg}}]{Masters-etal-2010}
{Masters}, K.~L., {Mosleh}, M., {Romer}, A.~K., {et~al.} 2010, \mnras, 405,
  783, \dodoi{10.1111/j.1365-2966.2010.16503.x}

\bibitem[{{McKinney} {et~al.}(2024){McKinney}, {Casey}, {Long}, {Cooper},
  {Manning}, {Franco}, {Akin}, {Lambrides}, {Gammon}, {Silva}, {Gentile},
  {Zavala}, {Amvrosiadis}, {Andika}, {Brinch}, {Champagne}, {Chartab},
  {Drakos}, {Faisst}, {Fujimoto}, {Gillman}, {Gozaliasl}, {Greve}, {Harish},
  {Hayward}, {Hirschmann}, {Ilbert}, {Kalita}, {Kartaltepe}, {Koekemoer},
  {Kokorev}, {Liu}, {Magdis}, {McCracken}, {Rhodes}, {Robertson}, {Talia},
  {Valentino}, \& {Vijayan}}]{McKinney-etal-2024}
{McKinney}, J., {Casey}, C.~M., {Long}, A.~S., {et~al.} 2024, arXiv e-prints,
  arXiv:2408.08346, \dodoi{10.48550/arXiv.2408.08346}

\bibitem[{{Meidt} {et~al.}(2014){Meidt}, {Schinnerer}, {van de Ven},
  {Zaritsky}, {Peletier}, {Knapen}, {Sheth}, {Regan}, {Querejeta},
  {Mu{\~n}oz-Mateos}, {Kim}, {Hinz}, {Gil de Paz}, {Athanassoula}, {Bosma},
  {Buta}, {Cisternas}, {Ho}, {Holwerda}, {Skibba}, {Laurikainen}, {Salo},
  {Gadotti}, {Laine}, {Erroz-Ferrer}, {Comer{\'o}n}, {Men{\'e}ndez-Delmestre},
  {Seibert}, \& {Mizusawa}}]{Meidt-etal-2014}
{Meidt}, S.~E., {Schinnerer}, E., {van de Ven}, G., {et~al.} 2014, \apj, 788,
  144, \dodoi{10.1088/0004-637X/788/2/144}

\bibitem[{{Melvin} {et~al.}(2014){Melvin}, {Masters}, {Lintott}, {Nichol},
  {Simmons}, {Bamford}, {Casteels}, {Cheung}, {Edmondson}, {Fortson},
  {Schawinski}, {Skibba}, {Smith}, \& {Willett}}]{Melvin-etal-2014}
{Melvin}, T., {Masters}, K., {Lintott}, C., {et~al.} 2014, \mnras, 438, 2882,
  \dodoi{10.1093/mnras/stt2397}

\bibitem[{{M{\'e}ndez-Abreu} {et~al.}(2023){M{\'e}ndez-Abreu}, {Costantin}, \&
  {Kruk}}]{Mendez-Abreu-etal-2023}
{M{\'e}ndez-Abreu}, J., {Costantin}, L., \& {Kruk}, S. 2023, \aap, 678, A54,
  \dodoi{10.1051/0004-6361/202346685}

\bibitem[{{Men{\'e}ndez-Delmestre} {et~al.}(2007){Men{\'e}ndez-Delmestre},
  {Sheth}, {Schinnerer}, {Jarrett}, \&
  {Scoville}}]{Menendez-Delmestre-etal-2007}
{Men{\'e}ndez-Delmestre}, K., {Sheth}, K., {Schinnerer}, E., {Jarrett}, T.~H.,
  \& {Scoville}, N.~Z. 2007, \apj, 657, 790, \dodoi{10.1086/511025}

\bibitem[{{Naiman} {et~al.}(2018){Naiman}, {Pillepich}, {Springel},
  {Ramirez-Ruiz}, {Torrey}, {Vogelsberger}, {Pakmor}, {Nelson}, {Marinacci},
  {Hernquist}, {Weinberger}, \& {Genel}}]{Naiman-etal-2018}
{Naiman}, J.~P., {Pillepich}, A., {Springel}, V., {et~al.} 2018, \mnras, 477,
  1206, \dodoi{10.1093/mnras/sty618}

\bibitem[{{Neeleman} {et~al.}(2023){Neeleman}, {Walter}, {Decarli}, {Drake},
  {Eilers}, {Meyer}, \& {Venemans}}]{Neeleman-etal-2023}
{Neeleman}, M., {Walter}, F., {Decarli}, R., {et~al.} 2023, \apj, 958, 132,
  \dodoi{10.3847/1538-4357/ad05d2}

\bibitem[{{Nelson} {et~al.}(2018){Nelson}, {Pillepich}, {Springel},
  {Weinberger}, {Hernquist}, {Pakmor}, {Genel}, {Torrey}, {Vogelsberger},
  {Kauffmann}, {Marinacci}, \& {Naiman}}]{Nelson-etal-2018}
{Nelson}, D., {Pillepich}, A., {Springel}, V., {et~al.} 2018, \mnras, 475, 624,
  \dodoi{10.1093/mnras/stx3040}

\bibitem[{{Nelson} {et~al.}(2019{\natexlab{a}}){Nelson}, {Pillepich},
  {Springel}, {Pakmor}, {Weinberger}, {Genel}, {Torrey}, {Vogelsberger},
  {Marinacci}, \& {Hernquist}}]{Nelson-etal-2019}
---. 2019{\natexlab{a}}, \mnras, 490, 3234, \dodoi{10.1093/mnras/stz2306}

\bibitem[{{Nelson} {et~al.}(2019{\natexlab{b}}){Nelson}, {Springel},
  {Pillepich}, {Rodriguez-Gomez}, {Torrey}, {Genel}, {Vogelsberger}, {Pakmor},
  {Marinacci}, {Weinberger}, {Kelley}, {Lovell}, {Diemer}, \&
  {Hernquist}}]{Nelson-etal-2019-Release}
{Nelson}, D., {Springel}, V., {Pillepich}, A., {et~al.} 2019{\natexlab{b}},
  Computational Astrophysics and Cosmology, 6, 2,
  \dodoi{10.1186/s40668-019-0028-x}

\bibitem[{{Oke} \& {Gunn}(1983)}]{Oke-Gunn1983}
{Oke}, J.~B., \& {Gunn}, J.~E. 1983, \apj, 266, 713

\bibitem[{{Olgu{\'\i}n-Iglesias} {et~al.}(2020){Olgu{\'\i}n-Iglesias},
  {Kotilainen}, \& {Chavushyan}}]{Olguin-Iglesias-Kotilainen-Chavushyan2020}
{Olgu{\'\i}n-Iglesias}, A., {Kotilainen}, J., \& {Chavushyan}, V. 2020, \mnras,
  492, 1450, \dodoi{10.1093/mnras/stz3549}

\bibitem[{{Pakmor} {et~al.}(2016){Pakmor}, {Springel}, {Bauer}, {Mocz},
  {Munoz}, {Ohlmann}, {Schaal}, \& {Zhu}}]{Pakmor-etal-2016}
{Pakmor}, R., {Springel}, V., {Bauer}, A., {et~al.} 2016, \mnras, 455, 1134,
  \dodoi{10.1093/mnras/stv2380}

\bibitem[{Peng {et~al.}(2010)Peng, {Ho}, {Impey}, \& {Rix}}]{Peng-et-al-2010}
Peng, C.~Y., {Ho}, L.~C., {Impey}, C.~D., \& {Rix}, H.-W. 2010, \aj, 139, 2097,
  \dodoi{10.1088/0004-6256/139/6/2097}

\bibitem[{{Peschken} \& {{\L}okas}(2019)}]{Peschken-etal-2019}
{Peschken}, N., \& {{\L}okas}, E.~L. 2019, \mnras, 483, 2721,
  \dodoi{10.1093/mnras/sty3277}

\bibitem[{{Pillepich} {et~al.}(2018){Pillepich}, {Springel}, {Nelson}, {Genel},
  {Naiman}, {Pakmor}, {Hernquist}, {Torrey}, {Vogelsberger}, {Weinberger}, \&
  {Marinacci}}]{Pillepich-etal-2018}
{Pillepich}, A., {Springel}, V., {Nelson}, D., {et~al.} 2018, \mnras, 473,
  4077, \dodoi{10.1093/mnras/stx2656}

\bibitem[{{Pillepich} {et~al.}(2019){Pillepich}, {Nelson}, {Springel},
  {Pakmor}, {Torrey}, {Weinberger}, {Vogelsberger}, {Marinacci}, {Genel}, {van
  der Wel}, \& {Hernquist}}]{Pillepich-etal-2019}
{Pillepich}, A., {Nelson}, D., {Springel}, V., {et~al.} 2019, \mnras, 490,
  3196, \dodoi{10.1093/mnras/stz2338}

\bibitem[{{Planck Collaboration} {et~al.}(2020){Planck Collaboration},
  {Aghanim}, {Akrami}, {Ashdown}, {Aumont}, {Baccigalupi}, {Ballardini},
  {Banday}, {Barreiro}, {Bartolo}, {Basak}, {Battye}, {Benabed}, {Bernard},
  {Bersanelli}, {Bielewicz}, {Bock}, {Bond}, {Borrill}, {Bouchet}, {Boulanger},
  {Bucher}, {Burigana}, {Butler}, {Calabrese}, {Cardoso}, {Carron},
  {Challinor}, {Chiang}, {Chluba}, {Colombo}, {Combet}, {Contreras}, {Crill},
  {Cuttaia}, {de Bernardis}, {de Zotti}, {Delabrouille}, {Delouis}, {Di
  Valentino}, {Diego}, {Dor{\'e}}, {Douspis}, {Ducout}, {Dupac}, {Dusini},
  {Efstathiou}, {Elsner}, {En{\ss}lin}, {Eriksen}, {Fantaye}, {Farhang},
  {Fergusson}, {Fernandez-Cobos}, {Finelli}, {Forastieri}, {Frailis},
  {Fraisse}, {Franceschi}, {Frolov}, {Galeotta}, {Galli}, {Ganga},
  {G{\'e}nova-Santos}, {Gerbino}, {Ghosh}, {Gonz{\'a}lez-Nuevo}, {G{\'o}rski},
  {Gratton}, {Gruppuso}, {Gudmundsson}, {Hamann}, {Handley}, {Hansen},
  {Herranz}, {Hildebrandt}, {Hivon}, {Huang}, {Jaffe}, {Jones}, {Karakci},
  {Keih{\"a}nen}, {Keskitalo}, {Kiiveri}, {Kim}, {Kisner}, {Knox},
  {Krachmalnicoff}, {Kunz}, {Kurki-Suonio}, {Lagache}, {Lamarre}, {Lasenby},
  {Lattanzi}, {Lawrence}, {Le Jeune}, {Lemos}, {Lesgourgues}, {Levrier},
  {Lewis}, {Liguori}, {Lilje}, {Lilley}, {Lindholm}, {L{\'o}pez-Caniego},
  {Lubin}, {Ma}, {Mac{\'\i}as-P{\'e}rez}, {Maggio}, {Maino}, {Mandolesi},
  {Mangilli}, {Marcos-Caballero}, {Maris}, {Martin}, {Martinelli},
  {Mart{\'\i}nez-Gonz{\'a}lez}, {Matarrese}, {Mauri}, {McEwen}, {Meinhold},
  {Melchiorri}, {Mennella}, {Migliaccio}, {Millea}, {Mitra},
  {Miville-Desch{\^e}nes}, {Molinari}, {Montier}, {Morgante}, {Moss}, {Natoli},
  {N{\o}rgaard-Nielsen}, {Pagano}, {Paoletti}, {Partridge}, {Patanchon},
  {Peiris}, {Perrotta}, {Pettorino}, {Piacentini}, {Polastri}, {Polenta},
  {Puget}, {Rachen}, {Reinecke}, {Remazeilles}, {Renzi}, {Rocha}, {Rosset},
  {Roudier}, {Rubi{\~n}o-Mart{\'\i}n}, {Ruiz-Granados}, {Salvati}, {Sandri},
  {Savelainen}, {Scott}, {Shellard}, {Sirignano}, {Sirri}, {Spencer},
  {Sunyaev}, {Suur-Uski}, {Tauber}, {Tavagnacco}, {Tenti}, {Toffolatti},
  {Tomasi}, {Trombetti}, {Valenziano}, {Valiviita}, {Van Tent}, {Vibert},
  {Vielva}, {Villa}, {Vittorio}, {Wandelt}, {Wehus}, {White}, {White},
  {Zacchei}, \& {Zonca}}]{Planck-etal-20}
{Planck Collaboration}, {Aghanim}, N., {Akrami}, Y., {et~al.} 2020, \aap, 641,
  A6, \dodoi{10.1051/0004-6361/201833910}

\bibitem[{{Pozzetti} {et~al.}(2010){Pozzetti}, {Bolzonella}, {Zucca},
  {Zamorani}, {Lilly}, {Renzini}, {Moresco}, {Mignoli}, {Cassata}, {Tasca},
  {Lamareille}, {Maier}, {Meneux}, {Halliday}, {Oesch}, {Vergani}, {Caputi},
  {Kova{\v{c}}}, {Cimatti}, {Cucciati}, {Iovino}, {Peng}, {Carollo}, {Contini},
  {Kneib}, {Le F{\'e}vre}, {Mainieri}, {Scodeggio}, {Bardelli}, {Bongiorno},
  {Coppa}, {de la Torre}, {de Ravel}, {Franzetti}, {Garilli}, {Kampczyk},
  {Knobel}, {Le Borgne}, {Le Brun}, {Pell{\`o}}, {Perez Montero},
  {Ricciardelli}, {Silverman}, {Tanaka}, {Tresse}, {Abbas}, {Bottini}, {Cappi},
  {Guzzo}, {Koekemoer}, {Leauthaud}, {Maccagni}, {Marinoni}, {McCracken},
  {Memeo}, {Porciani}, {Scaramella}, {Scarlata}, \&
  {Scoville}}]{Pozzetti-etal-2010}
{Pozzetti}, L., {Bolzonella}, M., {Zucca}, E., {et~al.} 2010, \aap, 523, A13,
  \dodoi{10.1051/0004-6361/200913020}

\bibitem[{{Ren} {et~al.}(2024){Ren}, {Liu}, {Li}, {Cui}, {Zhao}, {Li}, {Song},
  {Yesuf}, \& {Zheng}}]{Ren-etal-2024}
{Ren}, J., {Liu}, F.~S., {Li}, N., {et~al.} 2024, \apj, 969, 4,
  \dodoi{10.3847/1538-4357/ad4117}

\bibitem[{{Romano-D{\'\i}az} {et~al.}(2008){Romano-D{\'\i}az}, {Shlosman},
  {Heller}, \& {Hoffman}}]{Romano-Diaz-etal-2008}
{Romano-D{\'\i}az}, E., {Shlosman}, I., {Heller}, C., \& {Hoffman}, Y. 2008,
  \apjl, 687, L13, \dodoi{10.1086/593168}

\bibitem[{{Rosas-Guevara} {et~al.}(2020){Rosas-Guevara}, {Bonoli}, {Dotti},
  {Zana}, {Nelson}, {Pillepich}, {Ho}, {Izquierdo-Villalba}, {Hernquist}, \&
  {Pakmor}}]{Rosas-Guevara-etal-2020}
{Rosas-Guevara}, Y., {Bonoli}, S., {Dotti}, M., {et~al.} 2020, \mnras, 491,
  2547, \dodoi{10.1093/mnras/stz3180}

\bibitem[{{Rosas-Guevara} {et~al.}(2022){Rosas-Guevara}, {Bonoli}, {Dotti},
  {Izquierdo-Villalba}, {Lupi}, {Zana}, {Bonetti}, {Nelson}, {Springel},
  {Hernquist}, \& {Vogelsberger}}]{Rosas-Guevara-etal-2022}
---. 2022, \mnras, 512, 5339, \dodoi{10.1093/mnras/stac816}

\bibitem[{{Saha} \& {Naab}(2013)}]{Saha-Naab2013}
{Saha}, K., \& {Naab}, T. 2013, \mnras, 434, 1287,
  \dodoi{10.1093/mnras/stt1088}

\bibitem[{{Sakamoto} {et~al.}(1999){Sakamoto}, {Okumura}, {Ishizuki}, \&
  {Scoville}}]{Sakamoto-etal-1999}
{Sakamoto}, K., {Okumura}, S.~K., {Ishizuki}, S., \& {Scoville}, N.~Z. 1999,
  \apj, 525, 691, \dodoi{10.1086/307910}

\bibitem[{{Scannapieco} \& {Athanassoula}(2012)}]{Scannapieco-Athanassoula2012}
{Scannapieco}, C., \& {Athanassoula}, E. 2012, \mnras, 425, L10,
  \dodoi{10.1111/j.1745-3933.2012.01291.x}

\bibitem[{{Sellwood}(2016)}]{Sellwood2016}
{Sellwood}, J.~A. 2016, \apj, 819, 92, \dodoi{10.3847/0004-637X/819/2/92}

\bibitem[{{Shen} \& {Sellwood}(2004)}]{Shen-Sellwood2004}
{Shen}, J., \& {Sellwood}, J.~A. 2004, \apj, 604, 614, \dodoi{10.1086/382124}

\bibitem[{{Sheth} {et~al.}(2012){Sheth}, {Melbourne}, {Elmegreen}, {Elmegreen},
  {Athanassoula}, {Abraham}, \& {Weiner}}]{Sheth-etal-2012}
{Sheth}, K., {Melbourne}, J., {Elmegreen}, D.~M., {et~al.} 2012, \apj, 758,
  136, \dodoi{10.1088/0004-637X/758/2/136}

\bibitem[{{Sheth} {et~al.}(2008){Sheth}, {Elmegreen}, {Elmegreen}, {Capak},
  {Abraham}, {Athanassoula}, {Ellis}, {Mobasher}, {Salvato}, {Schinnerer},
  {Scoville}, {Spalsbury}, {Strubbe}, {Carollo}, {Rich}, \&
  {West}}]{Sheth-etal-2008}
{Sheth}, K., {Elmegreen}, D.~M., {Elmegreen}, B.~G., {et~al.} 2008, \apj, 675,
  1141, \dodoi{10.1086/524980}

\bibitem[{{Shlosman} {et~al.}(1989){Shlosman}, {Frank}, \&
  {Begelman}}]{Shlosman-etal-1989}
{Shlosman}, I., {Frank}, J., \& {Begelman}, M.~C. 1989, \nat, 338, 45,
  \dodoi{10.1038/338045a0}

\bibitem[{{Simmons} {et~al.}(2014){Simmons}, {Melvin}, {Lintott}, {Masters},
  {Willett}, {Keel}, {Smethurst}, {Cheung}, {Nichol}, {Schawinski},
  {Rutkowski}, {Kartaltepe}, {Bell}, {Casteels}, {Conselice}, {Almaini},
  {Ferguson}, {Fortson}, {Hartley}, {Kocevski}, {Koekemoer}, {McIntosh},
  {Mortlock}, {Newman}, {Ownsworth}, {Bamford}, {Dahlen}, {Faber},
  {Finkelstein}, {Fontana}, {Galametz}, {Grogin}, {Gr{\"u}tzbauch}, {Guo},
  {H{\"a}u{\ss}ler}, {Jek}, {Kaviraj}, {Lucas}, {Peth}, {Salvato}, {Wiklind},
  \& {Wuyts}}]{Simmons-etal-2014}
{Simmons}, B.~D., {Melvin}, T., {Lintott}, C., {et~al.} 2014, \mnras, 445,
  3466, \dodoi{10.1093/mnras/stu1817}

\bibitem[{{Skibba} {et~al.}(2012){Skibba}, {Masters}, {Nichol}, {Zehavi},
  {Hoyle}, {Edmondson}, {Bamford}, {Cardamone}, {Keel}, {Lintott}, \&
  {Schawinski}}]{Skibba-etal-2012}
{Skibba}, R.~A., {Masters}, K.~L., {Nichol}, R.~C., {et~al.} 2012, \mnras, 423,
  1485, \dodoi{10.1111/j.1365-2966.2012.20972.x}

\bibitem[{{Smail} {et~al.}(2023){Smail}, {Dudzevi{\v{c}}i{\={u}}t{\.{e}}},
  {Gurwell}, {Fazio}, {Willner}, {Swinbank}, {Arumugam}, {Summers}, {Cohen},
  {Jansen}, {Windhorst}, {Meena}, {Zitrin}, {Keel}, {Cheng}, {Coe},
  {Conselice}, {D'Silva}, {Driver}, {Frye}, {Grogin}, {Koekemoer}, {Marshall},
  {Nonino}, {Pirzkal}, {Robotham}, {Rutkowski}, {Ryan}, {Tompkins}, {Willmer},
  {Yan}, {Broadhurst}, {Diego}, {Kamieneski}, \& {Yun}}]{Smail-etal-2024}
{Smail}, I., {Dudzevi{\v{c}}i{\={u}}t{\.{e}}}, U., {Gurwell}, M., {et~al.}
  2023, \apj, 958, 36, \dodoi{10.3847/1538-4357/acf931}

\bibitem[{{Smethurst} {et~al.}(2019){Smethurst}, {Simmons}, {Lintott}, \&
  {Shanahan}}]{Smethurst-etal-2019}
{Smethurst}, R.~J., {Simmons}, B.~D., {Lintott}, C.~J., \& {Shanahan}, J. 2019,
  \mnras, 489, 4016, \dodoi{10.1093/mnras/stz2443}

\bibitem[{{Spinoso} {et~al.}(2017){Spinoso}, {Bonoli}, {Dotti}, {Mayer},
  {Madau}, \& {Bellovary}}]{Spinoso-etal-2017}
{Spinoso}, D., {Bonoli}, S., {Dotti}, M., {et~al.} 2017, \mnras, 465, 3729,
  \dodoi{10.1093/mnras/stw2934}

\bibitem[{{Springel}(2010)}]{Springel-2010}
{Springel}, V. 2010, \mnras, 401, 791, \dodoi{10.1111/j.1365-2966.2009.15715.x}

\bibitem[{{Springel} {et~al.}(2018){Springel}, {Pakmor}, {Pillepich},
  {Weinberger}, {Nelson}, {Hernquist}, {Vogelsberger}, {Genel}, {Torrey},
  {Marinacci}, \& {Naiman}}]{Springel-etal-2018}
{Springel}, V., {Pakmor}, R., {Pillepich}, A., {et~al.} 2018, \mnras, 475, 676,
  \dodoi{10.1093/mnras/stx3304}

\bibitem[{{Suess} {et~al.}(2022){Suess}, {Bezanson}, {Nelson}, {Setton},
  {Price}, {Dokkum}, {Brammer}, {Labb{\'e}}, {Leja}, {Miller}, {Robertson},
  {Wel}, {Weaver}, \& {Whitaker}}]{Suess-etal-2022}
{Suess}, K.~A., {Bezanson}, R., {Nelson}, E.~J., {et~al.} 2022, \apjl, 937,
  L33, \dodoi{10.3847/2041-8213/ac8e06}

\bibitem[{{Tamburri} {et~al.}(2014){Tamburri}, {Saracco}, {Longhetti},
  {Gargiulo}, {Lonoce}, \& {Ciocca}}]{Tamburri-etal-2014}
{Tamburri}, S., {Saracco}, P., {Longhetti}, M., {et~al.} 2014, \aap, 570, A102,
  \dodoi{10.1051/0004-6361/201424040}

\bibitem[{{van der Wel} {et~al.}(2014){van der Wel}, {Franx}, {van Dokkum},
  {Skelton}, {Momcheva}, {Whitaker}, {Brammer}, {Bell}, {Rix}, {Wuyts},
  {Ferguson}, {Holden}, {Barro}, {Koekemoer}, {Chang}, {McGrath},
  {H{\"a}ussler}, {Dekel}, {Behroozi}, {Fumagalli}, {Leja}, {Lundgren},
  {Maseda}, {Nelson}, {Wake}, {Patel}, {Labb{\'e}}, {Faber}, {Grogin}, \&
  {Kocevski}}]{Van-ver-Wel-etal-2014}
{van der Wel}, A., {Franx}, M., {van Dokkum}, P.~G., {et~al.} 2014, \apj, 788,
  28, \dodoi{10.1088/0004-637X/788/1/28}

\bibitem[{{Ward} {et~al.}(2024){Ward}, {de la Vega}, {Mobasher}, {McGrath},
  {Iyer}, {Calabr{\`o}}, {Costantin}, {Dickinson}, {Holwerda},
  {Huertas-Company}, {Hirschmann}, {Lucas}, {Pandya}, {Wilkins}, {Yung},
  {Arrabal Haro}, {Bagley}, {Finkelstein}, {Kartaltepe}, {Koekemoer},
  {Papovich}, \& {Pirzkal}}]{Ward-etal-2024}
{Ward}, E., {de la Vega}, A., {Mobasher}, B., {et~al.} 2024, \apj, 962, 176,
  \dodoi{10.3847/1538-4357/ad20ed}

\bibitem[{{Weinberg} \& {Katz}(2007)}]{Weinberg-etal-2007}
{Weinberg}, M.~D., \& {Katz}, N. 2007, \mnras, 375, 460,
  \dodoi{10.1111/j.1365-2966.2006.11307.x}

\bibitem[{{Weinzirl} {et~al.}(2009){Weinzirl}, {Jogee}, {Khochfar}, {Burkert},
  \& {Kormendy}}]{Weinzirl-etal-2009}
{Weinzirl}, T., {Jogee}, S., {Khochfar}, S., {Burkert}, A., \& {Kormendy}, J.
  2009, \apj, 696, 411, \dodoi{10.1088/0004-637X/696/1/411}

\bibitem[{{Wisnioski} {et~al.}(2019){Wisnioski}, {F{\"o}rster Schreiber},
  {Fossati}, {Mendel}, {Wilman}, {Genzel}, {Bender}, {Wuyts}, {Davies},
  {{\"U}bler}, {Bandara}, {Beifiori}, {Belli}, {Brammer}, {Chan}, {Davies},
  {Fabricius}, {Galametz}, {Lang}, {Lutz}, {Nelson}, {Momcheva}, {Price},
  {Rosario}, {Saglia}, {Seitz}, {Shimizu}, {Tacconi}, {Tadaki}, {van Dokkum},
  \& {Wuyts}}]{Wisnioski-etal-2019}
{Wisnioski}, E., {F{\"o}rster Schreiber}, N.~M., {Fossati}, M., {et~al.} 2019,
  \apj, 886, 124, \dodoi{10.3847/1538-4357/ab4db8}

\bibitem[{{Wozniak} {et~al.}(1995){Wozniak}, {Friedli}, {Martinet}, {Martin},
  \& {Bratschi}}]{Wozniak-etal-1995}
{Wozniak}, H., {Friedli}, D., {Martinet}, L., {Martin}, P., \& {Bratschi}, P.
  1995, \aaps, 111, 115

\bibitem[{{Zhou} {et~al.}(2015){Zhou}, {Cao}, \& {Wu}}]{Zhou-etal-2015}
{Zhou}, Z.-M., {Cao}, C., \& {Wu}, H. 2015, \aj, 149, 1,
  \dodoi{10.1088/0004-6256/149/1/1}

\end{thebibliography}




\end{document}